\documentclass[journal]{IEEEtran}

\usepackage{multirow}
\usepackage{subfigure}
\usepackage{xcolor}
\usepackage{cite}
\usepackage[pdftex]{graphicx}
\usepackage{amsmath}
\usepackage{todonotes}

\ifCLASSINFOpdf

\else
  
\fi

\begin{document}

\title{Neuro-memristive Circuits for Edge Computing: A Review}
 
\author{Olga Krestinskaya,~\IEEEmembership{Student Member,~IEEE,}
             Alex~James,~\IEEEmembership{Senior~Member,~IEEE}
          and~Leon O. Chua,~\IEEEmembership{Fellow,~IEEE}   
\thanks{O. Krestinskaya is a graduate student and A.P. James is a chair in Electrical and Computer engineering at Nazarbayev University. Email: apj@ieee.org}
\thanks{L.O Chua is Professor Emeritus in Electrical Engineering and Computer Sciences at University of California Berkeley.}

}


\maketitle

\begin{abstract}
The volume, veracity, variability and velocity of data produced from the ever increasing network of sensors connected to Internet pose challenges for power management, scalability and sustainability of cloud computing infrastructure. Increasing the data processing capability of edge computing devices at lower power requirements can reduce several overheads for cloud computing solutions.  This paper provides the review of neuromorphic CMOS-memristive architectures that can be integrated into edge computing devices. We discuss why the neuromorphic architectures are useful for edge devices and show the advantages, drawbacks and open problems in the field of neuro-memristive circuits for edge computing.
\end{abstract}


%
\IEEEpeerreviewmaketitle

\section{Introduction}

%
%
%
%
\IEEEPARstart{T}{he} increase in the number of edge devices such as mobile phones, and wearable electronics connected to Internet drives the scale-up of intelligent data applications. Edge computing is broadly defined as the method used for moving the control of data processing from centralized core computing nodes such as high-performance computing servers to the last edge nodes of the Internet where data is collected and connected to the physical world \cite{satyanarayanan2017emergence, sheltami2018fog}. The high velocity and volume of data generated lead to the need to scale up data centers, and puts added pressure on lowering energy consumption. However, the inability to linearly scale power with existing CMOS technology prompts us to look at neuromorphic computing architectures that can be used in edge devices and possibly useful for replacing hardware in cloud computing platforms. It is expected that in 2-5 years the edge computing technologies will be in the main stream \cite{panetta2017top}, along with machine learning, Internet of Things (IoT) and smart electronics, mutually contributing to each other areas growth \cite{8345562,8289317}.

The development of the neuro-memristive circuits that can be integrated to edge computing devices is an open research problem. Neuromorphic computing is inspired from the biological concepts of human brain processing that has the potential to replace traditional von Neumann computing paradigms. In the more than Moore's era of device scale-up and architectures, memristive circuits and computing architectures is one of the promising solutions \cite{8119503}. Memristors provide various advantages, such as scalability, small on-chip area, low power dissipation, efficiency and adaptability \cite{chua2012hodgkin,tcad}.

In this paper, the correlation between neuromorphic memristive architectures with the edge computing trends is illustrated, we discuss the different set of neuromorphic architectures for the edge computing that can be integrated directly to the edge devices. We illustrate the most recent approaches to implement neuron cell and synaptic connections and show the correlation with biological concepts. We present the clear overview of various neuromorphic architectures, such as different types of neural networks {\cite{learning18}}, Hierarchical Temporal Memory (HTM) {\cite{8471012,9}}, Long Short-Term Memory (LSTM) {\cite{smagulova2018memristor,li2018long}}, learning architectures and circuits for memory-based computing and storage. We discuss the advantages and primary challenges in the simulation and implementation of such architectures. Also, we present the main drawbacks and challenges that should be improved in existing neuromorphic architecture to use them in edge computing applications.



The paper is organized as the following. 
{
Section II provides the overview of edge computing on hardware and edge computing architectures.
}
Section \ref{sec2} provides the review of neuron models, relates the biological concepts to the existing neuron architectures and covers the most common circuits for hardware implementation of CMOS-memristive synapses and neuron cells. Section \ref{sec3} introduces various neuromorphic architectures that can be used for edge computing application. Section \ref{sec5} illustrates the advantages, issues and open problems of the CMOS-memristive architectures. Section VI concludes the paper.





\section{Edge Devices and Emerging Neural Computing}

{Figure 1 shows the overall concept of the edge computing system. The sensors in the edges of the concept map collect the data for processing in the edge devices, which in essence move part of information processing and computing tasks from cloud to edge devices. The increased demand on the edge devices to process information in intelligent and useful ways triggers the development of emerging hardware and edge AI computing.
}

{The real-time data produced by the ever increasing number of sensors in edge devices pushes for near-sensor computing for various intelligent information processing applications. There is an emerging market of artificial intelligence chips in edge devices for utilizing machine learning and neural networks \cite{appleinsider_2018,kirin980}.  The information from the sensor is converted to digital domain by analog to digital converter, followed by filtering methods and co-processors for implementing different neural network configurations \cite{appleinsider_2018,kirin980}.  However, with major issues in scaling the devices to sub-10nm range, emerging devices such as memristors become promising to increase the speed and on-chip area. Further, these emerging devices also promote the analog domain processing of information as many neural networks in hardware can be mapped to memristive array based computing architectures \cite{schuman2017survey}.}

{Mobile devices largely has driven the growth in high-performance logic and low-power digital logic chips in the last several years. And the limitation and challenges in device scaling has forced the community to move towards  neural computing solutions that can incorporate more than Moore's law \cite{kahng2010scaling} and beyond CMOS technologies \cite{bohr2017cmos, hu2018cross} as a key aspect of future hardware development. In edge devices, such as mobiles, the key computing drivers are towards having higher performance and more functionality at lower cost and energy which is constrained by battery. Several hardware technology aspects drives this development, they are: Logic technologies, Ground rule scaling, Performance boosters, Performance-power-area (PPA) scaling, 3D integration,  Memory technologies, DRAM technologies, Flash technologies and Emerging non-volatile-memory (NVM) technologies such as memristors. The key performance benchmarks for node scaling for edge devices in more than Moore's era integration in the next 2-3 years includes \cite{irds2017}: (1) increasing the operating frequency by 15\% relative to the scaled supply voltage, (2) for a given performance reduce the energy per switching by 35\%, (3) reduce the area on chip footprint by 35\%, and (4) reduce the scaled die cost by 20\%, while limit wafer cost to increase within 30\%.}

\begin{figure}[!t]
    \centering        
    \includegraphics[width=90mm]{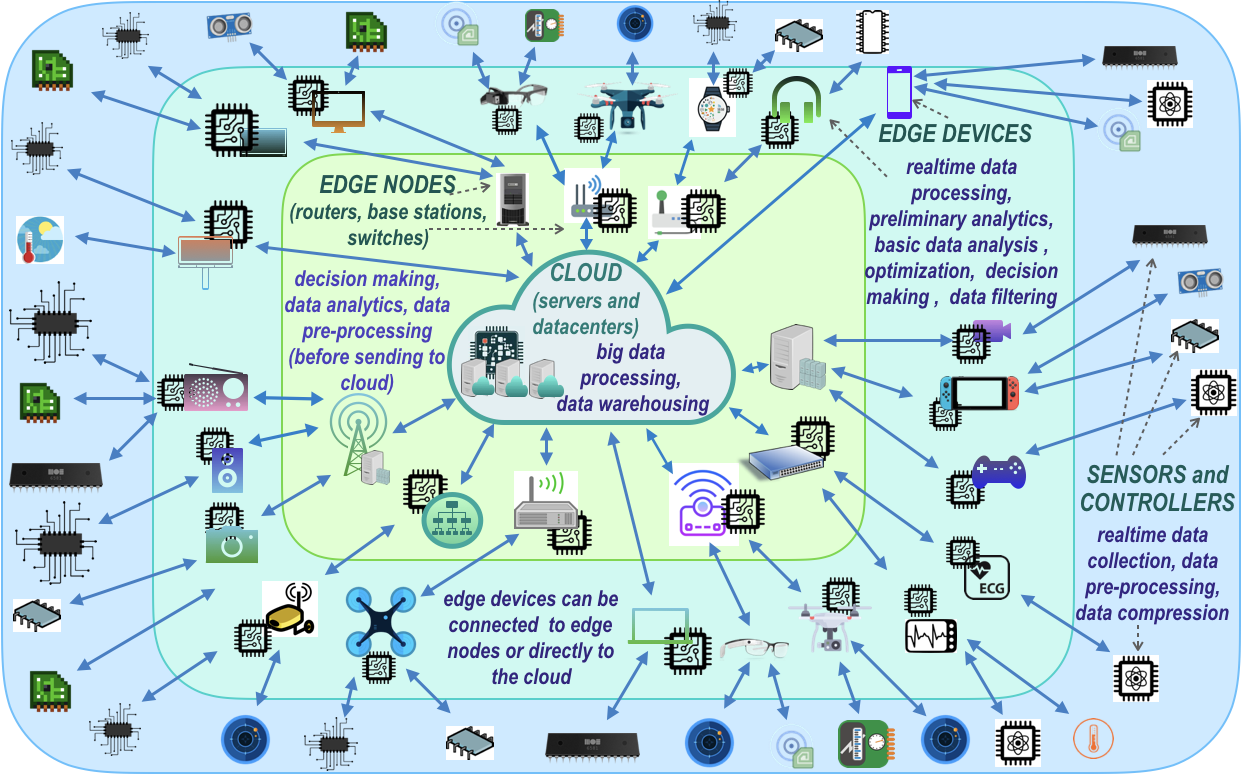}
    \caption{{Overall concept of edge computing system.}}
    \label{new1}
\end{figure}

{Memory and logic technologies together shape the development of near sensor neural computing solutions for edge devices. Memristive devices are emerging non-volatile memories that offer several potential features to support the growth in this field. There are several non-volatile variants of memristor devices such as  magnetic or MRAM \cite{wang2009spintronic}, phase-change or PCRAM \cite{kuzum2011nanoelectronic}, and resistive or ReRAM \cite{ho2009programmable} that can be used for building neural networks. The two-terminal resistive structure requires a selector device such as a transistor to program these devices in an array. The two common use of memristors in a neural computing paradigm is as a memory and as a dot-product computing unit \cite{kim2011functional}. The main purpose of memristor as a memory is as a storage unit for weights during the learning stages in digital or discrete analog domain processing. While, a memristor crossbar array can be used for computing the dot product between the input and weights in a neural network layer in analog domain.}

\begin{figure}[!t]
    \centering        
    \includegraphics[width=60mm]{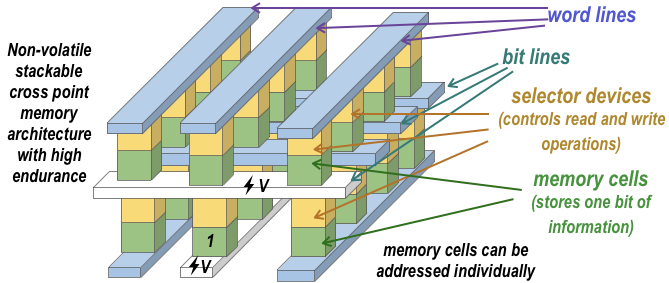}
    \caption{{3D XP Memory Architecture \cite{hady2017platform}.}}
    \label{new2}
\end{figure}

{The development of high density crossbar memristor architecture has been limited by the lack of a good and energy efficient selector device. Being a resistive device, memristors such as ReRAM require either bi- or unipoloar operation for programming to a particular state. The 3D XP memory shown in Fig. \ref{new2} \cite{hady2017platform} is been a promising direction to solve this bottleneck, and the major issue that remains is the device to device variability of the resistive state. Even with variability, the neural networks have shown robust performances, as during the learning phase, any variability in the states translates to the variability in weights, which are compensated by the learning algorithm to find the optimal set of weights that works best for the given neural network configuration.}


{The growth in hardware for edge computing is driven by Internet of things, where the sensors-humans-computers collaborate to provide efficient and useful intelligent application \cite{satyanarayanan2017emergence}. The data analysis for these application often needs to be fast, and also need to ensure security and privacy. Hardware level security is an essential advantage offered by the emerging devices \cite{jiang2018provable} that can be integrated into edge devices. The progress in NVM memristor devices and arrays due to its lower operating voltages, compatibility with CMOS devices and faster speeds allows to develop a large variety of energy and area efficient neural networks configurations\cite{maan2017survey,schuman2017survey}.}

{In the edge computing concept (Fig. 1), the data processing is shifted from the data centers to the edge devices \cite{shi2016edge}.
The edge computing relies on billions of various devices connected to the Internet. Each device collects the information and can process this data locally.
The data processed on edge level is collected in the aggregation nodes at the intermediate fog level that incorporates the networking devices, aggregation devices, and gateways required for sending processed data to the cloud data centers \cite{satyanarayanan2017emergence, sheltami2018fog}. Cloud is on the top level of the data processing containing data warehouses, which is responsible for large data processing. The edge computing is a basis for IoT systems, which incorporates the ideas of smart devices, smart vehicles, and connected systems and can be extended to a system of systems solutions involving big data analytics. The development of IoT networks and amount of data transferred and processed in cloud stresses the limits of the data centers. If the current trends continue developing at the same pace, in few decades, the amount of  energy required to process the ever growing data will overload the bandwidth requirements, and cloud computing requirements to a point that it would not be feasible to meet the demands of speed, and cost \cite{mao2016dynamic}.
}

{The main idea of edge computing is the local processing of the data, which does not require sending of significant amount of data to the servers.  All these decision making and processing mechanisms should be performed in low power levels. 
This removes the need to have complex data centers. 
The edge computing becomes more relevant because the power required for data processing in the data centers on servers increased significantly in the last few years. And if the growth of processed data continues, it would increase the costs for powering the data centers to support of same speed and amount of data processed on servers.
As all the edge devices are limited in terms of on-chip area and low power consumption requirements, the conventional von Neumann architectures with traditional CMOS devices become less feasible for such purposes in the long-term as transistor scalability is expensive and energy per computation saturates. The neuromorphic non von Neumann architectures discussed in Section \ref{sec3} are considered to be a promising solution for energy-related issues and optimization of such systems. Moreover, neuromorphic architectures can be used to solve the cloud computing energy-related issues in memory and processing units and to achieve energy-efficient computing \cite{chakma2018energy}.
}

{The distributed nature of the edge computing architectures allows to integrate neural chips as co-processing units within the edge devices. The neural chips make use of neuron models inspired from the biological understanding of neuronal behaviour and function. The neuron models are used to build different types of neural network configurations that can mimic functions and capacity of human brain. There are several neural architectures such as deep learning neural network (DNN) \cite{8060399, hasan2017chip}, convolutional neural network (CNN) \cite{7727302,7966055}, long short term memory (LSTM) \cite{smagulova2018memristor,li2018long},  hierarchical temporal memories (HTM) \cite{8471012,9} and generative adversarial networks (GAN) \cite{goodfellow2014generative} that has grown prominence in the last decade. 
}

{Edge processing often involves real-time localized data processing. Therefore, the primary goal of the edge computing is to make edge devices more intelligent, faster and less power hungry. 
Also, it is essential to consider the issues related to communication protocols, bandwidth, and correlation of data from all edge devices. It is important to make the device more intelligent and understand which information should be processed. Therefore, the learning process \cite{learning18} in neuromorphic systems is essential.}


{The memristive neuromorphic architectures aim to reduce the processing power, which allow integrating these architectures to the edge devices. The lower energy consumption increases the battery life, allows to pack more computing hardware modules and also decreases the overall cost of computation. The cost-effectiveness is achieved, because the memristor-based neuromorphic architectures require a smaller amount of memory for processing and networks can be learned to understand the information rather than storing and retrieving it using energy consuming hardware and software algorithms.
The learning process in neuromorphic architectures allows achieving faster processing time. In memristive hardware architectures, the learning process is slow, while the decision making and processing of data after learning is very fast. Once the memristive neuromorphic architecture is learned,  the information processing on local edge devices can be performed quickly. Also, the faster data processing can be achieved using analog learning architectures, which are useful for near-sensor processing. The analog neuromorphic architectures \cite{learning18} can be integrated directly to the sensors avoiding intermediate data conversion stage.}

{The data security issues are addressed in memristive neuromorphic architectures because the information processing is performed at a hardware level, where encryption level is high \cite{jiang2018provable, abunahla2018memristor}. There are growing incidents for hacking on chip data in digital hardware \cite{trust}. Neuromorphic architectures encode the data, and the memristive weights are learned, so it is impossible to predict the weights. Therefore, the natural encoding process is performed, and the system is more secured. 
The neuromorphic architectures are more robust to variations because of the learning process and weight adjustment. The interoperability can be ensured because the networks become more adaptable to the process variations in chip. Therefore, decision fusion and collaborative sensing also can be performed between the chips to ensure higher security levels. In addition, memristor based key generators can be incorporated into the chip to implement functional data security algorithms \cite{jiang2018provable,rajendran2012nano,mazady2015memristor, rose2013hardware, abunahla2018memristor}.}

\section{Neuron models}

\label{sec2}

In this section, we focus on the memristive models of neuron cells and synaptic connections that can be adapted and scaled for the edge computing applications. 

\subsection{Inspiration from biological concepts}

\begin{figure*}[!t]
    \centering        
    \subfigure[]
    	{
    \includegraphics[width=40mm]{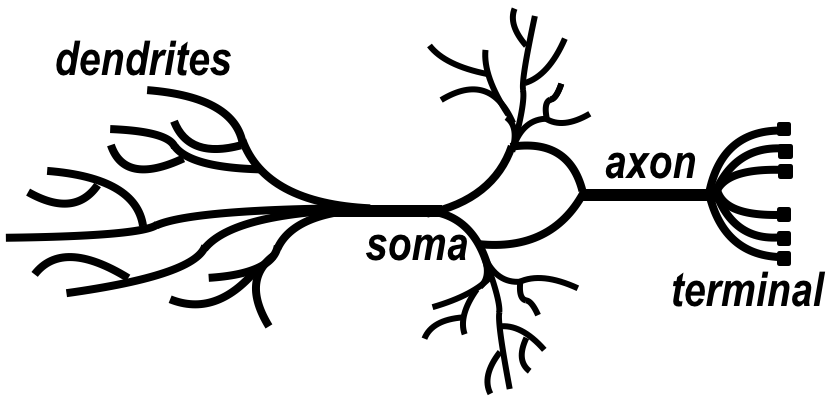}
		}       
     \subfigure[]
		{    	\includegraphics[width=30mm]{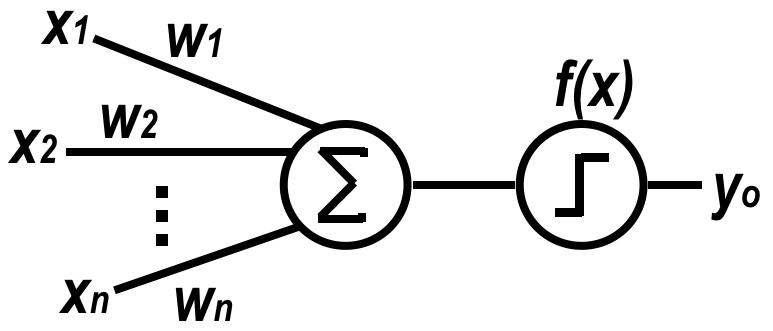}
		}
                     \subfigure[]
		{
    \includegraphics[width=42mm]{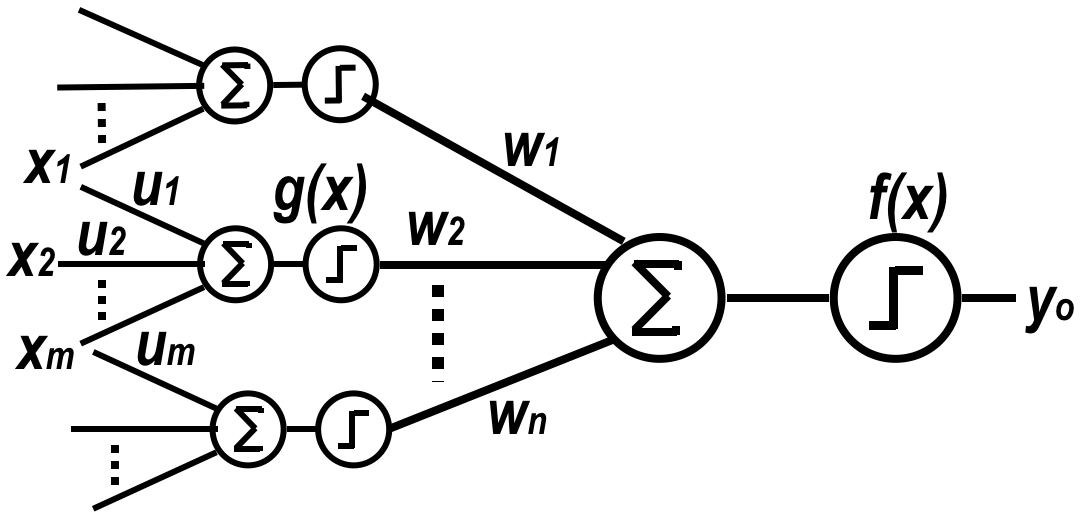}
		}
        \subfigure[]
		{
    \includegraphics[width=45mm]{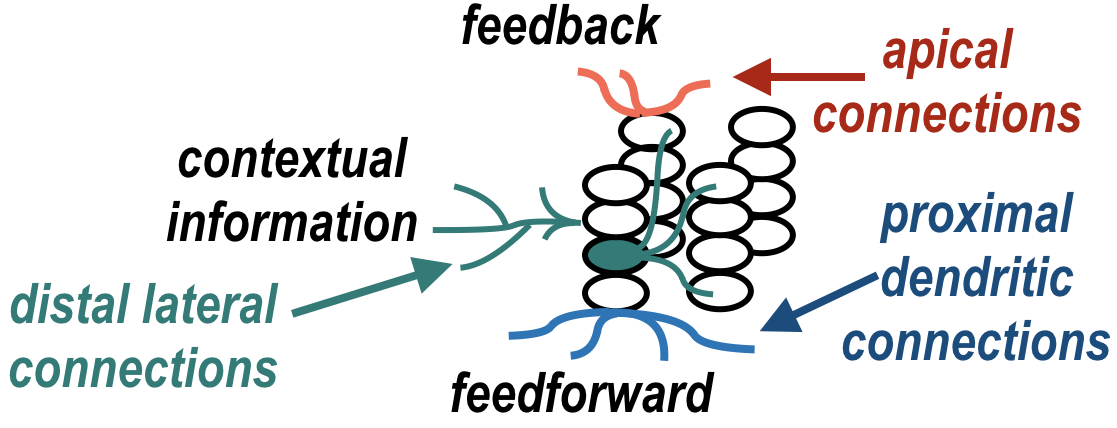}
		}  
    \caption{(a) Biological neuron \cite{tino2015artificial}, (b) threshold logic based linear neuron model \cite{6322959,tino2015artificial}, (c) dendritic threshold non-linear neuron model \cite{caze2014dendrites}, (d) HTM neuron \cite{hawkinsintelligence,hawkins2016neurons}.}
    \label{f0}
\end{figure*}

Neuromorphic circuits and architectures attempt to mimic different types of biological neural networks responsible for information processing in human brain \cite{rojas2013neural,tino2015artificial}.
The biological neuron architecture is shown in Fig. \ref{f0} (a). 
A biological neuron consists of the soma (cell body) with many dendrites that serve as connections to the other neurons and carry the information. The axon (output of the neuron) collects the information from all the dendrites and transmits it to the other neurons. The transmission of a signal from one neuron to another happens through the synapses. Synapses can either reinforce or inhibit the transmitted signals \cite{tino2015artificial}. The neuron fires (generates the output response), if the information that is collected in the axon exceeds the particular threshold \cite{6322959}. 

The equivalent structural and mathematical representation of biological neuron is shown in Fig. \ref{f0} (b) \cite{6889951} and Fig. \ref{f0} (c).
The neuron models can be divided into two categories: (1) simple threshold logic based linear neuron based models, where the neuron is presented as a most straightforward linear computing unit, and (2) dendritic threshold non-linear neuron based models, which has more complex computing units and is inspired by recent works \cite{caze2014dendrites}.

The simplest threshold logic based linear neuron model is known as McCulloch-Pitts neuron model \cite{mcculloch1943logical} and Rosenblatt`s perceptron \cite{rosenblatt1958perceptron}. Fig. \ref{f0} (b) and Eq. \ref{e1} shows the threshold logic based linear neuron model. The synapses are represented as weighted connections \cite{6322959}. 
The parameter $w_j$ represent the weights of the synapses, and $y_o$ is a neuron output.
The central concept of this model is that the weighted summation of the inputs $x_j$ is higher than the threshold $\theta$. This threshold determines the neuron firing. 
\begin{equation} 
y_o=f(\displaystyle\sum_{j=1}^{n} w_j x_j)
\label{e1}
\end{equation}
The particular case proposed in \cite{rosenblatt1958perceptron} is shown in Eq. \ref{e2}, where the hard threshold function is used an activation function.
\begin{equation}
y_o=
\begin{cases}
    +1       & \quad \displaystyle\sum_{j=1}^{n} w_j x_j \geq \theta \\
    -1  & \quad \text{otherwise}
  \end{cases}
  \label{e2}
\end{equation}

In the dendritic threshold non-linear neuron model the dendrites of the neuron can be nonlinear. Each dendritic unit in the neuron consists of various subunits (dendritic branches), and neurons are represented as complex computing unit \cite{caze2014dendrites}. Fig. \ref{f0} (c) and Eq. \ref{e3} shows the structure of non-linear dendritic neuron model. 
A single dendrite can have multiple inputs and specific threshold function. 

\begin{equation} 
y_o=f(\displaystyle\sum_{j=1}^{n} w_j g_j(\displaystyle\sum_{i=1}^{m} u_i x_i))
\label{e3}
\end{equation}
Comparing to threshold linear neuron model, like perceptron, which fails to compute particular functions, threshold non-linear neuron can compute linearly non-separable functions. 

 The volatility principle in human brain-inspired architectures is also important. The research work \cite{7568628} claims that it is of importance not only to remember important data but also forget the unnecessary information. An HTM neuron emulates this process. HTM neuron is a particular case of  dendritic threshold non-linear neuron model recently proposed to mimic functionality of pyramidal neurons \cite{polsky2009encoding} in human neocortex \cite{hawkinsintelligence,hawkins2016neurons}. 
The HTM neuron is shown in Fig, \ref{f0} (d). The neuron cell has three different inputs: feedforward, feedback, and contextual inputs. The feedforward input corresponds to the synapses of proximal soma known as proximal dendritic connections. The feedback inputs correspond to apical connections learned from the previous inputs, and the contextual inputs correspond to distal connections that connect different cells.

\subsection{Memristive circuit as a synapse}

\begin{figure*}[!t]
    \centering        
    \subfigure[]
    	{
    \includegraphics[width=24mm]{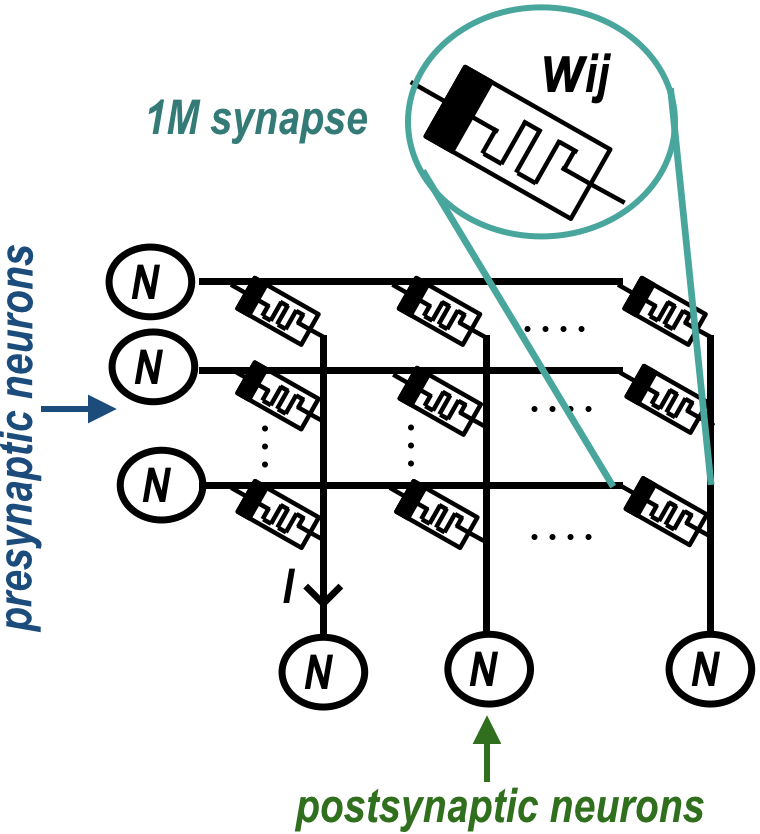}
		}       
     \subfigure[]
		{    	\includegraphics[width=23mm]{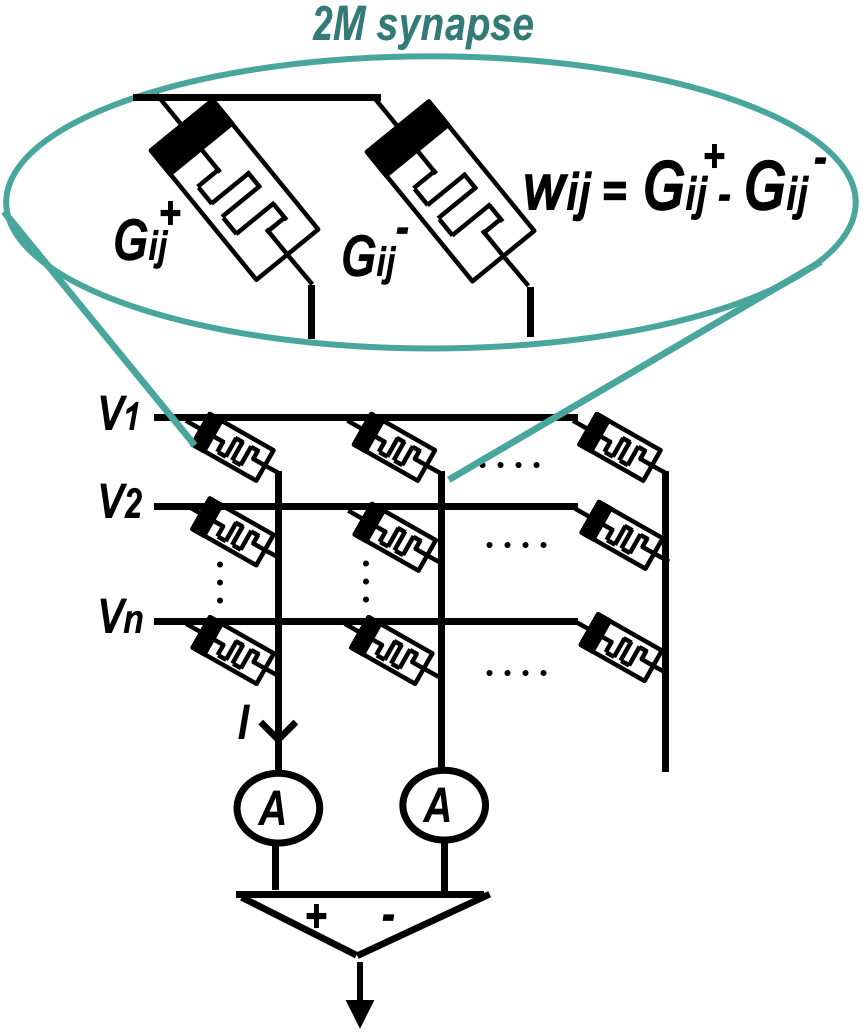}
		}
                     \subfigure[]
		{
    \includegraphics[width=21mm]{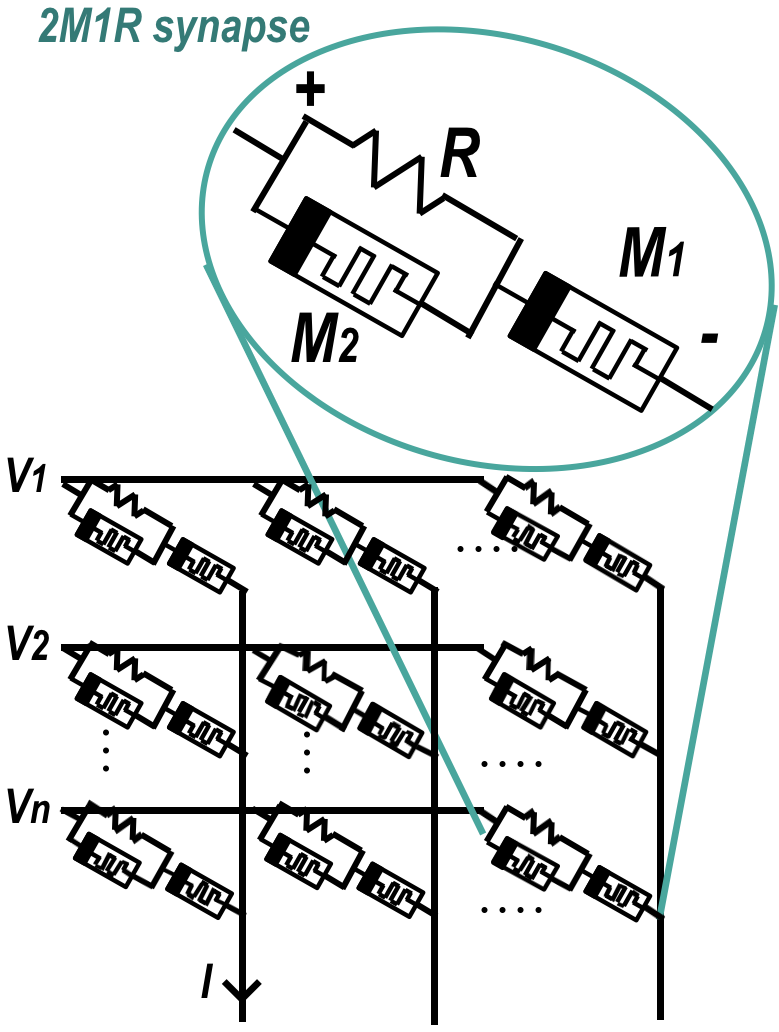}
		}
        \subfigure[]
		{
    \includegraphics[width=21mm]{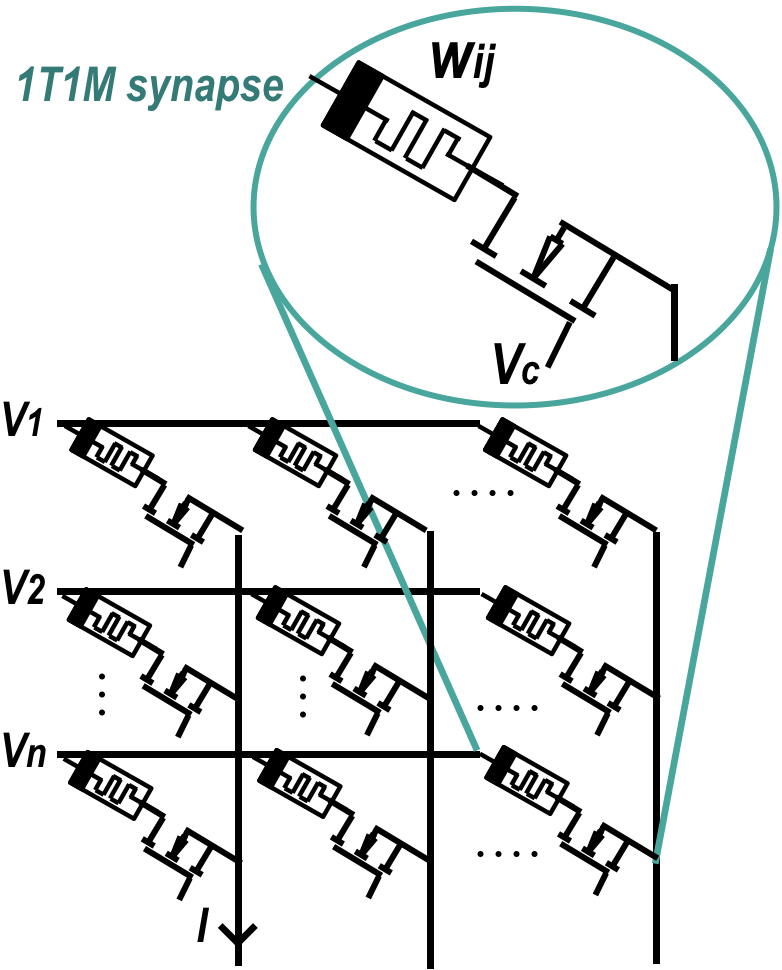}
		}  
               \subfigure[]
		{
    \includegraphics[width=21mm]{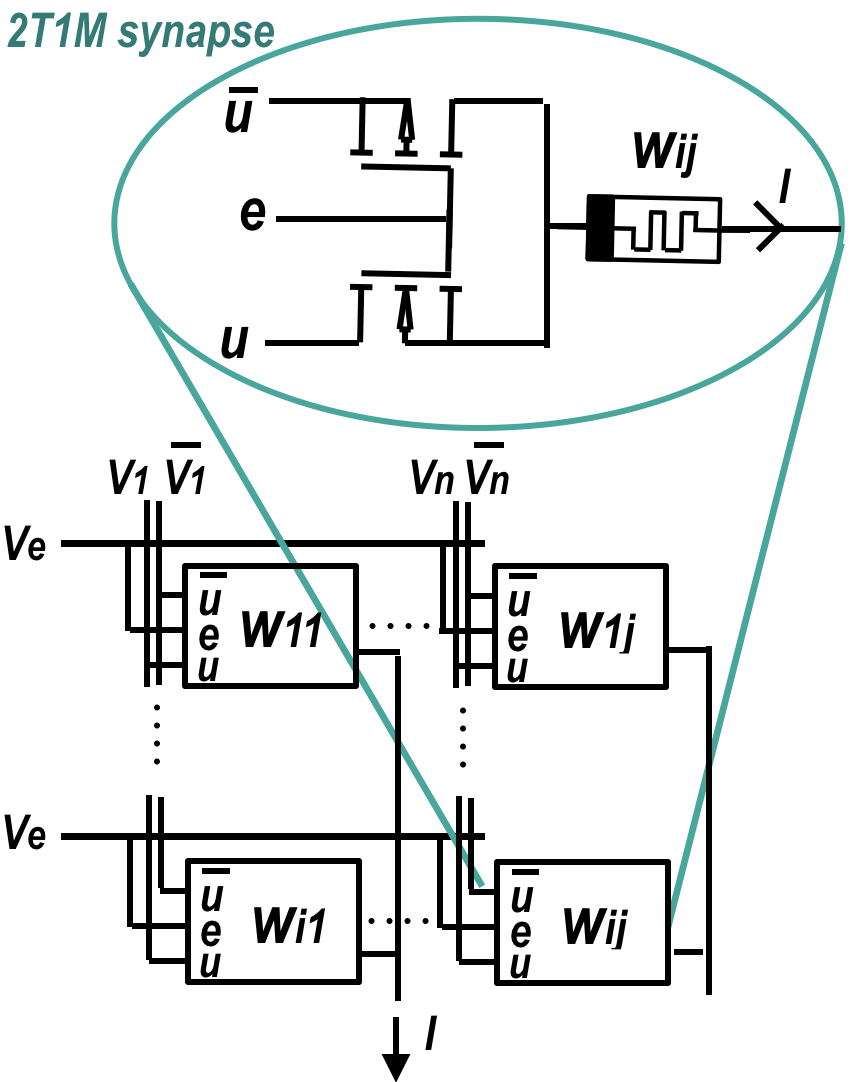}
		}
                \subfigure[]
		{
    \includegraphics[width=21mm]{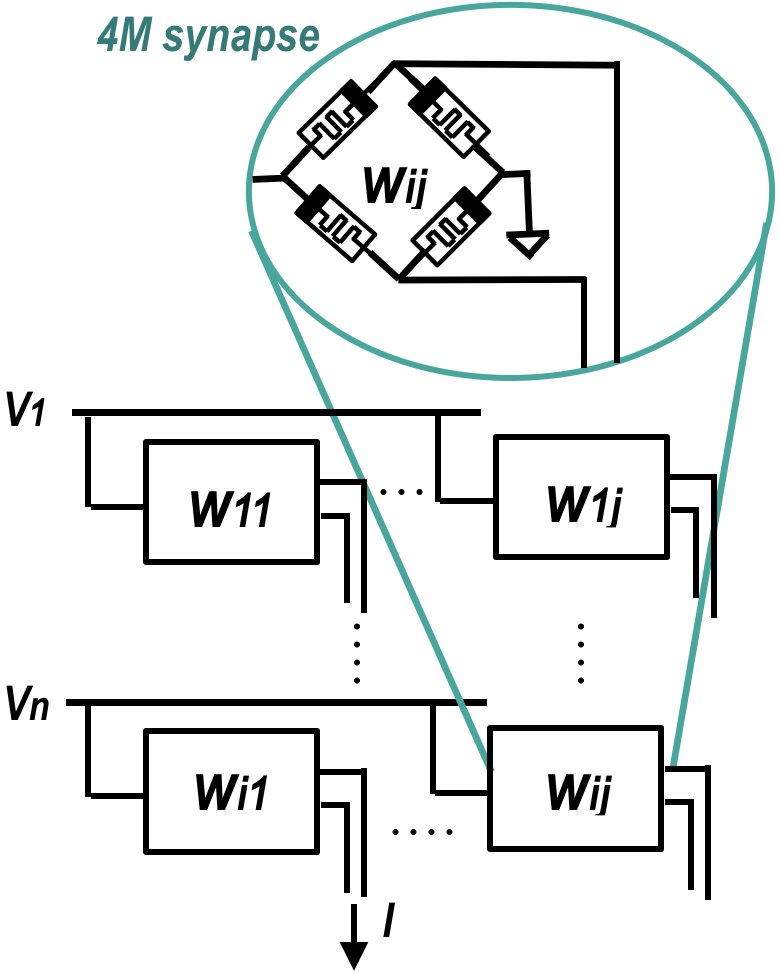}
		}
                     \subfigure[]
		{
    \includegraphics[width=21mm]{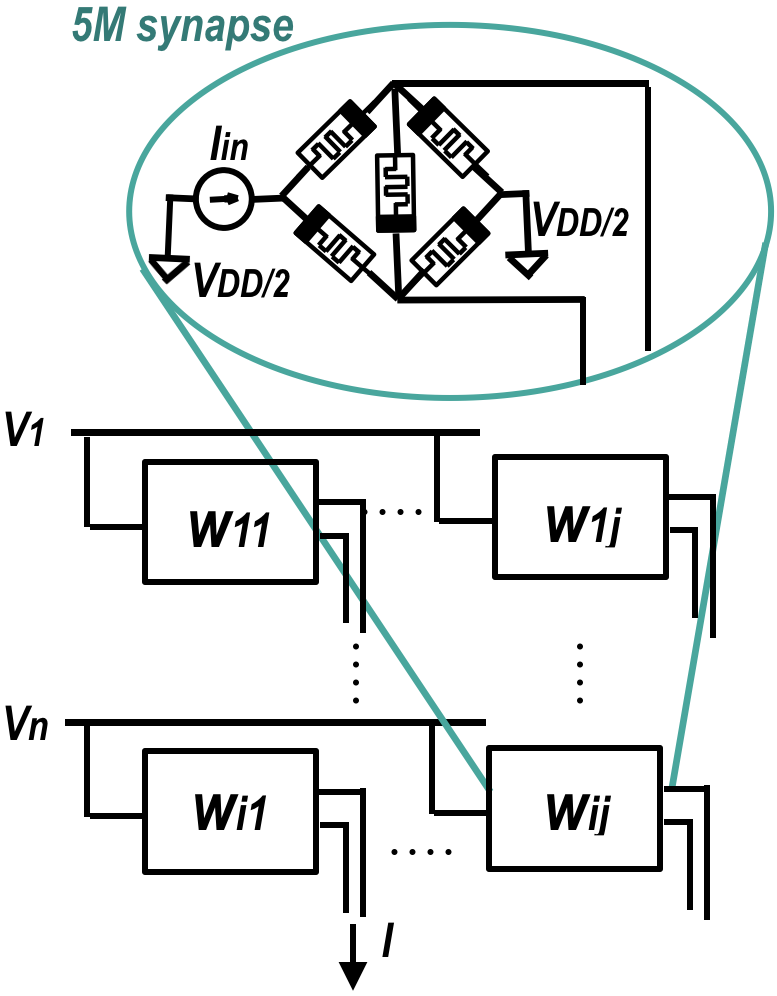}
		}
    \caption{Memristive synapses: (a) 1M synapse in a crossbar array, (b) 2M synapses \cite{prezioso2015training}, (c) 2M1R synapse \cite{7723927}, (d) 1T1M synapses \cite{yao2017face,6889951}, (e) 2T2M synapses \cite{7010034}, (f) 4M synapse \cite{6074916,6939735} and 5M synapse \cite{6331430}.}
    \label{f1}
\end{figure*}

\subsubsection{Single memristor as a synapse}
Most of the implementations of the neuron models propose to use memristor as a synapse.
The least complex representations of the synapse in memristive architectures is a single memristor (1M) structure. The single memristor synapses in a memristive crossbar array are shown in Fig. \ref{f1} (a).
The 1M structure is more efficient in terms of on-chip area and power consumption.
The recent works attempt to use 1M synapses for neural networks to avoid additional CMOS elements in the architectures \cite{zhang2017synaptic,zhang2017memristor,tcad}. However, the neuromorphic circuits with 1M synapses usually required additional control circuits and suffered from sneak path problems. Moreover, the update process of the memristor values in such structures requires complex switch circuits, which disconnect the memristors from presynaptic and postsynaptic neurons and connect the input signals used for memristor programming. Also, such configurations do not allow to obtain negative synaptic weights, and additional circuits should be involved to obtain the negative weights in neural networks.

\subsubsection{Synapses with two memristors}
The alternative to 1M synapses is the synapses with two memristors (2M) shown in Fig. \ref{f1} (b) \cite{prezioso2015training, alibart2013pattern,hasan2014enabling}. This architecture doubles the size of the crossbar and requires complex postsynaptic neurons. However, this allows implementing negative weights of the synapses.
In 2M structure the weight of the synapse is represented as $W_{ij}=G^{+}_{ij}-G^{-}_{ij}$, where $G^{\pm}_{ij}$ is an effective conductance of a memristor \cite{prezioso2015training,alibart2013pattern}. 

The alternative 2M synapse with PCMO memristors is shown in \cite{6573409}. 
{
In this particular example, memristors are connected to long-term depression (LTD) and long-term potentiation (LTP) neurons and correspond to LTD and LTP operations, which occur during particular periods of time. When the synapse is potentiated, only the LTP memristor conductance is increase, while LTP memristor remain unchanged, and vice versa. 
This allows to remove the effects of asymmetric changes of the resistance level from $R_{ON}$ to $R_{OFF}$ and $R_{OFF}$ to $R_{ON}$, avoiding abrupt changes in overall  resistance of the synapse comprised of the resistances of two devices.
}

{
The 2M synapses, where two memristors are connected in series, are presented in \cite{wang2017memristors}. In this work, the synapse is presented by two types of devices: diffusive memristor device $SiO_xNy:Ag$ (a device  based on silver nanoparticles in a dielectric film that can be used as a selector device \cite{wang2017memristors} or even neuron \cite{wang2018fully}) and drift memristor device $TaO_x$ (usual non-volatile device).  The synapse was designed to realize dynamic behavior, LTD and LTP of biological synapses.
}

The two memristor one resistor (2M1R) synapse is shown in Fig. \ref{f1} (c).
The research work \cite{7723927} proposes the modified dynamic synapse for SNN based on the two memristors and the resistor adjusted for $TaO_x$
devices, which includes temporal transformations and static weight and helps to realize the spiking behavior in large-scale simulations.

\subsubsection{Synapses with transistors}
The memristive synapses with transistors are also popular because the transistor is used as a switch, especially for read and update cycles. The synapse with one transistor and one memristor (1T1M) is shown in Fig. \ref{f1} (d) \cite{yao2017face,6889951}. This architecture is one of the possible solutions for sneak path problems. 
The synapse with two transistors and one memristor (2T1M) is illustrated in Fig. \ref{f1}
\cite{7010034,7527510}. 
{While 1T1M architecture is used to control memristor switching, program the memristor within a crossbar and eliminate sneak path problems, 2T1M also allows to control the sign of the memristor, as it is connected to two inputs: original and inverted input signal.}
The enabling signal $e$ controls the switching of the CMOS transistors. 
The transistors control the current flowing through the memristor and voltage across the memristor. The parameter $e$ represents the enable signal. If  $e=0$, the state variable of memristor does not change. If $e=V_{DD}$ or $e= - V_{DD}$, the current is flowing either through NMOS transistor or PMOS transistor, respectively. The enable signal is used to control the direction of current and to update the memristor value. 
This also allows achieving negative and positive sign of memristor weight. In this circuit, it is important to ensure that the transistor is in a linear state. The drawback of such circuit is a size of the synapse, which is appropriate for small-scale problems \cite{danial2017didactic}, and can be a critical issue for large-scale edge computing systems.

\subsubsection{Memristor bridge synapses}
The other type of synaptic weight implementations is a bridge arrangement.
The memristor-bridge synapse  with 4 memristors (4M) shown in Fig. \ref{f1} (f) was tested in various neural network architectures and applications \cite{6074916,6939735}. The circuit consists of 4 memristors that form Wheatstone bridge-like circuit and is able to represent zero, positive, and negative synaptic weights. To increase the resistance of $M_2$ and $M_3$ and decrease of resistance of $M_1$ and $M_4$, positive pulse should be applied as an input and vice versa. The weight is positive, if $\frac{M_2}{M_1} > \frac{M_4}{M_3}$.  The negative weight can be formed as $\frac{M_2}{M_1} < \frac{M_4}{M_3}$. A zero weight is formed as $\frac{M_2}{M_1} = \frac{M_4}{M_3}$. 
 This ensures the implementation of positive and negative weights and allows to change the weight sign, which depends on the direction of the current.
 
\subsection{Neuron cell models}


\begin{figure*}[!t]
    \centering        
    \subfigure[]
    	{
    \includegraphics[width=30mm]{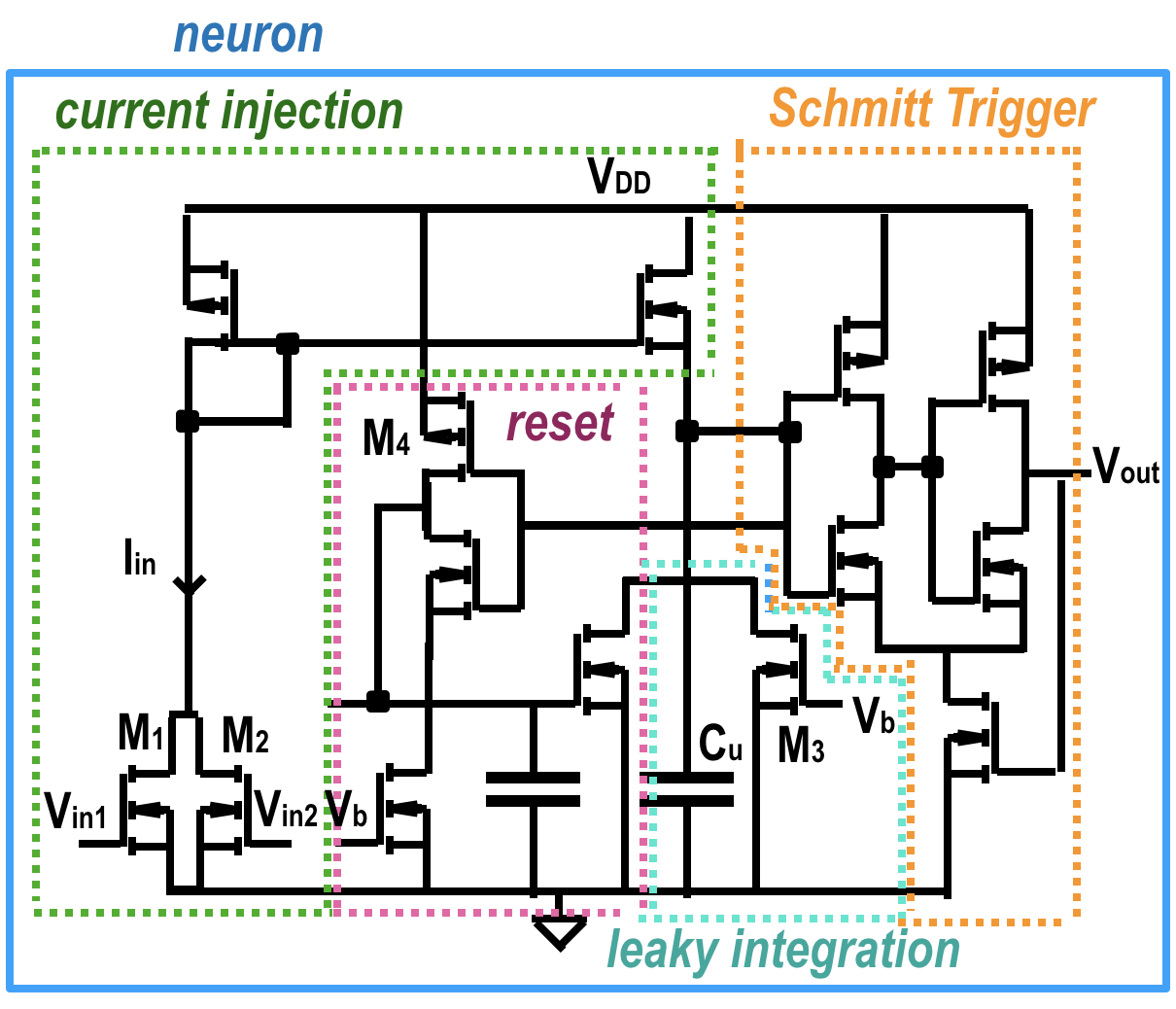}
		}       
		        \subfigure[]
		{
    \includegraphics[width=16.5mm]{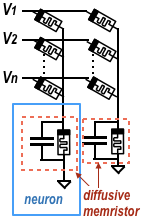}
		}
     \subfigure[]
		{    	\includegraphics[width=30mm]{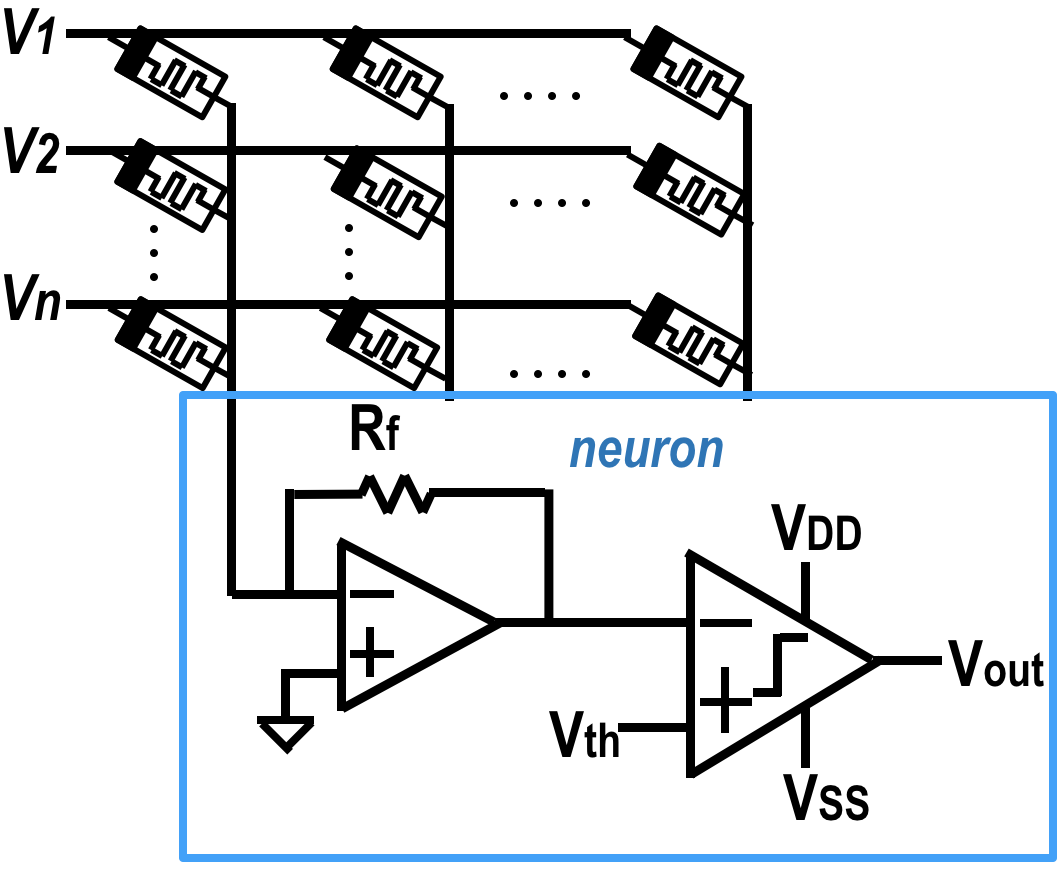}
		}
                     \subfigure[]
		{
    \includegraphics[width=35mm]{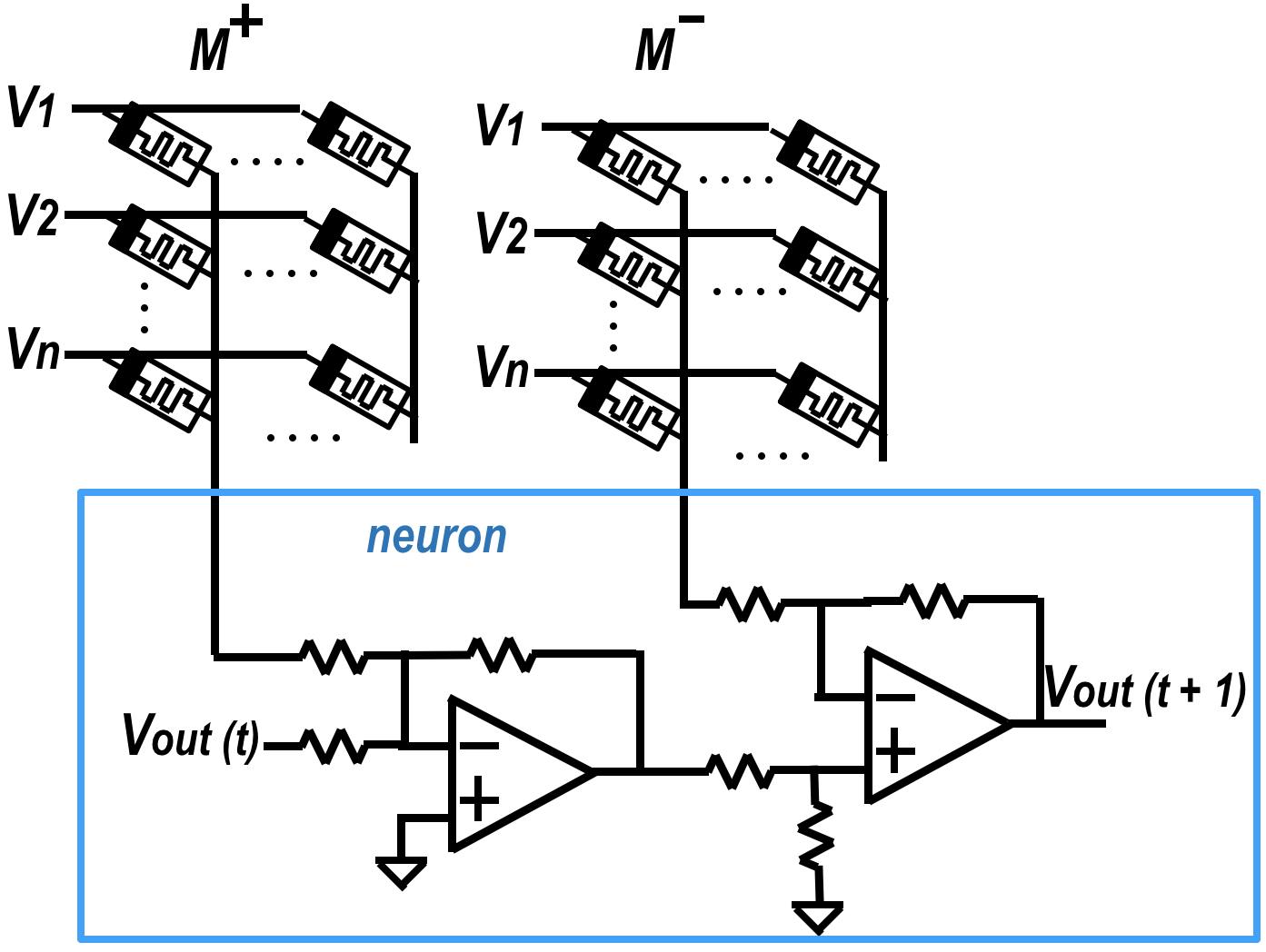}
		}
        \subfigure[]
		{
    \includegraphics[width=28mm]{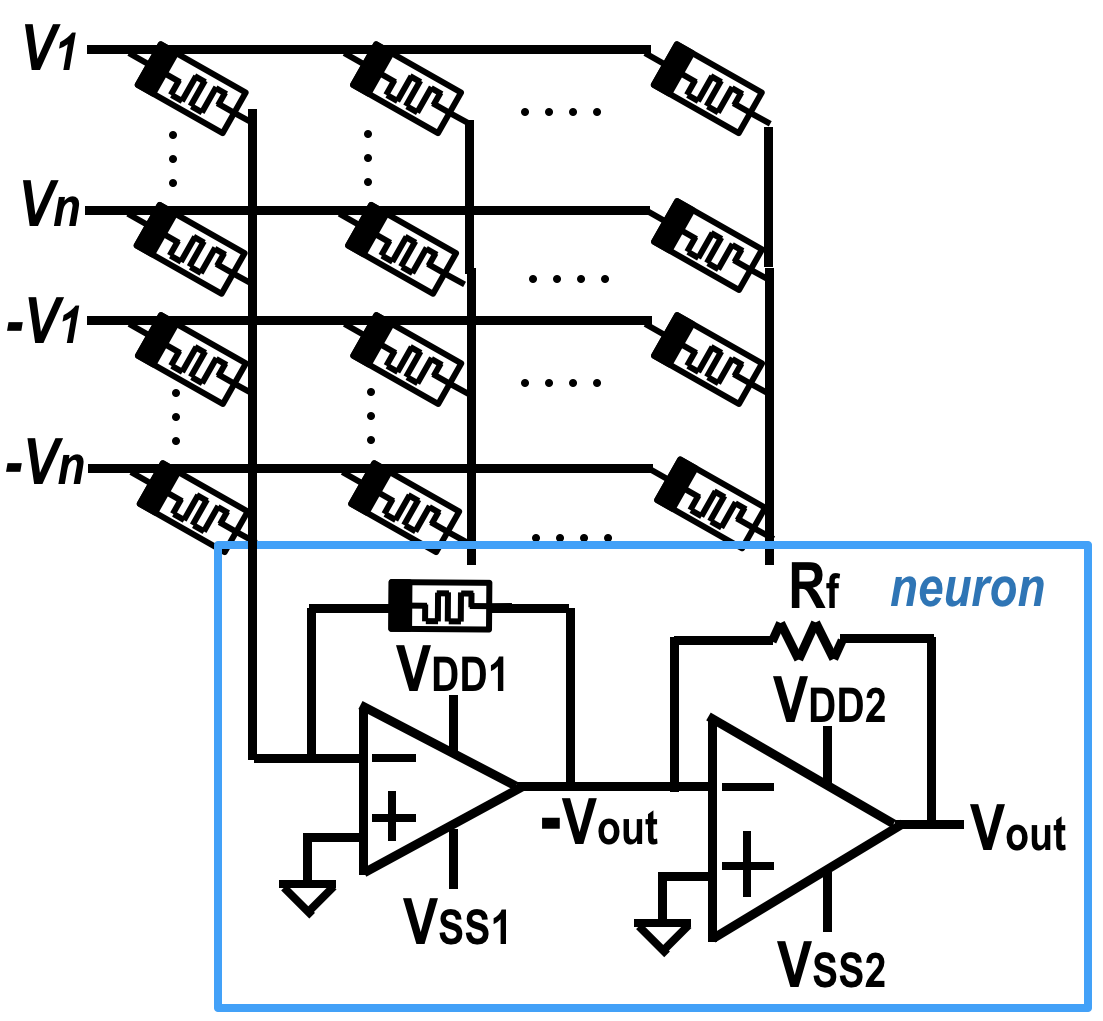}
		}  
                \subfigure[]
		{
    \includegraphics[width=30mm]{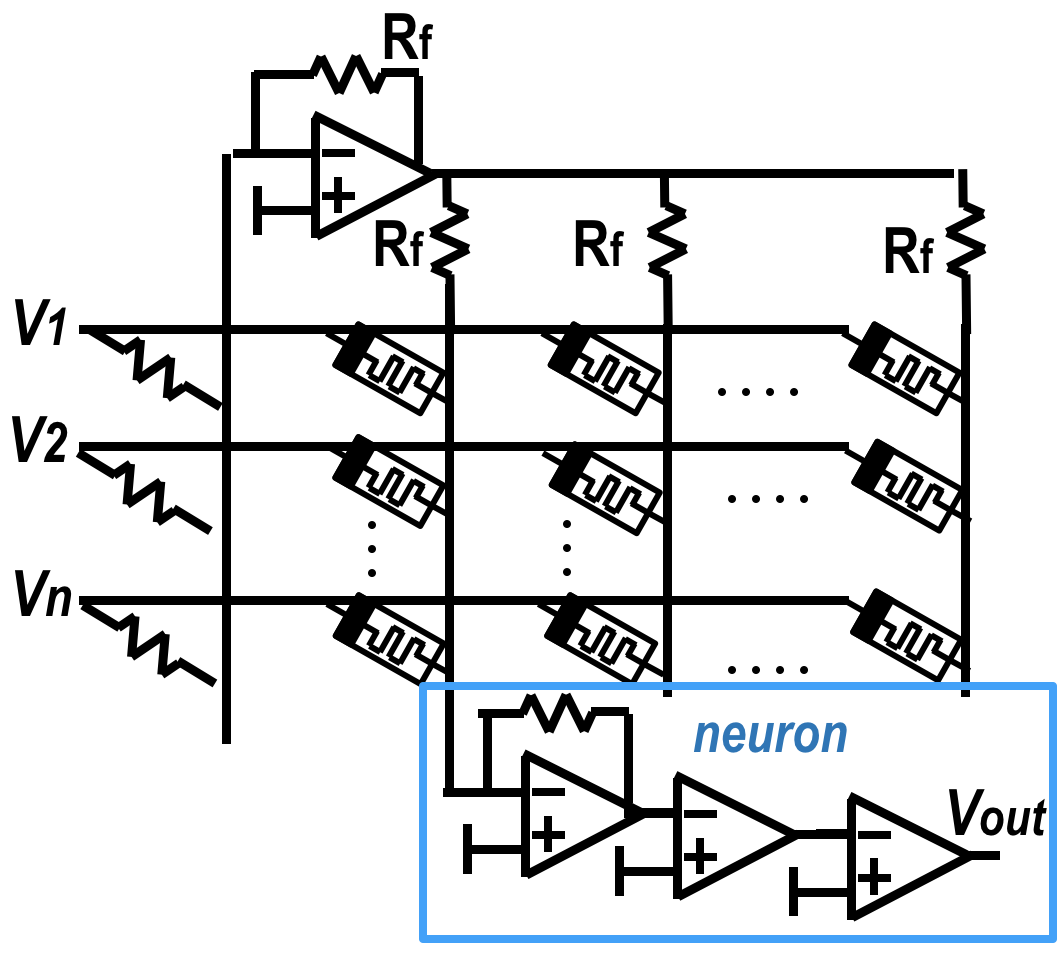}
		}
                     \subfigure[]
		{
    \includegraphics[width=30mm]{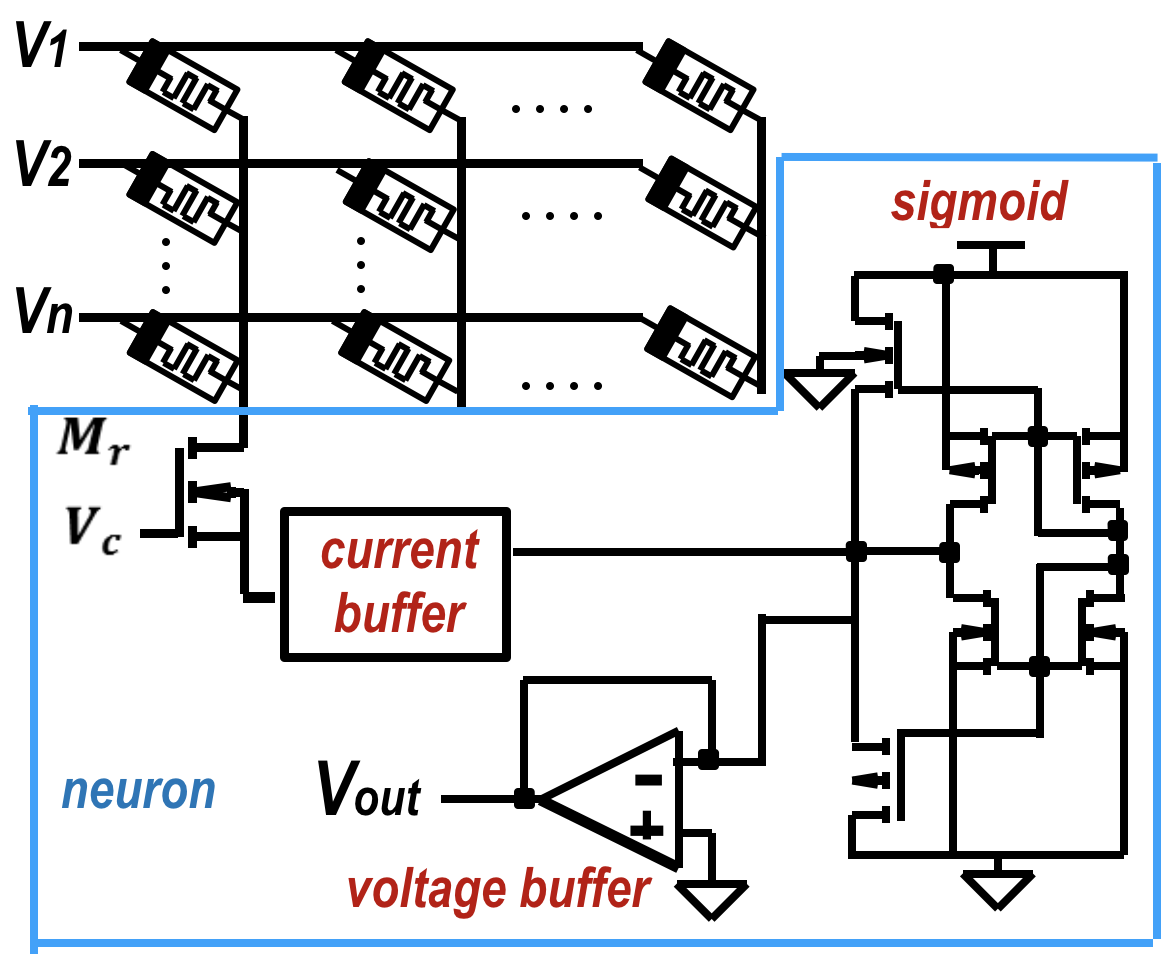}
		}
         \subfigure[]
		{
    \includegraphics[width=32mm]{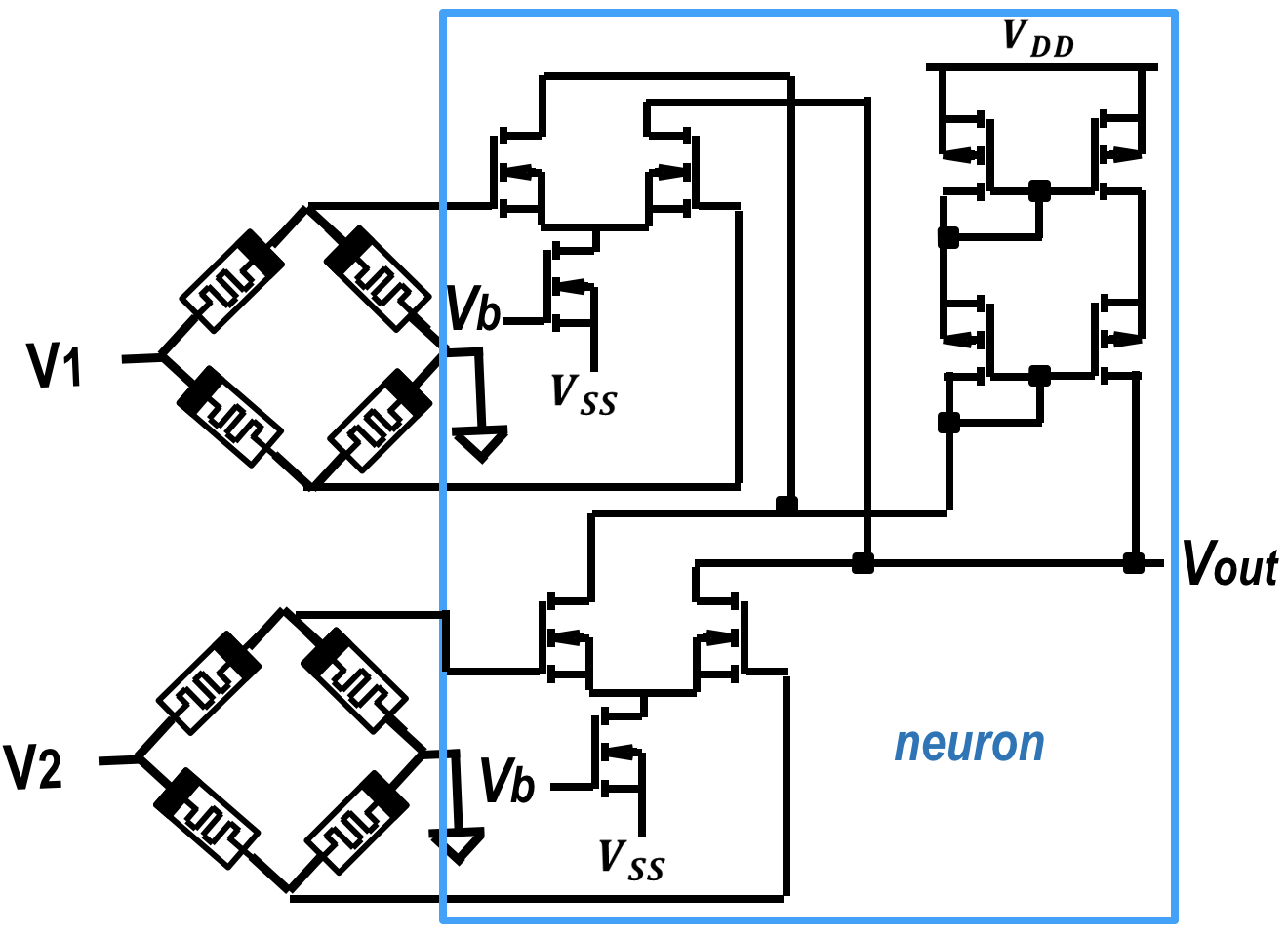}
		}
         \subfigure[]
		{
    \includegraphics[width=22mm]{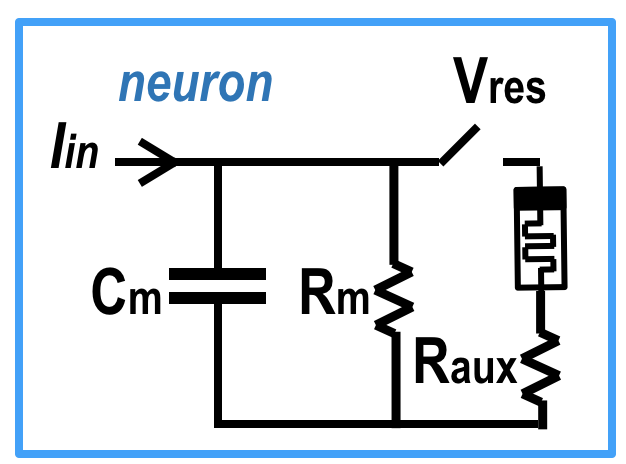}
		}
         \subfigure[]
		{
    \includegraphics[width=43mm]{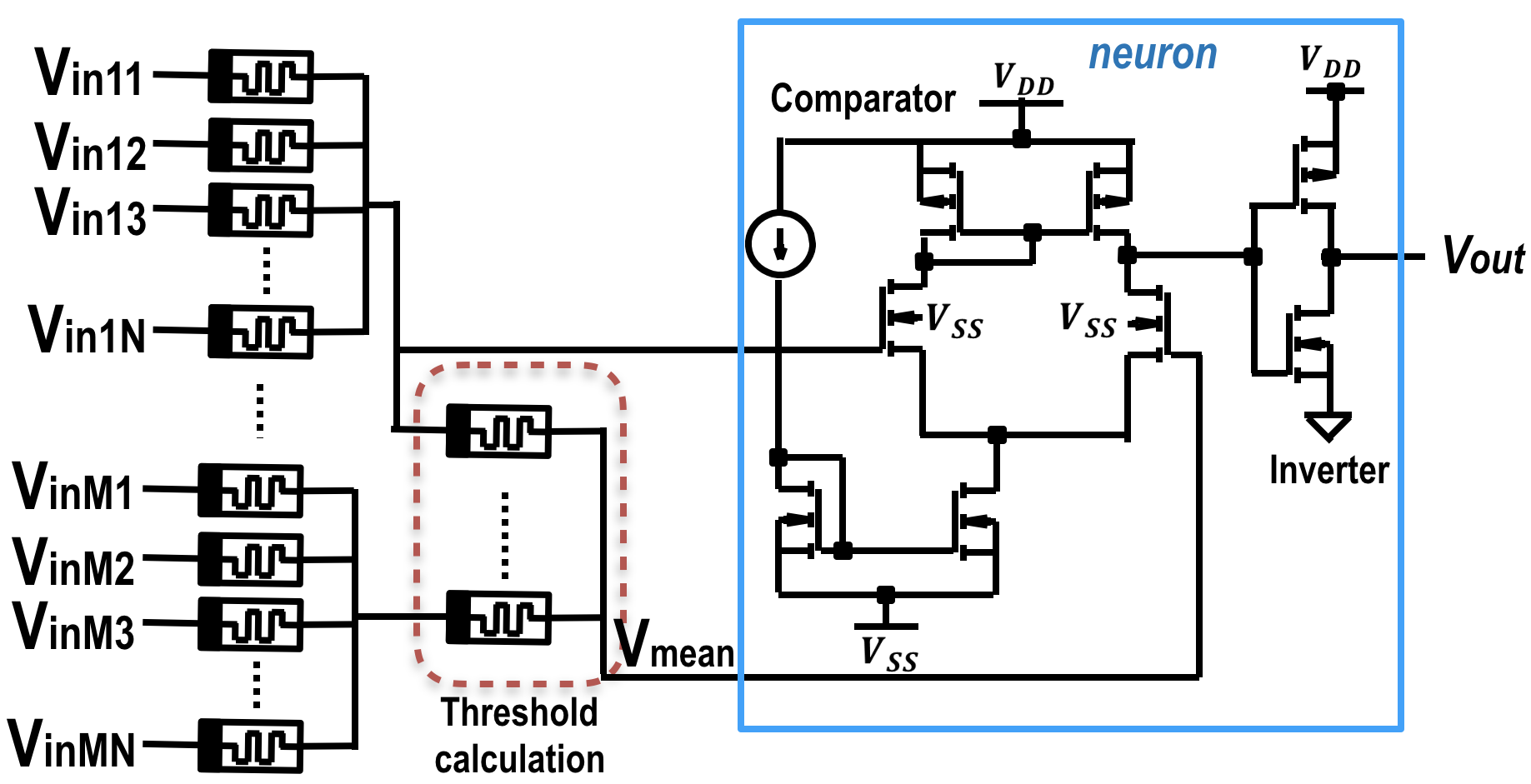}
		}
    \caption{Neuron cells: (a) Modified I\&F neuron \cite{shamsi2018hardware}; {(b) memristor-based capacitive neuron \cite{wang2018fully}; variations of neuron models based on summing amplifier and comparator: (c) \cite{8340055,8275126},\cite{hasan2017chip}, (d)  \cite{hu2014memristor}, (e) \cite{7727302},  and (f) \cite{zhang2017memristor}; }(g) neuron models with sigmoid activation function \cite{ISCASs}, (h) neuron model for memristor-bridge architectures \cite{6074916,6939735}; (i) stochastic neuron \cite{naous2016memristor}; and (j) HTM SP neuron \cite{tcad}.}
    \label{f2}
\end{figure*}

\subsubsection{Integrate and fire neuron model}
The earliest neuron cell models are based on capacitors that emulate the membrane of a biological neuron and integrate current \cite{chua2012hodgkin}. 
One of the basic and first neuron models is Integrate and Fire (I\&F) neuron model. In this model, single membrane capacitance sums the currents flowing into the neuron from all the synapses and membrane resistance causes the leakage of the membrane current \cite{7116617}. However, due to the large on-chip area and power consumption, such neurons are not applicable for large-scale circuits and edge devices, where the power consumption is limited. Even the novel I\&F neuron circuits proposed recently \cite{7280819,ebong2012cmos,8123644,zhang2017synchronization,jiang2016cyclical} cannot be extended for the use in the large-scale systems due to the number of the components. 

There are only a few attempts to use the I\&F based neuron models in large-scale architectures. 
The modified I\&F neuron used for neural network implementation is shown in Fig. \ref{f2} (a) \cite{shamsi2018hardware}. The neuron circuit consists of current integration part with capacitor $C_u$, spike generation Schmitt trigger circuit, reset circuit and control circuit for current input range and injection. When the voltage is applied to the terminals of transistors $M_1$ and $M_2$, the input current $I_{in}$ is injected to the leaky integration part of the neuron through the current mirror. This current is integrated and leaked though $M_3$. Then, the Schmitt trigger generates a spike, and the neuron is reset using $M_{4}$. The firing threshold of the neuron is determined by the Schmitt trigger circuit.

{
In one of the recent works, the integrate and fire effect was achieved by a neuron based on a single diffusive memristive device \cite{wang2018fully}, illustrated in Fig. \ref{f2} (b). The diffusive memristor exhibits capacitive effect and a temporal behavior due to the doping of $Ag$ nanoclusters between two electrodes of memristive material \cite{wang2017memristors, wang2018fully}. In the application of such memristor as a neuron \cite{wang2018fully}, it integrates the pre-synaptic signals, and when the memristor threshold is reached, the diffusive memristor changes its state and resistance of a memristor decreases causing a spike. The delay of a spike depends on the internal material properties and $Ag$ doping in the diffusive memristor.
}

\subsubsection{Neuron model based on summing amplifiers and comparators}

Most of the ANN implementations use the neuron structures based on the summing amplifiers and comparators \cite{8340055,danial2017didactic,8275126}. This model is usually used to represent threshold logic based linear neuron model. In most of the cases, this structure is used for postsynaptic neurons, while presynaptic neurons have various configurations depending on the application of the architectures, or are not even shown in several research works. Different variations of such neurons are shown in Fig. \ref{f2} (c), Fig. \ref{f2} (d), Fig. \ref{f2} (e) and Fig. \ref{f2} (f).

Fig. \ref{f2} (c) represents the conventional summing and thresholding neuron configuration \cite{8340055,8275126}. The summing amplifier sums the input currents and outputs the equivalent voltage. The comparator output the spike or pulse (depending on the configuration of the circuit), when the amplifier output is above the threshold \cite{8340055,8275126}. Fig. \ref{f2} (d) shows a similar configuration of the output neuron with the summing amplifier combining the outputs from negative and positive memristive arrays and comparator circuit \cite{hu2014memristor}. The other configuration is shown in Fig. \ref{f2} (e). The first amplifier is used to scale the output voltage and implement the sigmoid activation function, while the second unity gain amplifier inverts the output \cite{7727302}. 
Fig. \ref{f2} (f) shows a neuron consisting of three amplifiers \cite{zhang2017memristor} used to sum the currents, invert the output and calculate the error, which allows updating the synapses.

\subsubsection{Neuron models with different activation functions}
There are different ANN implementations which use various activation functions to implement the behavior of the neuron, such as sigmoid \cite{ISCASs} and tangent \cite{shamsi2015hyperbolic}. One of such sigmoid-based neurons is shown in Fig. \ref{f2} (g) \cite{ISCASs}. The neuron contains a sigmoid activation function with input current and output voltage and additional circuit to ensure the accurate performance and absence of loading effects. The currents from the memristive synapses are summed, and the current mirror is used to reduce the loading effect. The current is applied to the sigmoid activation function \cite{5719144}, and voltage buffer is used to normalize the sigmoid output. The voltage buffer is optional in this configuration.

\subsubsection{Neuron models for memristor bridge architecture}
The other possible implementation of the neuron is shown in Fig. \ref{f2} (h). These neurons correspond to the bridge synapse structure from \cite{6074916,6939735} and were proposed to be used only with those synapses. 
In this neuron, the voltage weighted by the memristor bridge synapses is converted to the current using differential amplifiers \cite{6939735}. 
Three transistors connected to the synapse represent voltage-to-current converter (VIC) acting as a current source. The neuron contains a self-biasing circuit to provide DC output current, an active load connected to all synaptic circuits which sum up the currents from all synaptic currents, and memristor load that converts output current into voltage. This circuit is used in various neural network architectures \cite{6232461,6939735}. Such configuration shows good performance for ideal simulations, however, if the circuit is constructed from the real memristors, the problems, such as switching response, switching time and connection issues of two memristors may occur. Also, if the number of connected synapses increases, the number of transistors in the neurons will increase significantly. Therefore, this is not the most efficient solution for very large architectures.

\subsubsection{Stochastic neurons}

In recent year, the exploration of the stochastic systems with added noise and memristor stochasticity gained the popularity. Such neuromorphic systems emulate the stochasticity in the cortex, where the biological noise helps the learning and information processing. In CMOS-memristive systems, stochasticity is introduced by ejecting the noise into the circuit. Either stochastic memristive synapses or stochastic neuron can be used for these purpose \cite{naous2016memristor,al2015inherently}. 
One of the possible implementations of a stochastic neuron is shown in Fig. \ref{f2} (i).
{
Memristor is arranged in parallel with original simple neuron circuit consisting of membrane resistor $R_{m}$ and capacitor $C_{m}$ \cite{7117477}. The variable threshold of the memristor allows to randomize the firing threshold of the neuron and ensures random neuron spiking behavior. 
}
This stochastic memristor based neuron model tested for the architectures with 16 and 32 stochastic neurons is proposed in \cite{naous2016memristor}. The stochastic neuron with memristor allows removing random number generator from the stochastic circuits.  However, the application of such neurons for large-scale arrays is still questionable because of the size of the neuron due to the capacitor.

The application of the stochastic neurons for digits recognition problem is investigated in \cite{naous2016memristor}. The accuracy that can be achieved is about 60 \% for a system with stochastic neurons and 65 \% for the stochastic synapses. 
The approach was tested for a small scale problem; however, it is mentioned that the 90 \% of recognition accuracy can be achieved using 300 neurons or  235200 synapses. However, such architecture will have a large area and power consumption. 
The simulation of the system with stochastic memristive synapses in \cite{querlioz2011simulation} allows achieving the recognition accuracy up to 82 \% for MNIST database. The other stochastic spiking WTA network used for handwritten digits recognition with 78 \% accuracy is shown in \cite{al2015inherently}.

\subsubsection{HTM Spatial Pooler neuron}
The implementation of HTM neuron is not fully explored in terms of hardware realization. The implementation of inhibition phase of HTM Spatial Pooler (SP) that can be considered as a neuron cell is shown in Fig. \ref{f2} (j) \cite{tcad}. The neuron consists of a comparator and inverter. 
This neuron of a part of modified HTM architecture, where the mean operation replaces the summation.
The comparator performs the comparison of the mean voltage with the threshold, and the inverter normalizes the comparator output and produces the binary output. The variations of HTM neuron based systems are shown in \cite{tcad}, \cite{fedorova2016htm} and \cite{fan2016hierarchical}.

\section{Neuromorphic architectures}
\label{sec3}

\subsection{Neural network architectures}

There are different memristive neuromorphic architectures that can be used for edge computing applications. The summary of these architectures is shown in Table \ref{table1}.
{
Also, there are several other memristive architectures proposed in the recent years, which are less common and not considered in this paper, such as Probabilistic Neural Networks \cite{krestinskaya2018approximate,serb2016unsupervised} and Binarized Neural Networks \cite{krestinskaya2018binary}.
}




\subsubsection{{One layer neural network} with learning}

\begin{figure}[!t]
    \centering        
    \includegraphics[width=55mm]{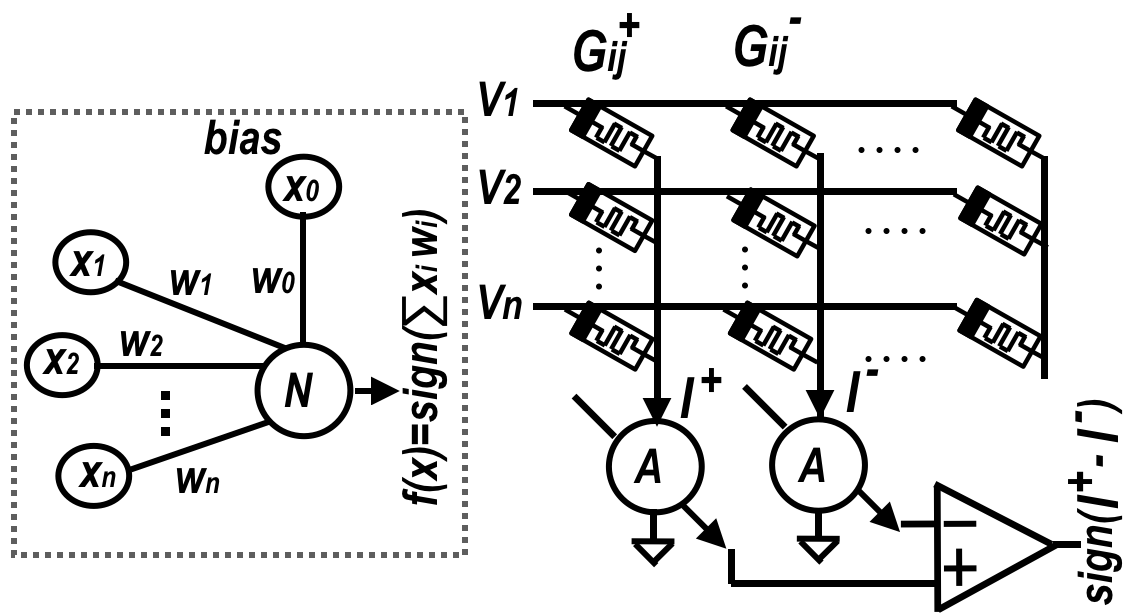}
    \caption{{One layer artificial neural} network \cite{alibart2013pattern}.}
    \label{arc2}
\end{figure}

The structure of {one-layer} ANN with learning is similar to the feed-forward neural network but contains the learning phase. Learning can be performed using various learning rules, like Hebbian learning, backpropagation and different modifications of them. 
One of the implementations of {one-layer} ANN is shown in Fig. \ref{arc2} \cite{alibart2013pattern}. The 2M $Pt/TiO_{2-x}/Pt$ memristor synapses are used to ensure the negative sign of synaptic weights. The output is calculated as a binary activation function of a sum of all the synapses, which is equivalent to a perceptron \cite{rosenblatt1958perceptron}. The learning is performed using a perceptron learning rule, where the memristive synapses are strengthened or weakened depending on the desired output: $\Delta w_i=\pm \eta x_i(y_i-y_o)$, where $y_i$ is an ideal output, $y_o$ is a real output,  $\eta$ is a learning rate and $x_i$ is an input. The architecture was tested for small-scale pattern classification problem.

The other implementation of {one-layer} ANN is proposed in \cite{soudry2013hebbian}. The architecture is designed as an array of 2T1M synapses (Fig. \ref{f1} (e)). The performance was tested for handwritten digits recognition, and the obtained accuracy is approximately 83\%. The implementation of {one-layer} ANN for face classification using single layer RRAM-based perceptron is shown in \cite{yao2017face}. The architecture is constructed using $TiN/TaO_x/HfAl_yO_x/TiN$ 1T1M synapses (Fig. \ref{f1} (d)). The achieved average face recognition accuracy for Yale Face Database \cite{47} is 88.08\%.

\subsubsection{{Two layer neural network}}

\begin{figure}[!t]
    \centering        
    \includegraphics[width=80mm]{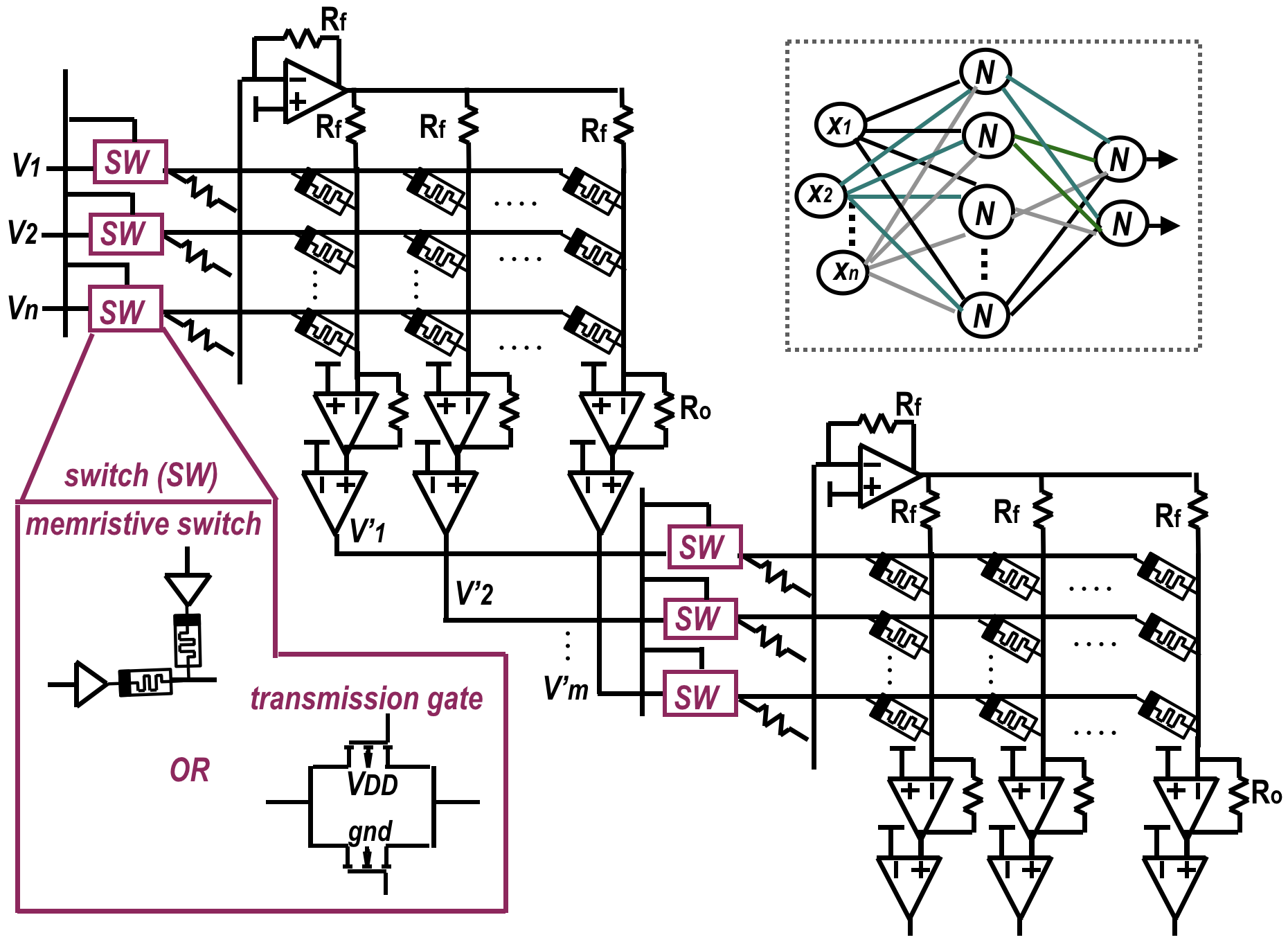}
    \caption{{Two layer neural network} \cite{zhang2017synaptic,zhang2017memristor}.}
    \label{arc3}
\end{figure}

 The typical example of{ two layer} neural network is a perceptron with a single hidden layer. Such architecture is shown in Fig. \ref{arc3} \cite{zhang2017synaptic,zhang2017memristor}. The architecture contains two crossbars with  $Ag/AgInSbTe/Ta$ 1M synapses \cite{zhang2017memristor} and neuron cells shown in Fig. \ref{f2} (f). The control cell in the architecture contains either transmission gate \cite{zhang2017synaptic} or memristive switch \cite{zhang2017memristor}.
 Both networks were tested for pattern recognition applications. The design is simulated for  digits recognition problem with the accuracy up to 100\% without noise \cite{zhang2017memristor}.

\begin{figure}[!t]
    \centering        
    \includegraphics[width=80mm]{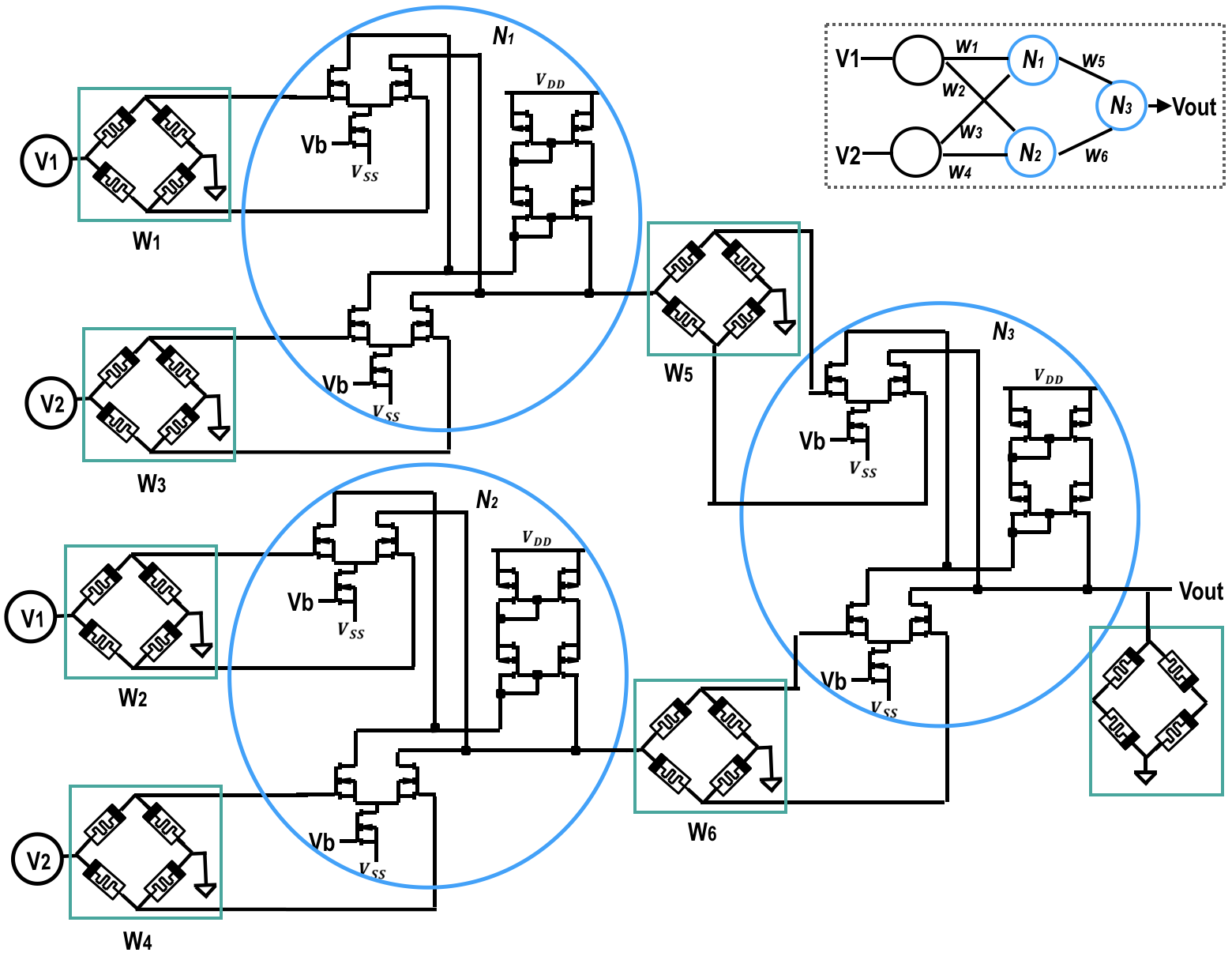}
    \caption{{Two layer neural} network with memristor bridge synapses \cite{6232461}.}
    \label{olgafig2a}
\end{figure}

 {
 Partially fabricated two-layer ANN with
 64 input, 54 hidden and 10 output neurons shown in \cite{li2018efficient}. The $128\times 64$ fabricated crossbar array was used in the network, while activation functions were implemented in software. The simulation was performed with rescaled images of size $8 \times 8$ pixels from MNIST database with the classification accuracy of 92\%. The training was performed online, the update values for memristors have been calculated in software, according to backpropagation algorithm, and the corresponding update pulses were applied to the crossbar.
 }

 The other architecture for {two-layer} ANN proposed in 
 \cite{6232461} is based on 4M bridge synapses (Fig. \ref{f1} (f) and Fig. \ref{f2} (h)). The architecture is shown in Fig. \ref{olgafig2a}. The architecture is similar to Radial Basis Network structure and consists of the artificial neurons with 7 CMOS transistors and memristor bridge synapses. 
 {
The network with 432 inputs, 10 hidden neurons and 1 output neuron was tested for car detection problem using images of size $24\times 18$ pixels. The results showed that the results obtained from circuit simulation are comparable with software simulation results.
}
A similar approach is proposed in
 \cite{6939735} with the implemented ANN is based on Random Weight Change (RWC) learning algorithm. The circuit implementation shows promising results in terms of processing time, which equals to $115ns$ in total for feedforward processing and the memristor programming.

\subsubsection{Deep Neural Networks}

Deep Neural Network (DNN) is a large class of the neural networks that consists of many cascaded layers and contains various activation functions between the layers. The number of layers in deep neural networks cause the scalability issues. Moreover, the application of memristive crossbars opens an opportunity to scale such networks staying at an acceptable level of power consumption.  
Therefore, memristor-based deep neural networks have been explored in the recent years. The architecture of memristive DNN is similar to {two-layer} neural networks but contains more crossbar arrays. 
{
The research work \cite{7727302} explores the deep memristive convolutional neural network with 5 layers and reports the accuracy of 91.8\% for MNIST handwritten digits classification.
While \cite{wijesinghe2017all} investigates the implementation of deep stochastic spiking convolutional 5 layer neural network with the MNIST classification accuracy of 97.84\%, selecting the output class based on the largest number of output spikes produced by the output neurons. The energy consumption and on-chip area of this memristive network is 6.4 and 8 times smaller than in equivalent CMOS-based design, respectively.
}

\subsubsection{Cellular Neural Network}

\begin{figure}[!t]
    \centering        
    \includegraphics[width=80mm]{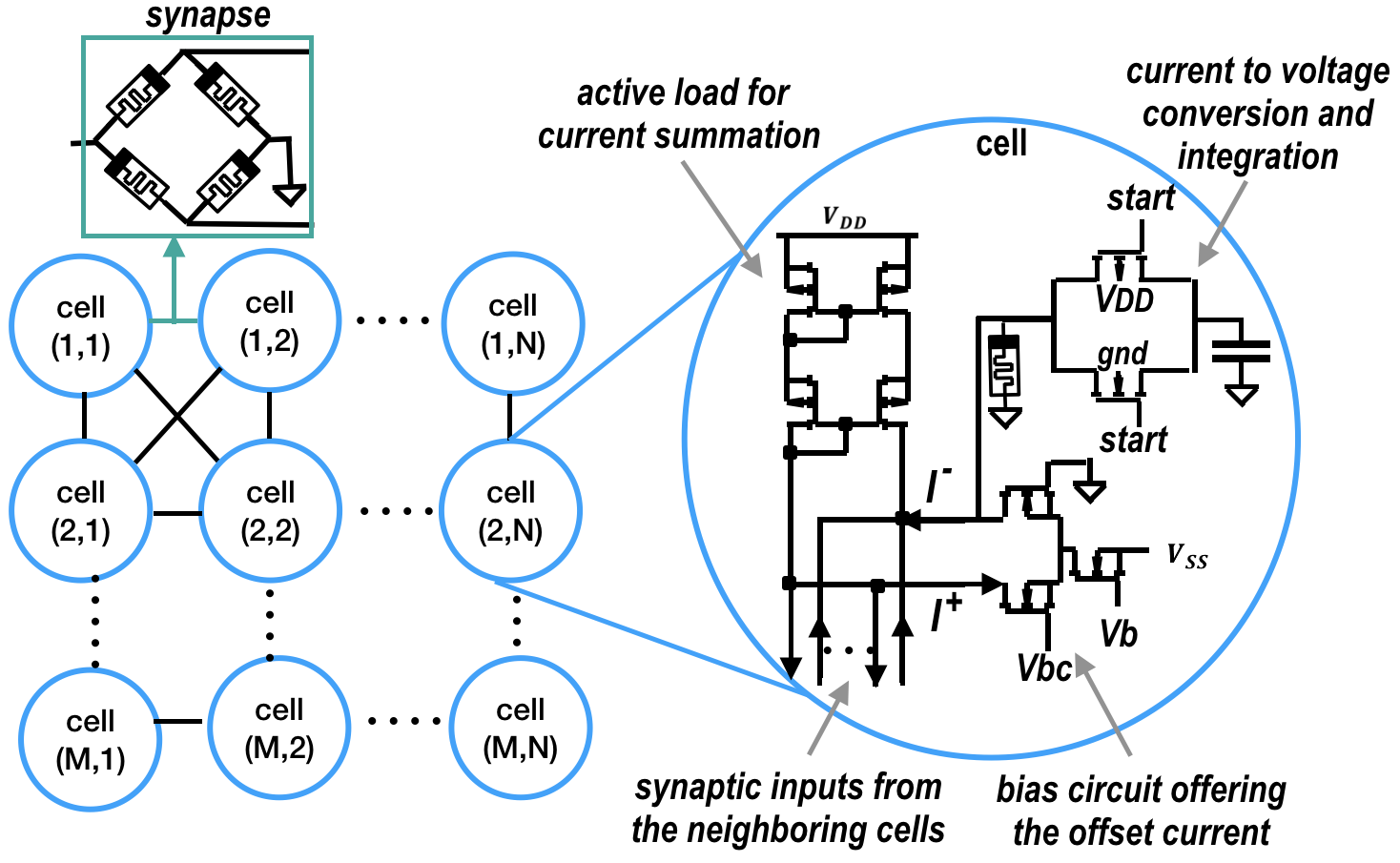}
    \caption{Cellular neural network \cite{6861426}.}
    \label{olgafig3}
\end{figure}

The architecture of the cellular neural network (CeNN) is illustrated in Fig. \ref{olgafig3}. The architecture implies that the cells are connected only to the closest neighbor cells in the network. The first analog hardware implementation of cellular neural networks was proposed in the 1980s. 
The cells were designed with the capacitor, current source and resistive elements \cite{7600}. In contrast to this early design, the architecture of recently proposed CeNN is based on the memristive-CMOS circuits as shown in \cite{6074916,6861426}.  
The most commonly implemented memristive CeNN architecture is based on 4M bridge synapses (Fig. \ref{f1} (f) and Fig. \ref{f2} (h)). The research work \cite{7168838} illustrates the use of the memristor bridge circuit application with 5M synapses.
 This architecture is useful for the image processing tasks, such as edge detection \cite{7469884,6861426} and image filtering \cite{7168838,6074916}. The two dimensional CeNN architecture in the flux-charge domain is described in \cite{di2017memristor}. The CeNN can also be used for noise removal, extraction of horizontal lines and hole filling tasks.




\subsubsection{Convolutional Neural Network}

Convolutional Neural Network (CNN) is a machine learning algorithm based on a convolution operation that has been proven to be an efficient solution for various classification tasks, image recognition problems \cite{8016501,8103129} and video analysis \cite{7444187}. Comparing to the software implementations of CNN, there are not many hardware implementations of CNN based on memristive circuits. 
Most of the hardware solutions for implementing CNN architecture are based on 1M memristive crossbar arrays or ReRAMs, while the processing units such as for implementing learning algorithm are digital \cite{7544367,7858419}.

\begin{figure}[!t]
    \centering        
    \includegraphics[width=90mm]{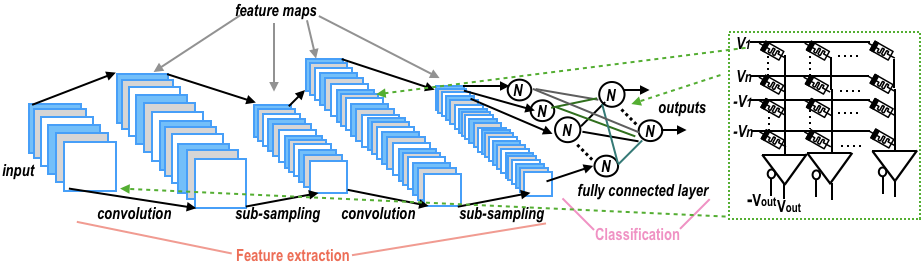}
    \caption{{Convolutional Neural Network \cite{7727302, 7966055}.}}
    \label{fig4}
\end{figure}

One of the hardware solutions for CNN is shown in Fig. \ref{fig4} \cite{7727302, 7966055}. The architecture is divided into feature extraction parts with convolution and sub-sampling (smoothing) layers and classification part. In CNN, the number of data features is reduced with the propagation through the network but the number of feature maps increases, which improves feature quality for inter-class discrimination. {The convolution layer is followed with a fully connected multi-layered neural network that act as the classifier. The learning in such system is performed on software and the values of the memristors in each layer are programmed. The testing and classification are performed on hardware. 
In convolution layer, memristors represent the convolution filters and perform dot product calculation, similar to the fully connected layer. The output current from the crossbar is converted into voltage using the system with two amplifiers (as in Fig. \ref{f2} (d)).}
The number of memristors in each layer is determined by the initial size of images and number of required feature maps in this layer \cite{7966055}. 
{
In \cite{7727302} and \cite{7966055}, the size of the input images is $28\times 28$. In the first convolution layer, the image is filtered by 6 convolution filters producing $24 \times 24$ feature maps, while sub-sampling layer reduces the size of feature maps to $12\times 12$. In the second convolution layer the feature maps are filtered by 12 convolution filters producing feature maps of size $8 \times 8$, which are reduced to the size of $4\times 4$ in the second sub-sampling layer.
}
The accuracy for handwritten digits recognition that can be achieved is 92\% \cite{7727302} and 94\% \cite{7966055}, comparing to 98.92\% of software simulation with MNIST database. 

The research work \cite{7551379} illustrates the memristive crossbar based accelerator for CNN implementation consisting of analog and digital components. In such systems, the analog components include only memristive crossbar; and most of the other components are digital. 
{
The power efficiency of such accelerator is $644.2$ giga-operations per seconds (GOPS) per Watt (GOPS/W).}
The CNN accelerator based on the crossbar architecture with digital ReRAM is shown in \cite{8009177} and \cite{7911265}. The accuracy results are 98.3\% and 91.4\% for MNIST and CIFAR-10 databases, respectively. The area and power consumption of the system are $1.02mm^2$ and $6.3mW$.
{
The system throughput is $792$ ($GOPS$) and energy efficiency is 126 tera-operations per second ($TOPS$) per Watt ($TOPS/W$).
}
The accuracy of CNN varies with the number of output feature maps from the convolution layer. The research work \cite{7280813} illustrates that the implementation of CNN for MNIST character recognition using memristive crossbar has the on-chip area of $0.5033947mm^2$ and power consumption of $0.001785W$, which is more efficient in comparison to the implementation of CNN {on the traditional RISC processor.}

{
One of the most recent works in memristive convolutional filtering is illustrated in \cite{li2018analogue}. In this work, parallel vector matrix multiplication of array of size $128 \times 64$ is implemented. The current from all fabricated crossbar columns are read in parallel, which illustrates the speed of 1.64 $TOPS$ for reading cycle. The power consumption of such crossbar is $13.7mW$, and the power efficiency is $119.7$ $TOPS/W$. Even though the image quality after convolution operation is worse comparing to software based convolution operation, memristive solution consumes 17 times less energy comparing to ASIC implementation. The recent work \cite{wang2018handwritten} illustrates the implementation of CNN in spike domain with digital memristor-based neuron using Time Division Multiplexing Access (TDMA) technique to reduce the number of required neurons. The classification accuracy of the network  for handwritten digits recognition is 97\%. However, the size and scalability of such network for the application on edge devices is an open problem.
} 

\subsubsection{Spiking Neural Network}
 
In Spiking Neural Networks (SNN) the data signals are transmitted as spikes of a specific shape. This emulates the brain processing and is based on the particular spike events \cite{6322959}. 
SNN focuses on the realization of plasticity rules and timing difference between pre- and postsynaptic spike.
The spike based architectures are mostly represented by Spike Timing Dependent Plasticity (STDP) implementation. STDP is based on biological concepts of presynaptic and postsynaptic impulses.
The implementation of the neuromorphic architectures with STDP are based on the memristive crossbar arrays. 
The crossbar represents synapses and connected with the neuron models \cite{6518253}. The possible implementation of such system is shown in \cite{7527393}.
Based on the correlation of the presynaptic and postsynaptic spikes, the synapse value between presynaptic and postsynaptic neurons represented by memristor is updated. 
Based on the postsynaptic neuron mode, the memristor is potentiated, depressed or stay unchanged. 
One of the advantages of SNN hardware implementation is that the power dissipation of such systems is smaller than in the pulse based systems. SNN can be used for handwritten digits recognition and letter recognition with the accuracy of up to 99\%  \cite{7527253}.

\begin{figure}[!t]
    \centering        
    \includegraphics[width=90mm]{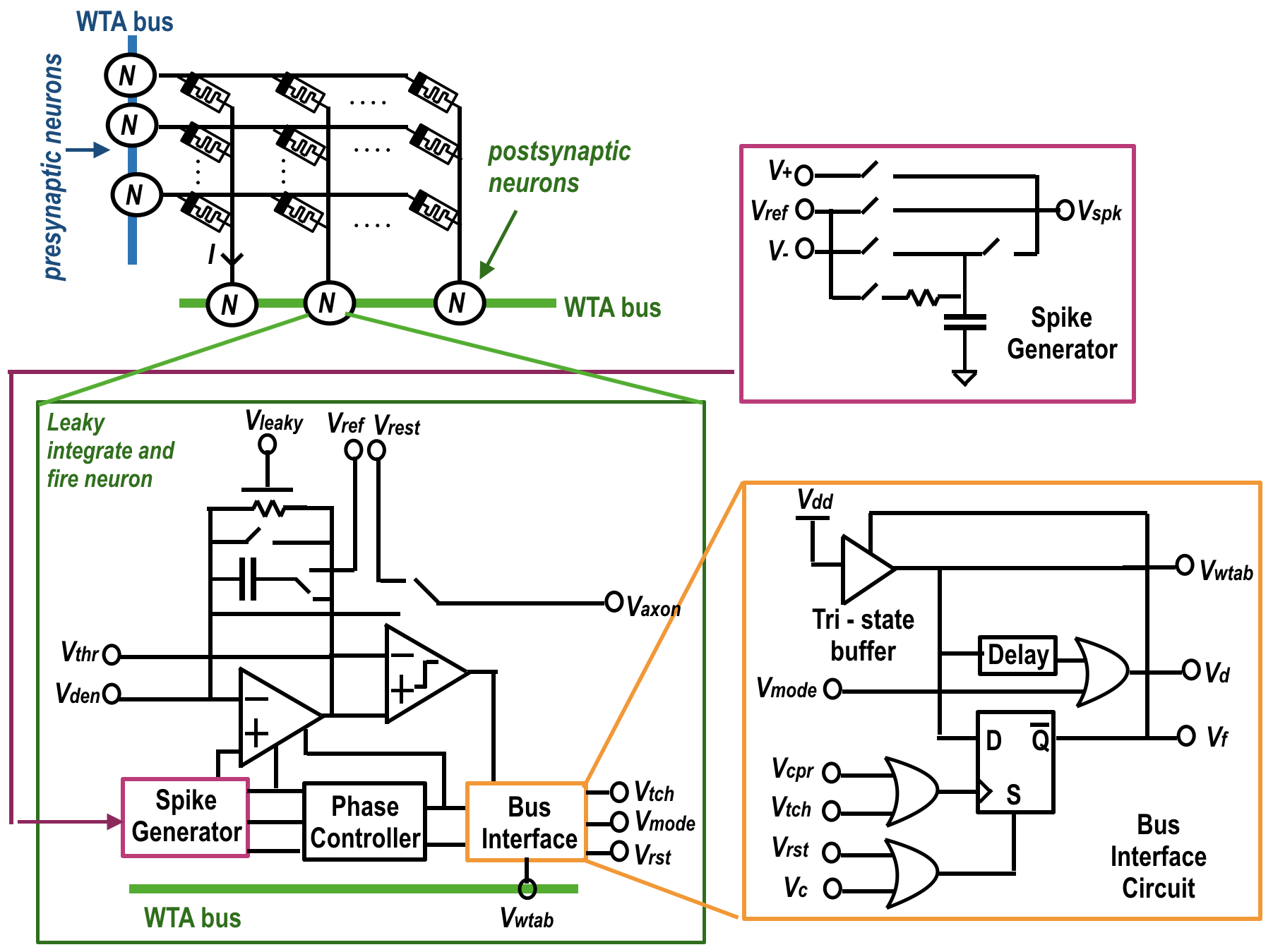}
    \caption{Spiking Neural Network \cite{7116617}.}
    \label{fig6-1}
\end{figure}

 The basic  SNN architecture is shown in Fig. \ref{fig6-1}; it consists of presynaptic neurons and postsynaptic neurons connected by 1M synapses \cite{7904675,7116617}. 
In most of the cases, the SNN is used with Winner-Takes-All (WTA) approach. One of such architectures for object position detection is introduced in \cite{6093706}. Each input neuron corresponds to a particular position of the object, and the output neuron determines the exact position of the object based on the spiking frequencies. If the object is located between the input neurons, the spiking frequencies of the output neurons are proportional to the exact position of the object in the input neurons. { Such position detector consisting of $5\times 5$ neurons has maximum power consumption of $15.6\mu W$, which is about 70\% less than equivalent CMOS design, and on-chip area of $6.1\times 10^{-5} cm^2$.}

\begin{figure}[!t]
    \centering        
    \includegraphics[width=70mm]{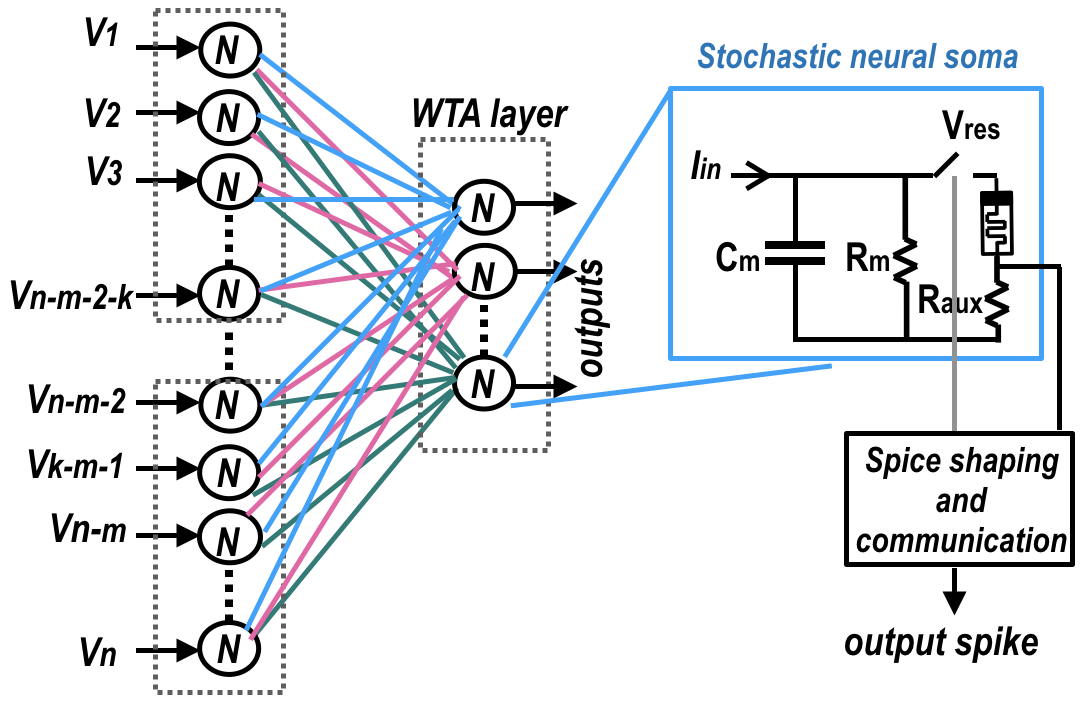}
    \caption{Stochastic spiking neural network \cite{7117477}.}
    \label{fig6-2}
\end{figure}

\begin{figure}[!t]
    \centering        
    \includegraphics[width=60mm]{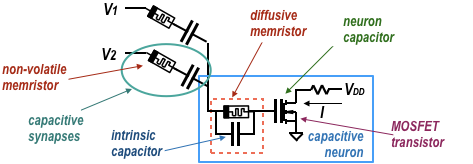}
    \caption{{Capacitive spiking neural network \cite{wang2018capacitive}.}}
    \label{fig6-2}
\end{figure}

The recent works introduce the stochasticity to SNN. 
The stochasticity implies the probabilistic behavior of neurons or synapses and represents the biological concept of the importance of neural noise during the information processing in the brain. In \cite{7117477}, the stochasticity is introduced to the simple Spiking WTA architecture shown in Fig. \ref{fig6-2}, where the output is determined by the first firing neuron from the output neurons. The simulation results from MNIST handwritten digits recognition vary with the size of the layer of output neurons and reach 78.4\% for 128 output neurons. 
{
The increase on the number of output neurons allows the network to capture more different patterns corresponding to the input data, which enhance the performance of majority voting procedure and allows to increase classification accuracy.
}

{
The alternative approach to implement SNN is a capacitive switching network presented in \cite{wang2018capacitive}. The resistive synapses are replaced with capacitive synapses concept. Capacitive synapses are based on non-volatile pseudo-memcapacitors formed by integrating non-volatile memristor  in series with a capacitor.  To form the capacitive neurons, the neuro-transistor is introduced, where dynamic pseudo-memcapacitors, formed by integrating recently proposed diffusive memristor with intrinsic capacitance \cite{wang2017memristors, wang2018fully} in series with capacitor, are integrated onto the gate of a MOSFET. If the neuron is triggered by high capacitive state synapses, the post-synaptic neuron fires.
The learning in such network is performed if presynaptic and postsynaptic neurons fire together causing the potentiation of low capacitance state in the synapses.
Capacitive spiking network has an advantage of sneak-path free outputs.
}

\subsubsection{Recurrent Neural Network and Long Short Term Memory}

\begin{figure}[!t]
    \centering  
    \subfigure[]{
    \includegraphics[width=22mm]{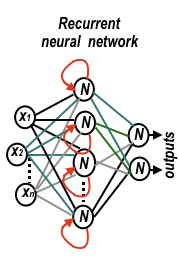}
    }
      \subfigure[]{
    \includegraphics[width=60mm]{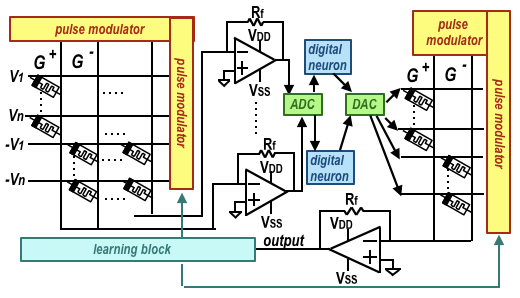}
    }  
    \caption{{(a) Recurrent neural network, and (b) Mixed Signal Implementation of one layer RNN \cite{deng2015complex}.}}
    \label{pn6}
\end{figure}

Recurrent Neural Network (RNN) is a neural network type, which involves the feedback calculation and the output of the layer effects the consequent outputs \cite{557671}. There are various architectures for RNN implemented in software; however, memristor-based hardware implementations of RNN is an open problem. There are several modifications of RNN, and simple RNN architecture is shown in Fig.\ref{pn6} (a).
{
Fig. \ref{pn6} (b) shows the implementation of one layer RNN with fabricated iron oxide memristive synapses \cite{deng2015complex}, containing two parallel memristors representing positive and negative conductance, as in Fig. \ref{f1} (b). RNN design involves digital neuron, ADC and DAC, pulse modulator and learning block based on recursive least-squares algorithm.
}

The RNN architecture, {especially} in analog domain, has not been fully explored yet. Most of the works on memristive RNN focus on a mathematical analysis of system stability \cite{bao2018region,7553910}. The hardware implementation of memristive RNN is presented in \cite{xavier2018memristive}. The work illustrates analog implementations of the RNN using $0.5\mu m$ CMOS technology and applied for combinatorial optimization problems.
Even though there are FPGA-based implementations of RNN \cite{1528541}, and some of the works show the possibility to integrate RNN with the memristive crossbar \cite{xavier2018memristive}, the implementation of a full memristor based RNN architectures is an open problem. One of the main problems in analog implementations of RNN is the implementation of feedback and complexity of the architectures. 

\begin{figure}[!t]
    \centering        
    \includegraphics[width=70mm]{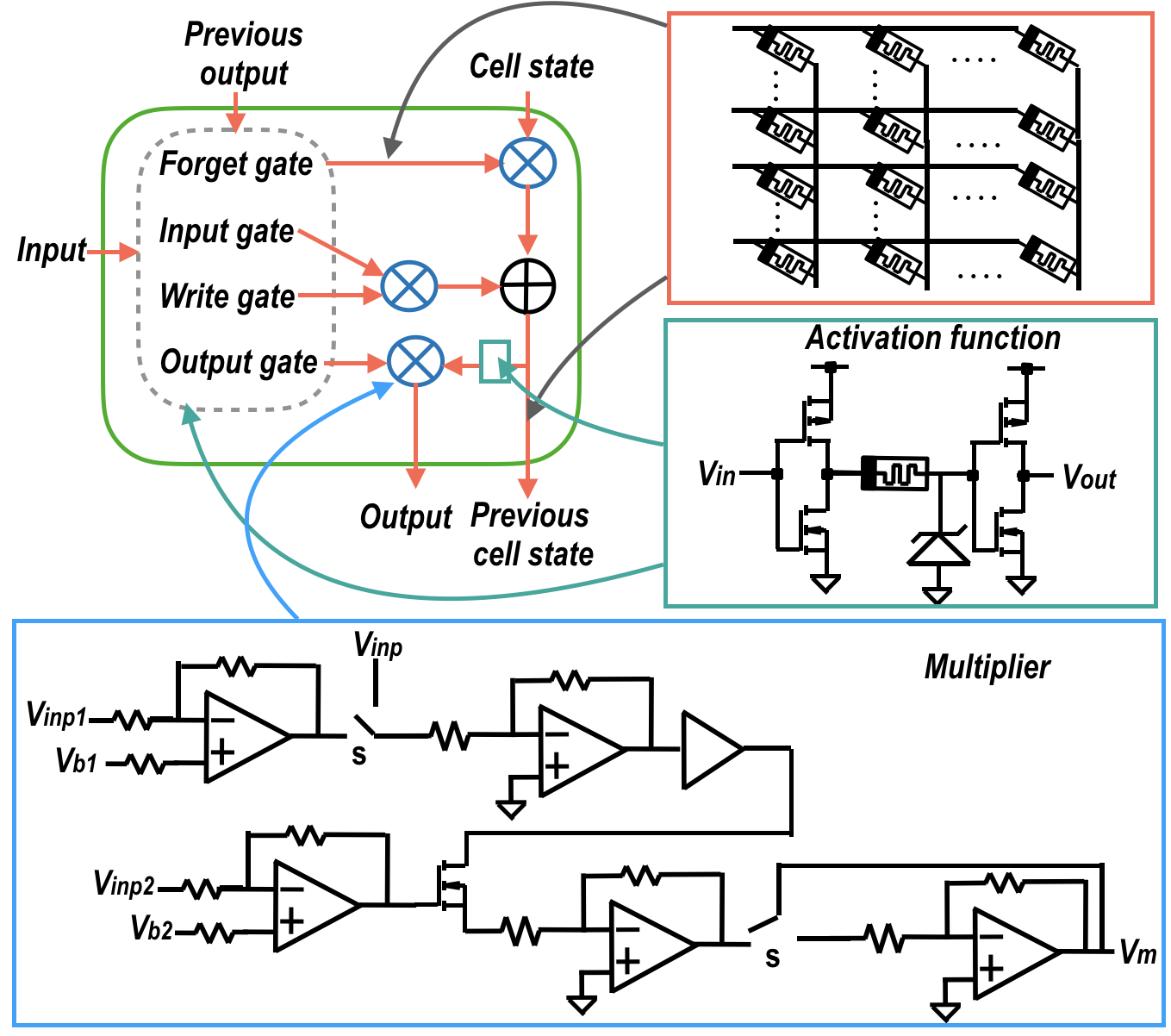}
    \caption{Long Short Term memory \cite{smagulova2018memristor}.}
    \label{LSTM}
\end{figure}

LSTM is a modification of RNN. One of the main features of LSTM is the feedback and selection of the information that effects future outputs. LSTM is based on modified dendritic threshold non-linear neuron model, where the output depends on the current input and previous outputs. The LSTM structure shown in Fig. \ref{LSTM} includes output gate, input gate, write gate and forget gate.
These gates are responsible for how much the current output should be affected by the current inputs and previous outputs of LSTM. Memristor-based implementation of LSTM is proposed in \cite{smagulova2018memristor}. The LSTM weights are presented as crossbars with 1M synapses, and the implementation of the activation function circuit that can be used as sigmoid or tangent is shown in Fig. \ref{LSTM}. As LSTM algorithm requires multiplication, the analog multiplier is used.
{
While the research work \cite{smagulova2018memristor} shows the implementation of separate LSTM components, the full implementation of LSTM system is illustrated in \cite{smagulova2018design} and \cite{adam2018memristive}. Both systems are tested for the prediction of the number of airline passengers, and show the successful prediction of a trend, where LSTM system in \cite{smagulova2018design} achieved the accuracy of 75\%.
The implementation RNN for edge inference with fabricated LSTM units based on memristive crossbar are shown in \cite{li2018long}. The implemented RNN consists of 15 LSTM units followed by fully connected layer. The LSTM is tested for prediction of the number of airline passengers and classification of an individual human by the person’s gait, showing the precise prediction results and classification accuracy of 79\%.
}

\subsubsection{Hierarchical Temporal Memory} 

\begin{figure}[!t]
    \centering        
    \includegraphics[width=90mm]{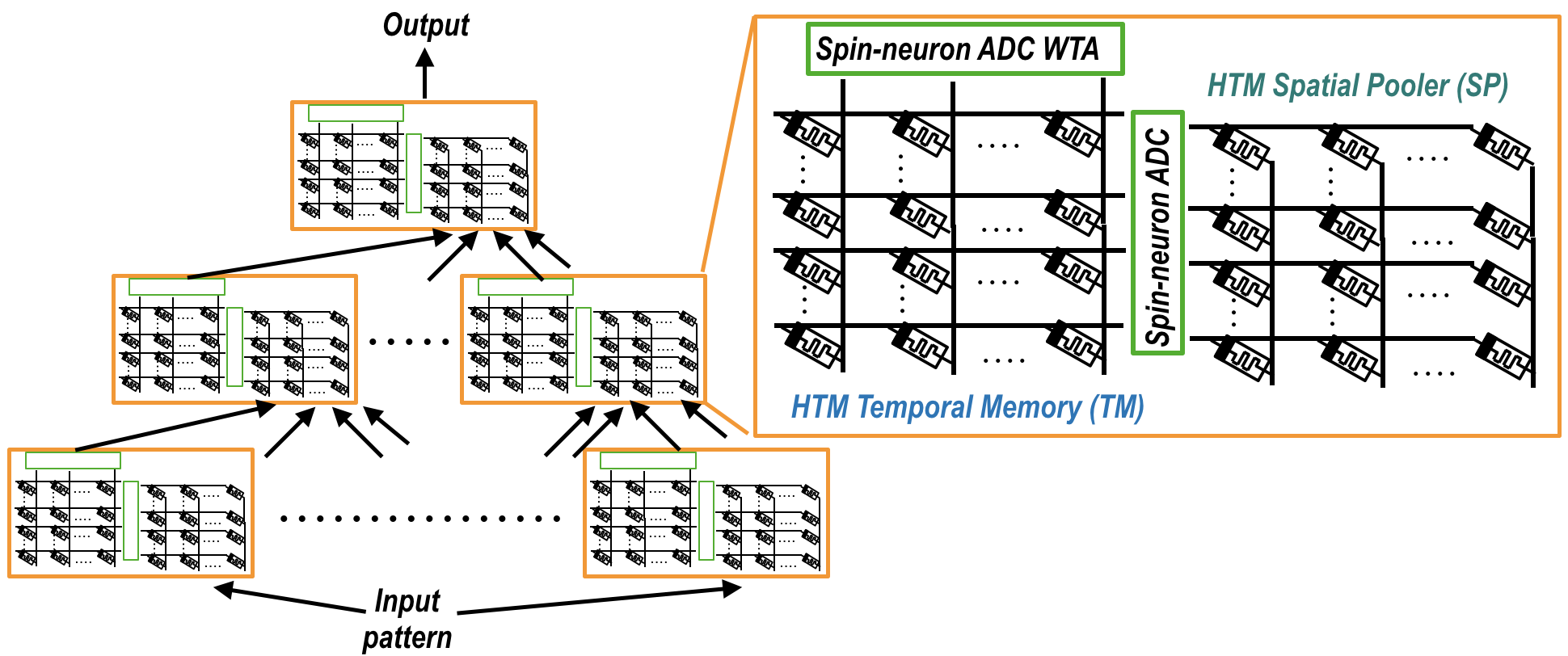}
    \caption{Hierarchical Temporal memory \cite{fan2016hierarchical}.}
    \label{HTMroy}
\end{figure}

HTM is a machine learning algorithm and architecture mimicking the structure and functionality of human neocortex \cite{james2018introduction,8471012}. 
{
HTM consists of HTM Spatial Pooler, which encodes the input patterns and produces sparse distributed representation of input data useful for visual data processing, and HTM Temporal Memory (TM), which can be used for prediction making \cite{8471012}. Both HTM SP and HTM TM involve learning process.
}
There are several CMOS-memristive hardware implementations of HTM proposed in recent years \cite{tcad, fan2016hierarchical, fedorova2016htm}. The mixed signal design of HTM is shown in \cite{fan2016hierarchical}, and the hierarchical structure of the proposed circuit is illustrated in Fig. \ref{HTMroy}. In this architecture, each level of HTM is presented by the memristive crossbar and spin-neuron devices are used as neurons for the processing. 
{
This architecture was used for MNIST handwritten digits recognition with the maximum accuracy of 95 \% \cite{fan2016hierarchical}.
}.
The architecture in \cite{fedorova2016htm} shows the alternative implementation of crossbar-based analog HTM SP circuit,
{
which was tested for face recognition with AR \cite{martinez1998ar} and speech recognition with TIMIT database \cite{garofolo1993darpa} and achieved the accuracy of 86 \% and 70\%, respectively.
}

\begin{figure}[!t]
    \centering        
    \includegraphics[width=90mm]{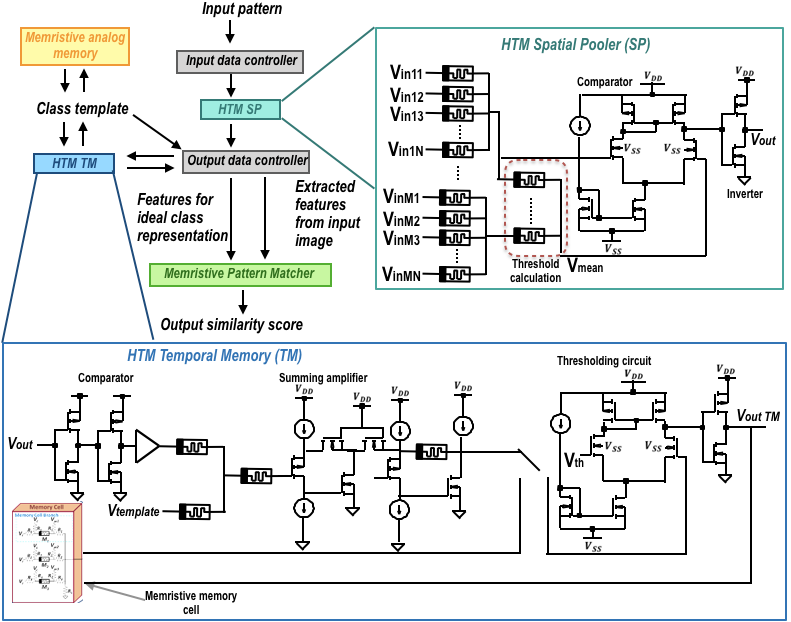}
    \caption{{Modified HTM Spatial Pooler and HTM Temporal Memory \cite{tcad}.}}
    \label{HTMmod}
\end{figure}

{
HTM
architecture in \cite{tcad} proposes the analog circuit level implementation of the modified HTM SP and HTM TM, inspired from HTM neuron shown in Fig. \ref{f0} (d). This architecture is shown in Fig. \ref{HTMmod}. In this system, HTM is used in combination with traditional supervised classification methods. 
Also, the implementation of HTM is modified to reduce the hardware level complexity. In this HTM system, HTM SP is used for encoding the input data patterns and presenting the inputs as sparse distributed binary patterns. 
In comparison with the traditional HTM algorithm, the HTM system uses HTM TM only for generation of the templates for all data classes stored in the memristive memory array. The classification is performed by the memristive pattern matcher, which compare the inputs processed by HTM SP with the ideal image templates.
}
 The synaptic weights are represented as separate memristors. The HTM SP part is based on the hardware implementation of HTM neuron shown in Fig. \ref{f2} (j).
While the HTM TM part consists of the comparator and summing amplifier. The output of the HTM TM is used to update the training template stored in a memristive memory array \cite{irmanova2018neuron}.
{ The system is tested for face with AR database and speech recognition with TIMIT database, achieving the accuracy of 87\%  and 95\%, respectively.
There are several other HTM related works that propose different variations of hardware for memristive HTM \cite{krestinskaya2018feature, ibrayev2017chip}. 
}





\begin{table*}[]
\centering
\caption{\textcolor{black}{Memristive neuromorphic architectures}}
\label{table1}
\begin{tabular}{|l|c|c|c|c|}
\hline
\textbf{Architectures}                                                  & \textbf{\begin{tabular}[c]{@{}c@{}}Applications \\ and simulation results\end{tabular}}                                                   & \textbf{Scalability}                                                                                      & \textbf{Open problems}                                                                                                                                                                                     & \textbf{\begin{tabular}[c]{@{}c@{}}Drawbacks to improve\\  for application \\ in edge computing\end{tabular}}                                                                                                                                          \\ \hline

\begin{tabular}[c]{@{}l@{}}{One} \\ layer \\ ANN\end{tabular}  & \begin{tabular}[c]{@{}c@{}}handwritten\\  digits recognition (83\%) \cite{soudry2013hebbian},\\ face recognition (88.08\%) \cite{yao2017face}\end{tabular}                                                       & \begin{tabular}[c]{@{}c@{}}Scalable \\ with 1M devices\end{tabular}

& \begin{tabular}[c]{@{}c@{}}Investigation of the scalability\\  of the system with 2T1M synapses\end{tabular}                                                                      &

\multirow{2}{*}{\begin{tabular}[c]{@{}c@{}}Investigation of the performance \\ with real devices,  processing \\ speed, scalability,  on-chip \\ area and power  dissipation \\ for large scale systems, \\ improvement of CMOS \\ components\end{tabular}} \\



\cline{1-4}
\begin{tabular}[c]{@{}l@{}} {Two}\\  layer \\ ANN\end{tabular}           & \begin{tabular}[c]{@{}c@{}}simple \\ digits recognition (100\%) \cite{zhang2017memristor}\end{tabular}                                                              & \begin{tabular}[c]{@{}c@{}}Scalable \\ with 1M devices,\\  not scalable for\\  bridge neuron\end{tabular} & \begin{tabular}[c]{@{}c@{}}Investigation of the scalability\\  of bridge neuron based\\  systems and reduction of\\  power dissipation of CMOS\\  components\end{tabular}                                  &                                                                                                                                                                                                                                                        \\ \hline
\begin{tabular}[c]{@{}l@{}}Deep\\  neural \\ networks\end{tabular}      & various applications                                                                                                                      & \begin{tabular}[c]{@{}c@{}}Scalable \\ with 1M \\ devices\end{tabular}                                    & \begin{tabular}[c]{@{}c@{}}Investigation of the possibility \\ of application for various problems,\\  investigation of the effects \\ of real memristors\end{tabular}                                     & \begin{tabular}[c]{@{}c@{}}Improvement of power \\ dissipation and \\ scalability issues\end{tabular}                                                                                                                                                  \\ \hline
CeNN                                                                     & image filtering                                                                                                                           & Not scalable                                                                                              & \begin{tabular}[c]{@{}c@{}}Investigation of the possibility to \\ improve architecture for\\  large scale simulations \\ and to create the multilayer \\ architectures\end{tabular}                        & \begin{tabular}[c]{@{}c@{}}Investigation of \\ the possibility to use\\  with 1M devices \\ to ensure the \\ scalability of the system\end{tabular}                                                                                                    \\ \hline
CNN                                                                  & \begin{tabular}[c]{@{}c@{}}handwritten\\  digits recognition (94\%) \cite{7966055},\end{tabular}                                                         & \begin{tabular}[c]{@{}c@{}}Partially \\ scalable\end{tabular}                                             & \begin{tabular}[c]{@{}c@{}}Investigate the possibility of\\  implementation of \\ fully on-chip system without \\ software part\end{tabular}                                                               & \begin{tabular}[c]{@{}c@{}}As the number \\ of layers is large, \\ the scalability \\ should be investigated\end{tabular}                                                                                                                              \\ \hline
SNN                                                                     & \begin{tabular}[c]{@{}c@{}}handwritten\\  digits recognition(78.4\%) \cite{7117477},\\ letter \\ recognition \cite{7527253} (99\%)\end{tabular}         & Scalable                                                                                                  & \begin{tabular}[c]{@{}c@{}}Investigation of the advantages \\ over pulse-based systems\\ and possibility to replace pulse\\  based systems with spike based\end{tabular}                                   & \begin{tabular}[c]{@{}c@{}}Design of the scalable \\ neurons producing spikes\\ with small of chip\\  area and power dissipation\end{tabular}                                                                                                          \\ \hline
RNN                                                                     & pattern recognition                                                                                                                       & \begin{tabular}[c]{@{}c@{}}Scalable \\ with 1M \\ devices\end{tabular}                                    & \multicolumn{2}{c|}{\multirow{2}{*}{\begin{tabular}[c]{@{}c@{}}Full circuit level design of the architecture, \\ investigation of scalability  and different applications\end{tabular}}}                                                                                                                                                                                                                                                                          \\ \cline{1-3}
LSTM                                                                    & prediction making                                                                                                                         & Scalable                                                                                                  & \multicolumn{2}{c|}{}                                                                                                                                                                                                                                                                                                                                                                                                                                               \\ \hline
HTM                                                                     & \begin{tabular}[c]{@{}c@{}}face recognition (98\%) \cite{fedorova2016htm,georgia}, \\ speech recognition (95\%) \cite{tcad},\\ handwritten\\  digits recognition (95\%) \cite{fan2016hierarchical}\end{tabular} & \begin{tabular}[c]{@{}c@{}}Partially\\  scalable\end{tabular}                                             & \begin{tabular}[c]{@{}c@{}}Implementation of full system \\ performance,  implementation of the exact\\  algorithm for HTM SP and HTM TM, \\ implementation\\  of sequence learning in HTM TM\end{tabular} & \begin{tabular}[c]{@{}c@{}}Improvement of CMOS \\ components\\  to ensure scalability\end{tabular}                                                                                                                                                     \\ \hline
\end{tabular}
\end{table*}





\subsection{Neural Network learning architectures}

\begin{figure}[!t]
    \centering        
    \includegraphics[width=70mm]{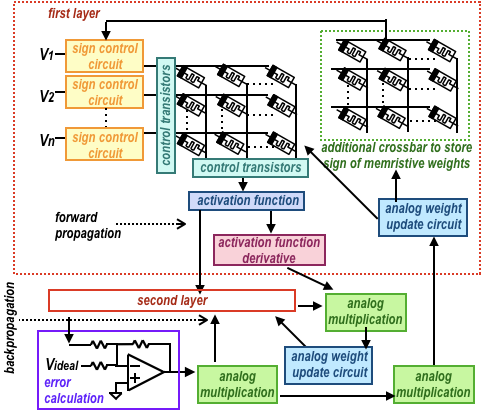}
    \caption{{Online backpropagation training architecture for memristive ANN \cite{learning18}.}}
    \label{newTCAS}
\end{figure}

The learning process in the neural networks is important, especially for large-scale edge computing architectures. 
{
In memristive architectures for edge computing, the concept of online training is important \cite{7010034,xiao2018gst}.
In most of the designs, the learning and  online training of memristive architectures is performed on software.
For example, partially fabricated neural network with online backpropagation training on software and online update of memristive weights of the crossbar is shown in \cite{li2018efficient}. 
}
However, it is important to ensure the scalability and low power dissipation in edge devices;
therefore, separate software components and training units are not efficient for edge devices, and development of the architectures with online {on-chip} digital, mixed-signal and analog training architectures is important.

{
The online digital training and learning architectures based on the combination of memristive crossbars with digital training circuits for neural network implementation have been recently proposed in \cite{7527510}, \cite{7010034}, \cite{8060399}.  In \cite{8060399}, the digital training architecture for memristive DNN is proposed to accelerate the learning process and transfer it to hardware. The work \cite{hasan2017chip} illustrates a mixed-signal design of neural network with analog neurons and digital error calculation and on-chip training. 
}

For near sensor processing, it is essential to use analog systems that can be easily integrated with analog sensors without additional stages of analog to digital and digital to analog conversion. 
{
Several works investigate the analog learning circuits for neural networks \cite{ISCASs}, \cite{learning18}, \cite{zhang2017memristor} and HTM \cite{tcad}.
In the implementation of backpropagation shown in Fig. \ref{arc3} \cite{zhang2017memristor}, the errors from the output neurons in the second layer are propagated back, and the memristors of the second and first neural network layer are updated sequentially. The memristors of the layer, which is not currently updated, are isolated by the memristive switch. The amount of the update value is proposed to be calculated on FPGA or using Look-Up Table (LUT).
}

{
The other recently proposed training architecture is illustrated in Fig. \ref{newTCAS} \cite{learning18}, showing the completer hardware implementation of the calculation of backpropagation of error with the derivative of the activation function, relevant multiplication circuits,  control transistors and analog weight update circuit. Also, the research work \cite{learning18} illustrates the application of analog backpropagation circuit for different analog learning architecture, such as Multiple Neural Network (MNN) containing several neural networks processing different types of data and ANN decision layer, Binary Neural Network (BNN) based only on two state memristors, DNN, LSTM and HTM.
Even though, several memristive analog implementations of neural networks has been proposed recently, the optimization and testing of fully analog learning systems with control circuitry 
without digital processing is still an open problem.
}

{
In the online training, one of the main issues of the learning process in memristor-based architectures is the update speed of the memristive weights. 
To update weights in a memristive crossbar, different update techniques can be used. The memristive synapses containing only memristors (1M,2M) and memristive synapses with transistors (1T1M and 2T1M) can be updated one at a time, which is a slow process. In this scheme, 1T1M and 2T1M allow to disconnect the memristors which are not involved in the update process  completely and eliminate the leakage currents and effect on those memristors. To speed up the learning process, memristors in a crossbar can be updated in 2 steps: 1) update all memristive weights requiring the change from $R_{ON}$ to $R_{OFF}$, and 2) update the others requiring the change from $R_{OFF}$ to $R_{ON}$ \cite{8329223}. This method can be more efficient for the small crossbars with negligible leakage current and for modular crossbar approach, which is proposed to reduce the leakage currents in the memristive crossbar by dividing a large crossbar into smaller sub-crossbars \cite{mikhailenko2018m}, where all sub-crossbars can be updated in parallel reducing the training time.
}

\section{Discussion}

\label{sec5}

This section includes the discussion of the advantages of memristive neuromorphic architectures, challenges that may occur during the simulation and implementation of the real system and open problems that should be addressed for efficient implementation and integration of neuromorphic architectures into the edge devices.
{
In the simulation of such large neuro-memristive networks, the selection of memristor model is one of the challenging tasks, which is discussed in Appendix A.
}

\subsection{Advantages of memristive architectures}

The main advantages of the memristor-based systems for edge computing applications are the small on-chip area, low power dissipations, and scalability of the memristor-based systems. Therefore, the memristor circuits are a promising solution for edge-computing devices, where the computation is performed on the device without sending information into the cloud.

{
\subsubsection{Push from market and users}
The increased number of edge devices in Internet of things and Cyber Physical System frameworks is driven by the needs from the users for applications such as for gaming, object detection, augmented reality, artificial intelligence, video analytic, and mobile computing \cite{gubbi2013internet, iot2017}. This demands devices and chips that consume low energy, smaller area, and can provide higher computational capacity. The memristive architectures is envisaged to have this potential to achieve these objectives promoting more than Moore's law integration, and emerging intelligent applications \cite{irds2017}. }

\subsubsection{On-chip area and power dissipation}
The advantages of the implementation of memristive circuits include the significant reduction of on-chip area and power dissipation. In several systems, memristor is proposed to be used instead of resistors due to the small on-chip area and low power dissipation.
For example, in comparison to CMOS-based design,for the memristive CAM array design, on-chip area and average power consumption are reduces by 45\% and 96\%, respectively \cite{5471063}.
The area of memristive devices varies based on the used materials and the required resistive levels. 
{
The area of memristive devices of various materials can vary from micron to sub-10 nm depending of the required device properties \cite{7527253,prezioso2015training,6889951,jiang2016sub}.
}

\subsubsection{Scalability}
The application of memristive devices allows scaling the systems because memristor does not exhibit leakage current problems, comparing to transistors and resistors. One of the most efficient solutions is scalable memristive crossbar structures. However, large crossbars can exhibit sneak path problems and the small variability of crossbar outputs. As a solution to this problem,
the scalability of the memristive circuits and arrays can also be achieved by dividing the large memristive arrays into smaller sub-arrays \cite{7024182}. Other well known solutions are to use selector devices along with memristors as outlined in previous sections, which however increases the cell area.

\subsection{Major issues, open problems, and future work prospective}

Even though there are a lot of benefits of memristor-based systems for edge computing applications, the research field of memristive circuits is not mature enough for commercial chip design solutions. Therefore, there is many drawbacks and open problems that can be investigated in future, such as compatibility issues, unstable switching behavior, limitations in the range of resistance and number of resistive levels, the complexity of fabrication of memristive systems and various issues of implementation of large-scale complex systems.


\subsubsection{Memristor materials and compatibility issues}
One of the major issues of the memristive circuits based design is the compatibility of memristive elements with the CMOS technology and fabrication issues. Several memristive devices are proven to be compatible with the CMOS fabrication process \cite{5471063,chen2018cmos}. 
While $TiO_{2-x}$ memristors were quite popular, there are other growing list of memristors based on materials such as $HfO_x$, $TaO_x$, $MoO_x$, $La_{1−x}Sr_xMnO_3$, $InGaZnO$ \cite{kim2017analog}, organic memristors with electrografted  redox  thin film \cite{6968169}, ferroelectric tunnel memristors (FTM), $Ge_2Sb_2Te_5$ (GST) memristors \cite{xiao2018gst}, $SiO_x$ \cite{chang2016demonstration}, $SiN_x$ \cite{kim2017analog} and $Pr_{0.7}Ca_{0.3}MnO_{3}$  (PCMO) \cite{6573409}.
As the memristor technology is only at early stages of development, the properties, stability issues, switching behavior and compatibility with CMOS devices of various memristive elements and selection of most stable material stack is an open problem.

 


\subsubsection{Variability in switching behavior}

The variability issues are common in the memristive devices due to the immaturity of the memristive technology. The switching behavior of the memristive devices may vary, which affects the performance accuracy of many architectures \cite{alibart2012high}. Even though most of the memristor models {used for simulations} illustrate the ideal switching behavior, the real devices show the variability in switching behavior. Several works investigate the probability of switching of the memristive devices and apply this property in the stochastic systems \cite{naous2016stochasticity}. While the stochasticity in switching may be useful for some systems, the effects of this behavior on various neuromorphic architectures and learning systems have not been investigated yet. {While, there have been works that have shown that the use of learning can compensate for variability at system level in digital neural architectures, implementation of learning algorithms with memristors for analog neural network remains a challenging problem.
In addition, the effect on the learning process and training speed should also be explored. It should be also noted that there are several memristor devices proposed in the last decade, device to device variability is high and large majority of them are still in its infancy for industrial use. }

\subsubsection{Range of resistance and number of stable states in memristive devices}
According to the material and physical properties, different memristive devices can be programmed into different ranges of resistance different 
{
and the number of stable resistive states. In most of the cases, the neuromorphic architectures are designed for a specific range of resistance and do not take into account the restricted number of resistive levels when simulating the overall system. In the real devices, depending on the material and fabrication process by changing the width of active layer, these parameters can vary, and the number of resistive states is finite. The recent research works show the memristive devices  can achieve up to 64 stable resistive states \cite{li2018analogue}.
One of the solution to increase the number of resistive level is to use parallel and series combination of memristive devices, which also implies that the issue of memristor interconnection should be considered. The issue of limited number of the resistive levels can be mitigated adding the additional circuits and components, however this increases complexity, number of components and power dissipation. 
} 


From the device perspective, the open problems include the investigation of the possibility to improve the number of  resistive states and the investigation of possible materials that can be used for such purposes. From a mathematical modeling perspective, the model of the memristor incorporating the limited number of stable resistive states and non-linearity of the switching between different states that reflect a realistic memristor is still an open problem. From circuit and system design perspective, it is essential to consider the limited number of stable resistive states in the design and investigate the effect of this issue on the overall system performance.


{
\subsubsection{Endurance of the memristor}
Lifetime and reliability of memristive devices is a subject for the investigation, as there is a large number of memristive materials, in which endurance properties may vary. For example, \cite{yang2010high} reports that $TiO_x$ and $TaO_x$ devices have an endurance of $10^5$ and $10^9$ cycles for $1\mu s$ applied voltage pulses, respectively.
}
{
The endurance and
reliability of memristive devices depend on process variability, including device-to-device and cycle-to-cycle variations, and endurance degradation referring to limited number of update cycles \cite{amat2018memristive}. 
Cycle-to-cycle variability depends on the material of a memristor, while device-to-device variability refers to time-varying device stability depending on the manufacturing process and operation parameters, such as voltage, temperature and duration of applied voltage pulses \cite{fantini2013intrinsic,amat2018memristive,kim2016voltage}.
}

In the edge computing architectures, especially involving the learning and training process, the lifetime of the memristor and number of possible update {cycles} is critical. 
{
For example, the online learning process to train simple two layer ANN for simple XOR problem requires 5000 training iterations \cite{learning18}, and involves the continuous update of the memristive synapses. 
Moreover, to achieve a high-performance accuracy of the neural network, usually the learning rate is decreased, which leads to the requirement to increase the number of update cycles \cite{learning18}.
}
 Therefore, it is important to investigate the endurance, reliability and lifetime limits of various memristive devices.

\subsubsection{Integration with CMOS devices and CMOS issues}

Considering the current trends in the technology market, it will be impossible to avoid the integration of the memristive devices into the CMOS architectures. Considering the importance of the implementation of the read and write circuits for the memristive devices, which are mostly based on the CMOS transistors, the number of CMOS devices per chip will be increased with the increase of the size of memristive architectures, primarily when the synapses or neurons in the neuromorphic architectures are based on hybrid CMOS-memristive designs. 

Even though the memristor is a two terminal vertical element \cite{jo2010nanoscale}, which ensures the reduction of the on-chip area of the CMOS-memristive circuits, the fabrication process of the complex neuromorphic architectures may still be difficult. 
In a complex multilayer structure, where the memristive arrays are combined with CMOS circuits, the fabrication temperature is a critical issue. In combination with CMOS devices, high deposition temperature can damage the devices, while low temperature cannot guarantee the reliable connection between the elements.
This also may increase the cost of such memristor-based architectures and systems. There is a variety of materials that exhibit memristive behavior; however, not all of them can be used for the fabrication of the complex architectures.
The fabrication issues should be considered during the design stage and selection of the memristive elements.
{In the recent years, the successful integration of memristive devices into CMOS architectures is performed using Back End Of Line (BEOL) process and building a layer of memristors on top of the existing chip \cite{beolknowm}.}

{
In addition, the increase in the number of CMOS devices on a chip, especially for such complex architectures as neural networks, leads to high power consumption. To avoid this issue, the size of CMOS devices should be decreased leading to lower supply voltages.}
As it is impossible to decrease the size of the CMOS devices further and maintain an accurate and precise performance of the device at the same time, the replacements of the CMOS devices, such as FinFET devices, should be further investigated and used in the memristive circuits.


\subsubsection{Implementation of large scale systems}

The investigation of complex multilayer architectures and systems is essential to ensure the scalability and accuracy of edge computing devices.
Most of the recent works representing the complex multilayer systems are digital and based on Field Programmable Gate Arrays (FPGA). However, for edge devices that are restricted in terms of area and power consumption, FPGA is not an efficient solution. The number of complex mixed-signal and analog implementations of the neuromorphic systems and architectures is limited.
There are plenty of implementations of simple neural networks, such as perceptron and feedforward neural network, which proves the concepts and illustrates a solution of a particular problem using a particular database. However, more complex and generalized systems have not been investigated yet, and it is important to consider scalability and performance issues of multilayer systems.
{
Also, in a full chip design of a complex system, it is important to consider the interconnection of memristive circuits with the other  elements and complexity of datapath. These are design specific and application dependant issues. For signal processing in analog domain, the interconnection of elements can introduce parasitics having a significant impact on a system performance, comparing to digital signal processing, where the effect of signal integrity issues can be easily mitigated. The interconnect networks for memristive crossbars are studies in \cite{xie2015interconnect,du2017interconnect}.
}
As a memristor is a vertical device and the number of layers in the deep learning systems can be substantial, the possibility of implementation of vertical on-chip systems can also be investigated.


\section{Conclusion}

{In this paper, we presented an overview of a range of  neuro-memristive circuits and architectures that is suitable to be developed as integrated circuit chips in edge computing devices. The pressing hardware issues and challenges involving emerging memristive circuits are presented. The growth of Internet of things and its growing impact on applications for drives the need to have smarter and faster computing in edge devices. Neuro-memristive architectures aims to emulate algorithms such as that based on neural networks and information processing mechanisms in human brain. The ability to (1) have lower on-chip area and power requirements, and (2) incorporate analog dot-product computing with memristive arrays, enables a highly efficient and scalable implementation possibility for on-chip neural networks. } While these architectures can be a promising solution for efficiency and energy issues of edge devices, various challenges and drawbacks should be considered during the design to make their architectures applicable for edge devices. The open problems include various memristive device issues, the ability of integration and implementation of complex systems.

\appendices
\label{ap1}
\section{Selection of memristor model}

To move from theoretical designs and simulations of the neuromorphic architectures to the implementation of the real chips and integration on the neuromorphic designs into the existing sensors, it is important to consider side effects, nonlinearities and drawbacks of the memristive circuits during the simulation process \cite{mazumder2012memristors}. The selection of the memristor model can affect the simulation results significantly. The ideal memristor models will not consider non-linear effects of the real implementation, and the simulation results will not be reliable. While the memristor models that are not designed for large-scale simulations may cause the simulations errors and non-convergence issues in the SPICE simulation of large architectures. Table \ref{table2} illustrates the most commonly used memristor models and their characteristics.
{
More comprehensive review and consideration of the other memristor models is provided in \cite{bayat2015phenomenological, biolek2018modeling, singh2018comparative}.
}

\begin{table*}[]
\centering
\caption{Comparison of the memristor models}
\label{table2}
\begin{tabular}{|l|l|l|l|l|}
\hline
\textbf{Memristor model}                                                                                                                           & \multicolumn{1}{c|}{\textbf{Description}}                                                                                                                        & \multicolumn{1}{c|}{\textbf{Linearity}}                          & \multicolumn{1}{c|}{\textbf{\begin{tabular}[c]{@{}c@{}}Consideration of\\  physical parameters \\ of the memristor\end{tabular}}} & \multicolumn{1}{c|}{\textbf{\begin{tabular}[c]{@{}c@{}}Application\\  for large scale simulations\end{tabular}}}                        \\ \hline
\begin{tabular}[c]{@{}l@{}}Linear dopant \\ drift models \\ \cite{pickett2009switching, strukov2008missing}\end{tabular}                          & \begin{tabular}[c]{@{}l@{}}Emulate the switching behavior of the devices \\ and do not consider the effects of electric\\  field and nonlinearities\end{tabular} & Linear                                                           & \begin{tabular}[c]{@{}l@{}}partially\\ considered\end{tabular}                                                                    & \begin{tabular}[c]{@{}l@{}} {less computationally complex than} \\ {non-linear  models; however can only } \\{be used for a proof of concept \cite{biolek2018modeling} } \end{tabular}                                       \\ \hline
\begin{tabular}[c]{@{}l@{}}Nonlinear dopant\\  drift models\\  \cite{prodromakis2011versatile, joglekar2009elusive, biolek2009spice}\end{tabular} & \begin{tabular}[c]{@{}l@{}}Models with different window functions \\ and consider the non-linear switching behavior\end{tabular}                                 & Non-linear                                                       & not considered                                                                                                                    & \begin{tabular}[c]{@{}l@{}}reduced simulation speed due to the\\  complexity of window function\end{tabular}                            \\ \hline
\begin{tabular}[c]{@{}l@{}}TEAM model\\  \cite{6353604}\end{tabular}                                                                              & \begin{tabular}[c]{@{}l@{}}Generalized model containing various \\ window functions, nonlinear switching and \\ effect of.physical parameters\end{tabular}       & Non-linear                                                       & considered                                                                                                                        & \begin{tabular}[c]{@{}l@{}}difficult to use in extremely large\\  arrays due to the complexity\end{tabular}                             \\ \hline
\begin{tabular}[c]{@{}l@{}}Modified \\ Biolek`s models \\ \cite{biol,biolek2018modeling}\end{tabular}                                             & \begin{tabular}[c]{@{}l@{}}Modification of the existing models designed \\ for simulation improvement\end{tabular}                                               & \begin{tabular}[c]{@{}l@{}}Linear and \\ non-linear\end{tabular} & \begin{tabular}[c]{@{}l@{}}partially\\  considered\end{tabular}                                                                   & \begin{tabular}[c]{@{}l@{}}can be used for large scale simulations \\ without numerical problems and \\ convergence issues\end{tabular} \\ \hline

\begin{tabular}[c]{@{}l@{}}{Data driven} \\ {simplified model} \\  \cite{messaris2018data}\end{tabular}                                             & \begin{tabular}[c]{@{}l@{}} {Model contains a window function allowing the }\\ { derivation of a resistive state time-response }\\{ expression for constant bias voltage}\end{tabular}                                               & \begin{tabular}[c]{@{}l@{}}{Non-linear}\end{tabular} & \begin{tabular}[c]{@{}l@{}} {considered}\end{tabular}                                                                   & \begin{tabular}[c]{@{}l@{}} {includes data driven parameters }\\ {and can be used for large scale }\\{ simulations  without  convergence issues}\end{tabular} \\ \hline

\end{tabular}
\end{table*}



\subsubsection{Early linear approximations and equivalent circuits}

One of the earliest works on linear approximations and equivalent circuits for current and voltage-based memristor models is proposed in \cite{12}. The circuits introduce the basic memristor concepts and are not used in the recent memristive architectures and systems. The equivalent circuit based memristor macro models are shown in \cite{batas2011memristor, 5433753, benderli2009spice}. These models are rarely used in the large-scale system simulation due to the complexity and lack of consideration of non-linearity and physical parameters of real devices. 


\subsubsection{Linear memristor models}
The other major class of the memristor models is linear ideal models. Linear memristor model emulates the switching behavior of the devices and does not consider the effects of electric field on the device performance. The simplest linear memristor model is shown by Pickett at al. in \cite{pickett2009switching}.
This model is based on drift mechanism of ionized dopants and emulates $TiO_2$ memristor \cite{7117477}. The linear relationship between the voltage and current in the memristor can be described as $v(t)=(R_{ON} \times x(t)+R_{OFF}(1− x(t)))\times i(t)$, where $x(t)=w(t)/D$ and $D$ is a width of the device and $w(t)$ is a width of a doped region at a particular time \cite{keshmiri2014study}. The linear window function can be shown as: $f (w) = w(D − w)/D$ \cite{strukov2008missing}. Even though this memristor model is frequently used in the simulations of neuromorphic circuits \cite{5976989}, it does not show various effects of non ideal behavior of real memristive devices.

\subsubsection{Nonlinear memristor models}


In the real device, the drift, diffusion, and thermophoresis due to ionic motion cause the nonlinear relationship between memristor current and voltage as well as nonlinear dynamical switching behavior \cite{strachan2013state}.
In comparison to the ideal linear memristor models, like as Pickett model \cite{pickett2009switching}, that was used earlier to simulated the memristive architectures and prove the concept of the design of various neuromorphic architectures, the recent research works focus on memristor model containing nonlinearity effects. It is vital to consider non-idealities of the memristive devices because the memristive technology is not mature yet. The lack of stability makes the nonlinearity factor to be relevant to investigate, primarily when the large-scale simulations are performed.

One of the known nonlinear memristor models is Joglekar`s model that allows controlling a non-linearity windowing function \cite{joglekar2009elusive}.
The main parameters causing non-linearities are $W/D$ ratio and $p$, where $W$ are the actual width, $D$ is a width of the thin film, and $p$ is a parameter of the window-function for modeling of nonlinear boundary conditions. The Joglekar's window function is represented as $f(x,p)=1-(2x-1)^{2 \times p}$. According to this equation, the parameter $p$ is responsible for the linearity of the memristor model that increases with the increase of $p$. The example of the application of this model is the CeNN architecture shown in \cite{di2017memristor}.


The memristor shown in research work \cite{biolek2009spice} is used in several neuromorphic architectures. The macro model is based on the study of the behavior of the $TiO_2$ memristor illustrated in \cite{strukov2008missing}. The model allows modifying the nonlinear boundary conditions that are not considered by the simplified linear memristor models. The model is based on the modification of non-linear window function from \cite{joglekar2009elusive} demonstrating non-linearities caused by non-linear dopant drift. The main parameters causing non-linearities are $i$, $W/D$ ratio and $p$, where $i$ is a current flowing through the memristor. The window function is represented as: $f(x,i,p)=1-(x-step(-i))^{2 \times p}$, where $step(-i)=0$ for $i<0$ and $step(-i)=1$ for $i\geq 0$.  The window function is involved into the calculation of the resistance value of the memristor and the speed of the movement of the boundary between the
doped and undoped regions of the memristor, which determines how fast the resistive state of a memristor changes. The main difference between the Joglekar`s and Biolek's memristor models \cite{biolek2009spice} is the ability of Biolek's model to reversely change the memristance after a reaching one of the resistance boundary \cite{6573409}.

The non-linear model that can be adjusted and scaled is proposed in \cite{prodromakis2011versatile}. This window function for this model is the following: 
$f (w)= j(1-[(w-0.5)^2+0.75]^p)$, where $j$ represents a control parameter to specify highest value of window function \cite{prodromakis2011versatile,keshmiri2014study}.


Even the memristor models proposed in \cite{joglekar2009elusive}, \cite{prodromakis2011versatile} and \cite{biolek2009spice} contains the nonlinear switching behavior of the memristor, the nonlinearities of the device, parasitic effects, leakages and other physical imperfections are not considered. The physical imperfections of the device are considered in \cite{memristormodel} and \cite{6353604}.

The memristor model that is used in many neuromorphic architectures is ThrEshold Adaptive Memristor (TEAM) model \cite{6861426}. The model is proposed in \cite{6353604}. The model represents a generalized solution for Simmons Tunnel Barrier model \cite{simmons1963generalized, memristormodel} that is complicated and designed for a particular memristor type \cite{keshmiri2014study}. This model illustrates a generalized approach and can be adjusted for the behavior of different memristive devices and different window functions considering various physical effects \cite{memristormodel}.
The model is used in different system level simulations and is proven to be useful for fast digital systems \cite{xavier2018memristive}.


\subsubsection{Memristor models for large-scale simulations}

Most of the architectures are realized on memristive crossbars and the amount of processed data is significant, especially in edge computing when the data is not sent to the cloud.
While the proof of the concept and overall ability of the system to perform a certain task can be tested using ideal memristors, the real performance of the memristive systems require more accurate non-linear memristor models. Therefore, one of the most important { aspect in memristor modeling is to} take into account the non-linearity problem and physical effects. However, the large-scale simulations of memristive systems cause various numerical problems and make it impossible to check the performance of the large-scale system.
{ 
The large number of internal equations in the models and the mathematical form of those equations, especially the ones incorporating non-linear dynamics of memristive devices, can cause the problems of data overflow and convergence issues 
\cite{biolek2018modeling}.
Therefore, it is important to find a trade-off between the accuracy of memristor models and computational complexity.
}

The modification of memristor model that can be used for large-scale simulations is shown in \cite{biol} and \cite{biolek2018modeling}. The model does not contain the window functions and allows to avoid different numerical problems and non-convergence issues \cite{biolek2018modeling}.  The research work \cite{biolek2018modeling} proposes the modification of complex physical–phenomenological nonlinear models appropriate for large-scale simulations of multilayer architectures for edge computing.
{
The other recent model suitable for large scale simulations is shown in \cite{messaris2018data}. This model simplifies the data fitting process, introduces a window function allowing the derivation of a resistive state time-response expression for constant bias voltage, and provides the possibility to perform computationally efficient simulations of the designed architectures for more realistic conditions. The model is data driven and provides the example of fitting parameters for $TiO_x$ and $TaO_x$ devices. 
}

\label{sec6}





\ifCLASSOPTIONcaptionsoff
  \newpage
\fi



%

\bibliographystyle{IEEEtran}
\bibliography{ref}

\begin{thebibliography}{100}
\providecommand{\url}[1]{#1}
\csname url@samestyle\endcsname
\providecommand{\newblock}{\relax}
\providecommand{\bibinfo}[2]{#2}
\providecommand{\BIBentrySTDinterwordspacing}{\spaceskip=0pt\relax}
\providecommand{\BIBentryALTinterwordstretchfactor}{4}
\providecommand{\BIBentryALTinterwordspacing}{\spaceskip=\fontdimen2\font plus
\BIBentryALTinterwordstretchfactor\fontdimen3\font minus
  \fontdimen4\font\relax}
\providecommand{\BIBforeignlanguage}[2]{{%
\expandafter\ifx\csname l@#1\endcsname\relax
\typeout{** WARNING: IEEEtran.bst: No hyphenation pattern has been}%
\typeout{** loaded for the language `#1'. Using the pattern for}%
\typeout{** the default language instead.}%
\else
\language=\csname l@#1\endcsname
\fi
#2}}
\providecommand{\BIBdecl}{\relax}
\BIBdecl

\bibitem{satyanarayanan2017emergence}
M.~Satyanarayanan, ``The emergence of edge computing,'' \emph{Computer},
  vol.~50, no.~1, pp. 30--39, 2017.

\bibitem{sheltami2018fog}
T.~R. Sheltami, E.~Q. Shahra, and E.~M. Shakshuki, ``Fog computing: Data
  streaming services for mobile end-users,'' \emph{Procedia computer science},
  vol. 134, pp. 289--296, 2018.

\bibitem{panetta2017top}
C.~Panetta, ``Top trends in the gartner hype cycle for emerging technologies.
  2017,'' \emph{Enterprises should explain the business potential of
  blockchain, artificial intelligence and augmented reality}, 2017.

\bibitem{8345562}
M.~Gusev and S.~Dustdar, ``Going back to the roots x2014;the evolution of edge
  computing, an iot perspective,'' \emph{IEEE Internet Computing}, vol.~22,
  no.~2, pp. 5--15, Mar 2018.

\bibitem{8289317}
G.~Premsankar, M.~D. Francesco, and T.~Taleb, ``Edge computing for the internet
  of things: A case study,'' \emph{IEEE Internet of Things Journal}, vol.~5,
  no.~2, pp. 1275--1284, April 2018.

\bibitem{8119503}
G.~Chakma, M.~M. Adnan, A.~R. Wyer, R.~Weiss, C.~D. Schuman, and G.~S. Rose,
  ``Memristive mixed-signal neuromorphic systems: Energy-efficient learning at
  the circuit-level,'' \emph{IEEE Journal on Emerging and Selected Topics in
  Circuits and Systems}, vol.~8, no.~1, pp. 125--136, March 2018.

\bibitem{chua2012hodgkin}
L.~Chua, V.~Sbitnev, and H.~Kim, ``Hodgkin--huxley axon is made of
  memristors,'' \emph{International Journal of Bifurcation and Chaos}, vol.~22,
  no.~03, p. 1230011, 2012.

\bibitem{tcad}
O.~Krestinskaya, T.~Ibrayev, and A.~P. James, ``Hierarchical temporal memory
  features with memristor logic circuits for pattern recognition,'' \emph{IEEE
  Transactions on Computer-Aided Design of Integrated Circuits and Systems},
  vol.~PP, no.~99, pp. 1--1, 2017.

\bibitem{learning18}
O.~Krestinskaya, K.~N. Salama, and A.~P. James, ``Learning in memristive neural
  network architectures using analog backpropagation circuits,'' \emph{IEEE
  Transactions on Circuits and Systems I: Regular Papers}, pp. 1--14, 2018.

\bibitem{8471012}
O.~Krestinskaya, I.~Dolzhikova, and A.~P. James, ``Hierarchical temporal memory
  using memristor networks: A survey,'' \emph{IEEE Transactions on Emerging
  Topics in Computational Intelligence}, vol.~2, no.~5, pp. 380--395, Oct 2018.

\bibitem{9}
\BIBentryALTinterwordspacing
D.~George and J.Hawkins, ``Hierarchical temporal memory: Concepts, theory and
  terminology,'' Tech. Rep., 2006. [Online]. Available:
  \url{http://www-edlab.cs.umass.edu/cs691jj/hawkins-and-george-2006.pdf}
\BIBentrySTDinterwordspacing

\bibitem{smagulova2018memristor}
K.~Smagulova, O.~Krestinskaya, and A.~P. James, ``A memristor-based long short
  term memory circuit,'' \emph{Analog Integrated Circuits and Signal
  Processing}, pp. 1--6, 2018.

\bibitem{li2018long}
C.~Li, Z.~Wang, M.~Rao, D.~Belkin, W.~Song, H.~Jiang, P.~Yan, Y.~Li, P.~Lin,
  M.~Hu \emph{et~al.}, ``Long short-term memory networks in memristor
  crossbars,'' \emph{arXiv preprint arXiv:1805.11801}, 2018.

\bibitem{appleinsider_2018}
\BIBentryALTinterwordspacing
AppleInsider, ``iphone xs a12 bionic chip features 7nm design, next-gen neural
  engine,'' Sep 2018. [Online]. Available:
  \url{https://appleinsider.com/articles/18/09/12/iphone-xs-a12-bionic-chip-features-7nm-design-next-gen-neural-engine}
\BIBentrySTDinterwordspacing

\bibitem{kirin980}
\BIBentryALTinterwordspacing
``The world's first 7nm process mobile ai chipset,'' Nov 2018. [Online].
  Available: \url{https://consumer.huawei.com/en/campaign/kirin980/}
\BIBentrySTDinterwordspacing

\bibitem{schuman2017survey}
C.~D. Schuman, T.~E. Potok, R.~M. Patton, J.~D. Birdwell, M.~E. Dean, G.~S.
  Rose, and J.~S. Plank, ``A survey of neuromorphic computing and neural
  networks in hardware,'' \emph{arXiv preprint arXiv:1705.06963}, 2017.

\bibitem{kahng2010scaling}
A.~B. Kahng, ``Scaling: More than moore's law,'' \emph{IEEE Design \& Test of
  Computers}, vol.~27, no.~3, pp. 86--87, 2010.

\bibitem{bohr2017cmos}
M.~T. Bohr and I.~A. Young, ``Cmos scaling trends and beyond,'' \emph{IEEE
  Micro}, vol.~37, no.~6, pp. 20--29, 2017.

\bibitem{hu2018cross}
X.~S. Hu, ``A cross-layer perspective for energy efficient processing:-from
  beyond-cmos devices to deep learning,'' in \emph{Proceedings of the 2018 on
  Great Lakes Symposium on VLSI}.\hskip 1em plus 0.5em minus 0.4em\relax ACM,
  2018, pp. 7--7.

\bibitem{irds2017}
F.~Balestra, M.~Graef, B.~Huizing, Y.~Hayashi, H.~Ishiuchi, T.~Conte, and
  P.~Gargini, ``Executive summary,'' \emph{International Roadmap for Devices
  and Systems}, 2017.

\bibitem{wang2009spintronic}
X.~Wang, Y.~Chen, H.~Xi, H.~Li, and D.~Dimitrov, ``Spintronic memristor through
  spin-torque-induced magnetization motion,'' \emph{IEEE electron device
  letters}, vol.~30, no.~3, pp. 294--297, 2009.

\bibitem{kuzum2011nanoelectronic}
D.~Kuzum, R.~G. Jeyasingh, B.~Lee, and H.-S.~P. Wong, ``Nanoelectronic
  programmable synapses based on phase change materials for brain-inspired
  computing,'' \emph{Nano letters}, vol.~12, no.~5, pp. 2179--2186, 2011.

\bibitem{ho2009programmable}
C.~Ho, E.~K. Lai, and K.~Y. Hsieh, ``Programmable resistive ram and
  manufacturing method,'' Sep.~29 2009, uS Patent 7,595,218.

\bibitem{kim2011functional}
K.-H. Kim, S.~Gaba, D.~Wheeler, J.~M. Cruz-Albrecht, T.~Hussain, N.~Srinivasa,
  and W.~Lu, ``A functional hybrid memristor crossbar-array/cmos system for
  data storage and neuromorphic applications,'' \emph{Nano letters}, vol.~12,
  no.~1, pp. 389--395, 2011.

\bibitem{hady2017platform}
F.~T. Hady, A.~Foong, B.~Veal, and D.~Williams, ``Platform storage performance
  with 3d xpoint technology,'' \emph{Proceedings of the IEEE}, vol. 105, no.~9,
  pp. 1822--1833, 2017.

\bibitem{jiang2018provable}
H.~Jiang, C.~Li, R.~Zhang, P.~Yan, P.~Lin, Y.~Li, J.~J. Yang, D.~Holcomb, and
  Q.~Xia, ``A provable key destruction scheme based on memristive crossbar
  arrays,'' \emph{Nature Electronics}, vol.~1, no.~10, p. 548, 2018.

\bibitem{maan2017survey}
A.~K. Maan, D.~A. Jayadevi, and A.~P. James, ``A survey of memristive threshold
  logic circuits,'' \emph{IEEE transactions on neural networks and learning
  systems}, vol.~28, no.~8, pp. 1734--1746, 2017.

\bibitem{shi2016edge}
W.~Shi, J.~Cao, Q.~Zhang, Y.~Li, and L.~Xu, ``Edge computing: Vision and
  challenges,'' \emph{IEEE Internet of Things Journal}, vol.~3, no.~5, pp.
  637--646, 2016.

\bibitem{mao2016dynamic}
Y.~Mao, J.~Zhang, and K.~B. Letaief, ``Dynamic computation offloading for
  mobile-edge computing with energy harvesting devices,'' \emph{IEEE Journal on
  Selected Areas in Communications}, vol.~34, no.~12, pp. 3590--3605, 2016.

\bibitem{chakma2018energy}
G.~Chakma, N.~D. Skuda, C.~D. Schuman, J.~S. Plank, M.~E. Dean, and G.~S. Rose,
  ``Energy and area efficiency in neuromorphic computing for resource
  constrained devices,'' in \emph{Proceedings of the 2018 on Great Lakes
  Symposium on VLSI}.\hskip 1em plus 0.5em minus 0.4em\relax ACM, 2018, pp.
  379--383.

\bibitem{8060399}
M.~Cheng, L.~Xia, Z.~Zhu, Y.~Cai, Y.~Xie, Y.~Wang, and H.~Yang, ``Time: A
  training-in-memory architecture for memristor-based deep neural networks,''
  in \emph{2017 54th ACM/EDAC/IEEE Design Automation Conference (DAC)}, June
  2017, pp. 1--6.

\bibitem{hasan2017chip}
R.~Hasan, T.~M. Taha, and C.~Yakopcic, ``On-chip training of memristor crossbar
  based multi-layer neural networks,'' \emph{Microelectronics Journal},
  vol.~66, pp. 31--40, 2017.

\bibitem{7727302}
C.~Yakopcic, M.~Z. Alom, and T.~M. Taha, ``Memristor crossbar deep network
  implementation based on a convolutional neural network,'' in \emph{2016
  International Joint Conference on Neural Networks (IJCNN)}, July 2016, pp.
  963--970.

\bibitem{7966055}
------, ``Extremely parallel memristor crossbar architecture for convolutional
  neural network implementation,'' in \emph{Neural Networks (IJCNN), 2017
  International Joint Conference on}.\hskip 1em plus 0.5em minus 0.4em\relax
  IEEE, 2017, pp. 1696--1703.

\bibitem{goodfellow2014generative}
I.~Goodfellow, J.~Pouget-Abadie, M.~Mirza, B.~Xu, D.~Warde-Farley, S.~Ozair,
  A.~Courville, and Y.~Bengio, ``Generative adversarial nets,'' in
  \emph{Advances in neural information processing systems}, 2014, pp.
  2672--2680.

\bibitem{abunahla2018memristor}
H.~Abunahla and B.~Mohammad, \emph{Memristor Technology: Synthesis and Modeling
  for Sensing and Security Applications}.\hskip 1em plus 0.5em minus
  0.4em\relax Springer, 2018.

\bibitem{trust}
J.~Rajendran, O.~Sinanoglu, and R.~Karri, ``Regaining trust in vlsi design:
  Design-for-trust techniques,'' \emph{Proceedings of the IEEE}, vol. 102,
  no.~8, pp. 1266--1282, Aug 2014.

\bibitem{rajendran2012nano}
J.~Rajendran, G.~S. Rose, R.~Karri, and M.~Potkonjak, ``Nano-ppuf: A
  memristor-based security primitive,'' in \emph{VLSI (ISVLSI), 2012 IEEE
  Computer Society Annual Symposium on}.\hskip 1em plus 0.5em minus 0.4em\relax
  IEEE, 2012, pp. 84--87.

\bibitem{mazady2015memristor}
A.~Mazady, M.~T. Rahman, D.~Forte, and M.~Anwar, ``Memristor puf—a security
  primitive: Theory and experiment,'' \emph{IEEE Journal on Emerging and
  Selected Topics in Circuits and Systems}, vol.~5, no.~2, pp. 222--229, 2015.

\bibitem{rose2013hardware}
G.~S. Rose, J.~Rajendran, N.~McDonald, R.~Karri, M.~Potkonjak, and B.~Wysocki,
  ``Hardware security strategies exploiting nanoelectronic circuits,'' in
  \emph{Design Automation Conference (ASP-DAC), 2013 18th Asia and South
  Pacific}.\hskip 1em plus 0.5em minus 0.4em\relax IEEE, 2013, pp. 368--372.

\bibitem{tino2015artificial}
P.~Tino, L.~Benuskova, and A.~Sperduti, ``Artificial neural network models,''
  in \emph{Springer Handbook of Computational Intelligence}.\hskip 1em plus
  0.5em minus 0.4em\relax Springer, 2015, pp. 455--471.

\bibitem{6322959}
S.~Carrillo, J.~Harkin, L.~J. McDaid, F.~Morgan, S.~Pande, S.~Cawley, and
  B.~McGinley, ``Scalable hierarchical network-on-chip architecture for spiking
  neural network hardware implementations,'' \emph{IEEE Transactions on
  Parallel and Distributed Systems}, vol.~24, no.~12, pp. 2451--2461, Dec 2013.

\bibitem{caze2014dendrites}
R.~D. Caz{\'e}, M.~D. Humphries, and B.~S. Gutkin, ``Dendrites enhance both
  single neuron and network computation,'' in \emph{The Computing
  Dendrite}.\hskip 1em plus 0.5em minus 0.4em\relax Springer, 2014, pp.
  365--380.

\bibitem{hawkinsintelligence}
J.~Hawkins and S.~Blakeslee, ``On intelligence. 2004,'' \emph{New York St.
  Martin's Griffin}, pp. 156--8.

\bibitem{hawkins2016neurons}
J.~Hawkins and S.~Ahmad, ``Why neurons have thousands of synapses, a theory of
  sequence memory in neocortex,'' \emph{Frontiers in neural circuits}, vol.~10,
  p.~23, 2016.

\bibitem{rojas2013neural}
R.~Rojas, \emph{Neural networks: a systematic introduction}.\hskip 1em plus
  0.5em minus 0.4em\relax Springer Science \& Business Media, 2013.

\bibitem{6889951}
Z.~Wang, W.~Zhao, W.~Kang, Y.~Zhang, J.~O. Klein, and C.~Chappert,
  ``Ferroelectric tunnel memristor-based neuromorphic network with 1t1r
  crossbar architecture,'' in \emph{2014 International Joint Conference on
  Neural Networks (IJCNN)}, July 2014, pp. 29--34.

\bibitem{mcculloch1943logical}
W.~S. McCulloch and W.~Pitts, ``A logical calculus of the ideas immanent in
  nervous activity,'' \emph{The bulletin of mathematical biophysics}, vol.~5,
  no.~4, pp. 115--133, 1943.

\bibitem{rosenblatt1958perceptron}
F.~Rosenblatt, ``The perceptron: a probabilistic model for information storage
  and organization in the brain.'' \emph{Psychological review}, vol.~65, no.~6,
  p. 386, 1958.

\bibitem{7568628}
M.~Shahsavari, P.~Falez, and P.~Boulet, ``Combining a volatile and nonvolatile
  memristor in artificial synapse to improve learning in spiking neural
  networks,'' in \emph{2016 IEEE/ACM International Symposium on Nanoscale
  Architectures (NANOARCH)}, July 2016, pp. 67--72.

\bibitem{polsky2009encoding}
A.~Polsky, B.~Mel, and J.~Schiller, ``Encoding and decoding bursts by nmda
  spikes in basal dendrites of layer 5 pyramidal neurons,'' \emph{Journal of
  Neuroscience}, vol.~29, no.~38, pp. 11\,891--11\,903, 2009.

\bibitem{prezioso2015training}
M.~Prezioso, F.~Merrikh-Bayat, B.~Hoskins, G.~Adam, K.~K. Likharev, and D.~B.
  Strukov, ``Training and operation of an integrated neuromorphic network based
  on metal-oxide memristors,'' \emph{Nature}, vol. 521, no. 7550, p.~61, 2015.

\bibitem{7723927}
M.~Hu, Y.~Chen, J.~J. Yang, Y.~Wang, and H.~H. Li, ``A compact memristor-based
  dynamic synapse for spiking neural networks,'' \emph{IEEE Transactions on
  Computer-Aided Design of Integrated Circuits and Systems}, vol.~36, no.~8,
  pp. 1353--1366, Aug 2017.

\bibitem{yao2017face}
P.~Yao, H.~Wu, B.~Gao, S.~B. Eryilmaz, X.~Huang, W.~Zhang, Q.~Zhang, N.~Deng,
  L.~Shi, H.-S.~P. Wong \emph{et~al.}, ``Face classification using electronic
  synapses,'' \emph{Nature communications}, vol.~8, p. 15199, 2017.

\bibitem{7010034}
D.~Soudry, D.~D. Castro, A.~Gal, A.~Kolodny, and S.~Kvatinsky,
  ``Memristor-based multilayer neural networks with online gradient descent
  training,'' \emph{IEEE Transactions on Neural Networks and Learning Systems},
  vol.~26, no.~10, pp. 2408--2421, Oct 2015.

\bibitem{6074916}
H.~Kim, M.~P. Sah, C.~Yang, T.~Roska, and L.~O. Chua, ``Memristor bridge
  synapses,'' \emph{Proceedings of the IEEE}, vol. 100, no.~6, pp. 2061--2070,
  June 2012.

\bibitem{6939735}
S.~P. Adhikari, H.~Kim, R.~K. Budhathoki, C.~Yang, and L.~O. Chua, ``A
  circuit-based learning architecture for multilayer neural networks with
  memristor bridge synapses,'' \emph{IEEE Transactions on Circuits and Systems
  I: Regular Papers}, vol.~62, no.~1, pp. 215--223, Jan 2015.

\bibitem{6331430}
Y.~S. Kim and K.~S. Min, ``Synaptic weighting circuits for cellular neural
  networks,'' in \emph{2012 13th International Workshop on Cellular Nanoscale
  Networks and their Applications}, Aug 2012, pp. 1--6.

\bibitem{zhang2017synaptic}
Y.~Zhang, Y.~Li, X.~Wang, and E.~G. Friedman, ``Synaptic characteristics of
  ag/aginsbte/ta-based memristor for pattern recognition applications,''
  \emph{IEEE Transactions on Electron Devices}, vol.~64, no.~4, pp. 1806--1811,
  2017.

\bibitem{zhang2017memristor}
Y.~Zhang, X.~Wang, and E.~G. Friedman, ``Memristor-based circuit design for
  multilayer neural networks,'' \emph{IEEE Transactions on Circuits and Systems
  I: Regular Papers}, vol.~65, no.~2, pp. 677--686, 2018.

\bibitem{alibart2013pattern}
F.~Alibart, E.~Zamanidoost, and D.~B. Strukov, ``Pattern classification by
  memristive crossbar circuits using ex situ and in situ training,''
  \emph{Nature communications}, vol.~4, p. 2072, 2013.

\bibitem{hasan2014enabling}
R.~Hasan and T.~M. Taha, ``Enabling back propagation training of memristor
  crossbar neuromorphic processors,'' in \emph{Neural Networks (IJCNN), 2014
  International Joint Conference on}.\hskip 1em plus 0.5em minus 0.4em\relax
  IEEE, 2014, pp. 21--28.

\bibitem{6573409}
A.~M. Sheri, H.~Hwang, M.~Jeon, and B.~g.~Lee, ``Neuromorphic character
  recognition system with two pcmo memristors as a synapse,'' \emph{IEEE
  Transactions on Industrial Electronics}, vol.~61, no.~6, pp. 2933--2941, June
  2014.

\bibitem{wang2017memristors}
Z.~Wang, S.~Joshi, S.~E. Savel'ev, H.~Jiang, R.~Midya, P.~Lin, M.~Hu, N.~Ge,
  J.~P. Strachan, Z.~Li \emph{et~al.}, ``Memristors with diffusive dynamics as
  synaptic emulators for neuromorphic computing,'' \emph{Nature materials},
  vol.~16, no.~1, p. 101, 2017.

\bibitem{wang2018fully}
Z.~Wang, S.~Joshi, S.~Savel'ev, W.~Song, R.~Midya, Y.~Li, M.~Rao, P.~Yan,
  S.~Asapu, Y.~Zhuo \emph{et~al.}, ``Fully memristive neural networks for
  pattern classification with unsupervised learning,'' \emph{Nature
  Electronics}, vol.~1, no.~2, p. 137, 2018.

\bibitem{7527510}
E.~Rosenthal, S.~Greshnikov, D.~Soudry, and S.~Kvatinsky, ``A fully analog
  memristor-based neural network with online gradient training,'' in \emph{2016
  IEEE International Symposium on Circuits and Systems (ISCAS)}, May 2016, pp.
  1394--1397.

\bibitem{danial2017didactic}
L.~Danial, N.~Wainstein, S.~Kraus, and S.~Kvatinsky, ``Didactic: A
  data-intelligent digital-to-analog converter with a trainable integrated
  circuit using memristors,'' \emph{IEEE Journal on Emerging and Selected
  Topics in Circuits and Systems}, 2017.

\bibitem{shamsi2018hardware}
J.~Shamsi, K.~Mohammadi, and S.~B. Shokouhi, ``A hardware architecture for
  columnar-organized memory based on cmos neuron and memristor crossbar
  arrays,'' \emph{IEEE Transactions on Very Large Scale Integration (VLSI)
  Systems}, 2018.

\bibitem{8340055}
Y.~Jiang, P.~Huang, D.~Zhu, Z.~Zhou, R.~Han, L.~Liu, X.~Liu, and J.~Kang,
  ``Design and hardware implementation of neuromorphic systems with rram
  synapses and threshold-controlled neurons for pattern recognition,''
  \emph{IEEE Transactions on Circuits and Systems I: Regular Papers}, pp.
  1--13, 2018.

\bibitem{8275126}
A.~Chowdhury, A.~Ayman, S.~Dey, M.~Sarker, and A.~I. Arka, ``Simulations of
  threshold logic unit problems using memristor based synapses and cmos
  neuron,'' in \emph{2017 3rd International Conference on Electrical
  Information and Communication Technology (EICT)}, Dec 2017, pp. 1--4.

\bibitem{hu2014memristor}
M.~Hu, H.~Li, Y.~Chen, Q.~Wu, G.~S. Rose, and R.~W. Linderman, ``Memristor
  crossbar-based neuromorphic computing system: A case study,'' \emph{IEEE
  transactions on neural networks and learning systems}, vol.~25, no.~10, pp.
  1864--1878, 2014.

\bibitem{ISCASs}
O.~Krestinskaya, K.~N. Salama, and A.~P. James, ``Analog backpropagation
  learning circuits for memristive crossbar neural networks,'' in \emph{2018
  IEEE International Symposium on Circuits and Systems (ISCAS)}, May 2018, pp.
  1--5.

\bibitem{naous2016memristor}
R.~Naous, M.~AlShedivat, E.~Neftci, G.~Cauwenberghs, and K.~N. Salama,
  ``Memristor-based neural networks: Synaptic versus neuronal stochasticity,''
  \emph{AIP Advances}, vol.~6, no.~11, p. 111304, 2016.

\bibitem{7116617}
X.~Wu, V.~Saxena, and K.~Zhu, ``Homogeneous spiking neuromorphic system for
  real-world pattern recognition,'' \emph{IEEE Journal on Emerging and Selected
  Topics in Circuits and Systems}, vol.~5, no.~2, pp. 254--266, June 2015.

\bibitem{7280819}
W.~Xinyu, S.~V, and Z.~Kehan, ``A cmos spiking neuron for dense
  memristor-synapse connectivity for brain-inspired computing,'' in \emph{2015
  International Joint Conference on Neural Networks (IJCNN)}, July 2015, pp.
  1--6.

\bibitem{ebong2012cmos}
I.~E. Ebong and P.~Mazumder, ``Cmos and memristor-based neural network design
  for position detection,'' \emph{Proceedings of the IEEE}, vol. 100, no.~6,
  pp. 2050--2060, 2012.

\bibitem{8123644}
A.~Irmanova, O.~Krestinskaya, and A.~P. James, ``Neuromorphic adaptive
  edge-preserving denoising filter,'' in \emph{2017 IEEE International
  Conference on Rebooting Computing (ICRC)}, Nov 2017, pp. 1--6.

\bibitem{zhang2017synchronization}
J.~Zhang and X.~Liao, ``Synchronization and chaos in coupled memristor-based
  fitzhugh-nagumo circuits with memristor synapse,'' \emph{AEU-International
  Journal of Electronics and Communications}, vol.~75, pp. 82--90, 2017.

\bibitem{jiang2016cyclical}
H.~Jiang, W.~Zhu, F.~Luo, K.~Bai, C.~Liu, X.~Zhang, J.~J. Yang, Q.~Xia,
  Y.~Chen, and Q.~Wu, ``Cyclical sensing integrate-and-fire circuit for
  memristor array based neuromorphic computing,'' in \emph{Circuits and Systems
  (ISCAS), 2016 IEEE International Symposium on}.\hskip 1em plus 0.5em minus
  0.4em\relax IEEE, 2016, pp. 930--933.

\bibitem{shamsi2015hyperbolic}
J.~Shamsi, A.~Amirsoleimani, S.~Mirzakuchaki, A.~Ahmade, S.~Alirezaee, and
  M.~Ahmadi, ``Hyperbolic tangent passive resistive-type neuron,'' in
  \emph{Circuits and Systems (ISCAS), 2015 IEEE International Symposium
  on}.\hskip 1em plus 0.5em minus 0.4em\relax IEEE, 2015, pp. 581--584.

\bibitem{5719144}
G.~Khodabandehloo, M.~Mirhassani, and M.~Ahmadi, ``Analog implementation of a
  novel resistive-type sigmoidal neuron,'' \emph{IEEE Transactions on Very
  Large Scale Integration (VLSI) Systems}, vol.~20, no.~4, pp. 750--754, April
  2012.

\bibitem{6232461}
S.~P. Adhikari, C.~Yang, H.~Kim, and L.~O. Chua, ``Memristor bridge
  synapse-based neural network and its learning,'' \emph{IEEE Transactions on
  Neural Networks and Learning Systems}, vol.~23, no.~9, pp. 1426--1435, Sept
  2012.

\bibitem{al2015inherently}
M.~Al-Shedivat, R.~Naous, E.~Neftci, G.~Cauwenberghs, and K.~N. Salama,
  ``Inherently stochastic spiking neurons for probabilistic neural
  computation,'' in \emph{Neural Engineering (NER), 2015 7th International
  IEEE/EMBS Conference on}.\hskip 1em plus 0.5em minus 0.4em\relax IEEE, 2015,
  pp. 356--359.

\bibitem{7117477}
M.~Al-Shedivat, R.~Naous, G.~Cauwenberghs, and K.~N. Salama, ``Memristors
  empower spiking neurons with stochasticity,'' \emph{IEEE Journal on Emerging
  and Selected Topics in Circuits and Systems}, vol.~5, no.~2, pp. 242--253,
  June 2015.

\bibitem{querlioz2011simulation}
D.~Querlioz, O.~Bichler, and C.~Gamrat, ``Simulation of a memristor-based
  spiking neural network immune to device variations,'' in \emph{Neural
  Networks (IJCNN), The 2011 International Joint Conference on}.\hskip 1em plus
  0.5em minus 0.4em\relax IEEE, 2011, pp. 1775--1781.

\bibitem{fedorova2016htm}
A.~P. James, I.~Fedorova, T.~Ibrayev, and D.~Kudithipudi, ``Htm spatial pooler
  with memristor crossbar circuits for sparse biometric recognition,''
  \emph{IEEE Transactions on Biomedical Circuits and Systems}, vol.~PP, no.~99,
  pp. 1--12, 2017.

\bibitem{fan2016hierarchical}
D.~Fan, M.~Sharad, A.~Sengupta, and K.~Roy, ``Hierarchical temporal memory
  based on spin-neurons and resistive memory for energy-efficient
  brain-inspired computing,'' \emph{IEEE transactions on neural networks and
  learning systems}, vol.~27, no.~9, pp. 1907--1919, 2016.

\bibitem{krestinskaya2018approximate}
O.~Krestinskaya and A.~P. James, ``Approximate probabilistic neural networks
  with gated threshold logic,'' \emph{arXiv preprint arXiv:1808.00733}, 2018.

\bibitem{serb2016unsupervised}
A.~Serb, J.~Bill, A.~Khiat, R.~Berdan, R.~Legenstein, and T.~Prodromakis,
  ``Unsupervised learning in probabilistic neural networks with multi-state
  metal-oxide memristive synapses,'' \emph{Nature communications}, vol.~7, p.
  12611, 2016.

\bibitem{krestinskaya2018binary}
O.~Krestinskaya and A.~P. James, ``Binary weighted memristive analog deep
  neural network for near-sensor edge processing,'' \emph{arXiv preprint
  arXiv:1808.00737}, 2018.

\bibitem{soudry2013hebbian}
D.~Soudry, D.~Di~Castro, A.~Gal, A.~Kolodny, and S.~Kvatinsky, ``Hebbian
  learning rules with memristors,'' \emph{Israel Institute of Technology:
  Haifa, Israel}, 2013.

\bibitem{47}
R.~Senthilkumar and R.~K. Gnanamurthy, ``A detailed survey on 2d and 3d still
  face and face video databases part i,'' in \emph{Communications and Signal
  Processing (ICCSP), 2014 International Conference on}, April 2014, pp.
  1405--1409.

\bibitem{li2018efficient}
C.~Li, D.~Belkin, Y.~Li, P.~Yan, M.~Hu, N.~Ge, H.~Jiang, E.~Montgomery, P.~Lin,
  Z.~Wang \emph{et~al.}, ``Efficient and self-adaptive in-situ learning in
  multilayer memristor neural networks,'' \emph{Nature Communications}, vol.~9,
  no.~1, p. 2385, 2018.

\bibitem{wijesinghe2017all}
P.~Wijesinghe, A.~Ankit, A.~Sengupta, and K.~Roy, ``An all-memristor deep
  spiking neural computing system: A step toward realizing the low-power
  stochastic brain,'' \emph{IEEE Transactions on Emerging Topics in
  Computational Intelligence}, vol.~2, no.~5, pp. 345--358, Oct 2018.

\bibitem{6861426}
S.~Duan, X.~Hu, Z.~Dong, L.~Wang, and P.~Mazumder, ``Memristor-based cellular
  nonlinear/neural network: Design, analysis, and applications,'' \emph{IEEE
  Transactions on Neural Networks and Learning Systems}, vol.~26, no.~6, pp.
  1202--1213, June 2015.

\bibitem{7600}
L.~O. Chua and L.~Yang, ``Cellular neural networks: theory,'' \emph{IEEE
  Transactions on Circuits and Systems}, vol.~35, no.~10, pp. 1257--1272, Oct
  1988.

\bibitem{7168838}
S.~N. Truong, S.~Shin, J.~Song, H.~S. Mo, F.~Corinto, and K.~S. Min,
  ``Memristor-based cellular nanoscale networks: Theory, circuits, and
  applications,'' in \emph{2015 IEEE International Symposium on Circuits and
  Systems (ISCAS)}, May 2015, pp. 1134--1137.

\bibitem{7469884}
X.~Hu, G.~Feng, S.~Duan, and L.~Liu, ``A memristive multilayer cellular neural
  network with applications to image processing,'' \emph{IEEE Transactions on
  Neural Networks and Learning Systems}, vol.~28, no.~8, pp. 1889--1901, Aug
  2017.

\bibitem{di2017memristor}
M.~Di~Marco, M.~Forti, and L.~Pancioni, ``Memristor standard cellular neural
  networks computing in the flux-charge domain,'' \emph{Neural Networks}, 2017.

\bibitem{8016501}
W.~Rawat and Z.~Wang, ``Deep convolutional neural networks for image
  classification: A comprehensive review,'' \emph{Neural Computation}, vol.~29,
  no.~9, pp. 2352--2449, Sept 2017.

\bibitem{8103129}
M.~T. McCann, K.~H. Jin, and M.~Unser, ``Convolutional neural networks for
  inverse problems in imaging: A review,'' \emph{IEEE Signal Processing
  Magazine}, vol.~34, no.~6, pp. 85--95, Nov 2017.

\bibitem{7444187}
A.~Kappeler, S.~Yoo, Q.~Dai, and A.~K. Katsaggelos, ``Video super-resolution
  with convolutional neural networks,'' \emph{IEEE Transactions on
  Computational Imaging}, vol.~2, no.~2, pp. 109--122, June 2016.

\bibitem{7544367}
L.~Xia, T.~Tang, W.~Huangfu, M.~Cheng, X.~Yin, B.~Li, Y.~Wang, and H.~Yang,
  ``Switched by input: Power efficient structure for rram-based convolutional
  neural network,'' in \emph{2016 53nd ACM/EDAC/IEEE Design Automation
  Conference (DAC)}, June 2016, pp. 1--6.

\bibitem{7858419}
T.~Tang, L.~Xia, B.~Li, Y.~Wang, and H.~Yang, ``Binary convolutional neural
  network on rram,'' in \emph{2017 22nd Asia and South Pacific Design
  Automation Conference (ASP-DAC)}, Jan 2017, pp. 782--787.

\bibitem{7551379}
A.~Shafiee, A.~Nag, N.~Muralimanohar, R.~Balasubramonian, J.~P. Strachan,
  M.~Hu, R.~S. Williams, and V.~Srikumar, ``Isaac: A convolutional neural
  network accelerator with in-situ analog arithmetic in crossbars,'' in
  \emph{2016 ACM/IEEE 43rd Annual International Symposium on Computer
  Architecture (ISCA)}, June 2016, pp. 14--26.

\bibitem{8009177}
L.~Ni, Z.~Liu, W.~Song, J.~J. Yang, H.~Yu, K.~Wang, and Y.~Wang, ``An
  energy-efficient and high-throughput bitwise cnn on sneak-path-free digital
  reram crossbar,'' in \emph{2017 IEEE/ACM International Symposium on Low Power
  Electronics and Design (ISLPED)}, July 2017, pp. 1--6.

\bibitem{7911265}
L.~Ni, Z.~Liu, H.~Yu, and R.~V. Joshi, ``An energy-efficient digital
  reram-crossbar-based cnn with bitwise parallelism,'' \emph{IEEE Journal on
  Exploratory Solid-State Computational Devices and Circuits}, vol.~3, pp.
  37--46, Dec 2017.

\bibitem{7280813}
C.~Yakopcic, R.~Hasan, and T.~M. Taha, ``Memristor based neuromorphic circuit
  for ex-situ training of multi-layer neural network algorithms,'' in
  \emph{2015 International Joint Conference on Neural Networks (IJCNN)}, July
  2015, pp. 1--7.

\bibitem{li2018analogue}
C.~Li, M.~Hu, Y.~Li, H.~Jiang, N.~Ge, E.~Montgomery, J.~Zhang, W.~Song,
  N.~D{\'a}vila, C.~E. Graves \emph{et~al.}, ``Analogue signal and image
  processing with large memristor crossbars,'' \emph{Nature Electronics},
  vol.~1, no.~1, p.~52, 2018.

\bibitem{wang2018handwritten}
J.~Wang, S.~Hu, X.~Zhan, Q.~Yu, Z.~Liu, T.~P. Chen, Y.~Yin, S.~Hosaka, and
  Y.~Liu, ``Handwritten-digit recognition by hybrid convolutional neural
  network based on hfo 2 memristive spiking-neuron,'' \emph{Scientific
  reports}, vol.~8, no.~1, p. 12546, 2018.

\bibitem{6518253}
T.~Serrano-Gotarredona, T.~Prodromakis, and B.~Linares-Barranco, ``A proposal
  for hybrid memristor-cmos spiking neuromorphic learning systems,'' \emph{IEEE
  Circuits and Systems Magazine}, vol.~13, no.~2, pp. 74--88, Secondquarter
  2013.

\bibitem{7527393}
H.~Mostafa, C.~Mayr, and G.~Indiveri, ``Beyond spike-timing dependent
  plasticity in memristor crossbar arrays,'' in \emph{2016 IEEE International
  Symposium on Circuits and Systems (ISCAS)}, May 2016, pp. 926--929.

\bibitem{7527253}
E.~Covi, S.~Brivio, A.~Serb, T.~Prodromakis, M.~Fanciulli, and S.~Spiga,
  ``Hfo2-based memristors for neuromorphic applications,'' in \emph{2016 IEEE
  International Symposium on Circuits and Systems (ISCAS)}, May 2016, pp.
  393--396.

\bibitem{7904675}
N.~Panwar, B.~Rajendran, and U.~Ganguly, ``Arbitrary spike time dependent
  plasticity (stdp) in memristor by analog waveform engineering,'' \emph{IEEE
  Electron Device Letters}, vol.~38, no.~6, pp. 740--743, June 2017.

\bibitem{6093706}
I.~E. Ebong and P.~Mazumder, ``Cmos and memristor-based neural network design
  for position detection,'' \emph{Proceedings of the IEEE}, vol. 100, no.~6,
  pp. 2050--2060, June 2012.

\bibitem{wang2018capacitive}
Z.~Wang, M.~Rao, J.-W. Han, J.~Zhang, P.~Lin, Y.~Li, C.~Li, W.~Song, S.~Asapu,
  R.~Midya \emph{et~al.}, ``Capacitive neural network with neuro-transistors,''
  \emph{Nature communications}, vol.~9, no.~1, p. 3208, 2018.

\bibitem{deng2015complex}
L.~Deng, G.~Li, N.~Deng, D.~Wang, Z.~Zhang, W.~He, H.~Li, J.~Pei, and L.~Shi,
  ``Complex learning in bio-plausible memristive networks,'' \emph{Scientific
  reports}, vol.~5, p. 10684, 2015.

\bibitem{557671}
S.-W. Lee and H.-H. Song, ``A new recurrent neural-network architecture for
  visual pattern recognition,'' \emph{IEEE Transactions on Neural Networks},
  vol.~8, no.~2, pp. 331--340, Mar 1997.

\bibitem{bao2018region}
G.~Bao, Z.~Zeng, and Y.~Shen, ``Region stability analysis and tracking control
  of memristive recurrent neural network,'' \emph{Neural Networks}, vol.~98,
  pp. 51--58, 2018.

\bibitem{7553910}
H.~Liu, Z.~Wang, and B.~Shen, ``Discrete-time memristive recurrent neural
  networks with time-varying delays: Exponential stability analysis,'' in
  \emph{2016 35th Chinese Control Conference (CCC)}, July 2016, pp. 3584--3589.

\bibitem{xavier2018memristive}
G.~M.~T. Xavier, F.~G. Casta{\~n}eda, L.~M.~F. Nava, and J.~A.~M. Cadenas,
  ``Memristive recurrent neural network,'' \emph{Neurocomputing}, vol. 273, pp.
  281--295, 2018.

\bibitem{1528541}
Y.~Maeda and M.~Wakamura, ``Simultaneous perturbation learning rule for
  recurrent neural networks and its fpga implementation,'' \emph{IEEE
  Transactions on Neural Networks}, vol.~16, no.~6, pp. 1664--1672, Nov 2005.

\bibitem{smagulova2018design}
K.~Smagulova, K.~Adam, O.~Krestinskaya, and A.~P. James, ``Design of
  cmos-memristor circuits for lstm architecture,'' \emph{arXiv preprint
  arXiv:1806.02366}, 2018.

\bibitem{adam2018memristive}
K.~Adam, K.~Smagulova, and A.~P. James, ``Memristive lstm network hardware
  architecture for time-series predictive modeling problem,'' \emph{arXiv
  preprint arXiv:1809.03119}, 2018.

\bibitem{james2018introduction}
A.~James, T.~Ibrayev, O.~Krestinskaya, and I.~Dolzhikova, ``Introduction to
  memristive htm circuits,'' in \emph{Memristor and Memristive Neural
  Networks}.\hskip 1em plus 0.5em minus 0.4em\relax InTech, 2018.

\bibitem{martinez1998ar}
A.~Mart{\i}nez and R.~Benavente, ``The ar face database,'' \emph{Rapport
  technique}, vol.~24, 1998.

\bibitem{garofolo1993darpa}
J.~S. Garofolo, L.~F. Lamel, W.~M. Fisher, J.~G. Fiscus, and D.~S. Pallett,
  ``Darpa timit acoustic-phonetic continous speech corpus cd-rom. nist speech
  disc 1-1.1,'' \emph{NASA STI/Recon Technical Report N}, vol.~93, 1993.

\bibitem{irmanova2018neuron}
A.~Irmanova and A.~P. James, ``Neuron inspired data encoding memristive
  multi-level memory cell,'' \emph{Analog Integrated Circuits and Signal
  Processing}, pp. 1--6, 2018.

\bibitem{krestinskaya2018feature}
O.~Krestinskaya and A.~P. James, ``Feature extraction without learning in an
  analog spatial pooler memristive-cmos circuit design of hierarchical temporal
  memory,'' \emph{Analog Integrated Circuits and Signal Processing}, pp. 1--9,
  2018.

\bibitem{ibrayev2017chip}
T.~Ibrayev, U.~Myrzakhan, O.~Krestinskaya, A.~Irmanova, and A.~P. James,
  ``On-chip face recognition system design with memristive hierarchical
  temporal memory,'' \emph{arXiv preprint arXiv:1709.08184}, 2017.

\bibitem{georgia}
A.~James, T.~Ibrayev, and O.~Krestinskaya, ``Design and implication of a rule
  based weight sparsity module in htm spatial pooler,'' in \emph{Electronics ,
  Circuits and Systems (ICECS), 2017 24th IEEE International}.\hskip 1em plus
  0.5em minus 0.4em\relax IEEE, 2017.

\bibitem{xiao2018gst}
S.~Xiao, X.~Xie, S.~Wen, Z.~Zeng, T.~Huang, and J.~Jiang, ``Gst-memristor-based
  online learning neural networks,'' \emph{Neurocomputing}, vol. 272, pp.
  677--682, 2018.

\bibitem{8329223}
N.~Dastanova, S.~Duisenbay, O.~Krestinskaya, and A.~P. James, ``Bit-plane
  extracted moving-object detection using memristive crossbar-cam arrays for
  edge computing image devices,'' \emph{IEEE Access}, vol.~6, pp.
  18\,954--18\,966, 2018.

\bibitem{mikhailenko2018m}
D.~Mikhailenko, C.~Liyanagedera, A.~P. James, and K.~Roy, ``M 2 ca: Modular
  memristive crossbar arrays,'' in \emph{Circuits and Systems (ISCAS), 2018
  IEEE International Symposium on}.\hskip 1em plus 0.5em minus 0.4em\relax
  IEEE, 2018, pp. 1--5.

\bibitem{gubbi2013internet}
J.~Gubbi, R.~Buyya, S.~Marusic, and M.~Palaniswami, ``Internet of things (iot):
  A vision, architectural elements, and future directions,'' \emph{Future
  generation computer systems}, vol.~29, no.~7, pp. 1645--1660, 2013.

\bibitem{iot2017}
J.~Lin, W.~Yu, N.~Zhang, X.~Yang, H.~Zhang, and W.~Zhao, ``A survey on internet
  of things: Architecture, enabling technologies, security and privacy, and
  applications,'' \emph{IEEE Internet of Things Journal}, vol.~4, no.~5, pp.
  1125--1142, Oct 2017.

\bibitem{5471063}
K.~Eshraghian, K.~R. Cho, O.~Kavehei, S.~K. Kang, D.~Abbott, and S.~M.~S. Kang,
  ``Memristor mos content addressable memory (mcam): Hybrid architecture for
  future high performance search engines,'' \emph{IEEE Transactions on Very
  Large Scale Integration (VLSI) Systems}, vol.~19, no.~8, pp. 1407--1417, Aug
  2011.

\bibitem{jiang2016sub}
H.~Jiang, L.~Han, P.~Lin, Z.~Wang, M.~H. Jang, Q.~Wu, M.~Barnell, J.~J. Yang,
  H.~L. Xin, and Q.~Xia, ``Sub-10 nm ta channel responsible for superior
  performance of a hfo 2 memristor,'' \emph{Scientific reports}, vol.~6, p.
  28525, 2016.

\bibitem{7024182}
I.~Vourkas, D.~Stathis, G.~C. Sirakoulis, and S.~Hamdioui, ``Alternative
  architectures toward reliable memristive crossbar memories,'' \emph{IEEE
  Transactions on Very Large Scale Integration (VLSI) Systems}, vol.~24, no.~1,
  pp. 206--217, Jan 2016.

\bibitem{chen2018cmos}
L.~Chen, Z.-Y. He, T.-Y. Wang, Y.-W. Dai, H.~Zhu, Q.-Q. Sun, and D.~W. Zhang,
  ``Cmos compatible bio-realistic implementation with ag/hfo2-based synaptic
  nanoelectronics for artificial neuromorphic system,'' \emph{Electronics},
  vol.~7, no.~6, p.~80, 2018.

\bibitem{kim2017analog}
S.~Kim, H.~Kim, S.~Hwang, M.-H. Kim, Y.-F. Chang, and B.-G. Park, ``Analog
  synaptic behavior of a silicon nitride memristor,'' \emph{ACS applied
  materials \& interfaces}, vol.~9, no.~46, pp. 40\,420--40\,427, 2017.

\bibitem{6968169}
T.~Cabaret, L.~Fillaud, B.~Jousselme, J.~O. Klein, and V.~Derycke,
  ``Electro-grafted organic memristors: Properties and prospects for artificial
  neural networks based on stdp,'' in \emph{14th IEEE International Conference
  on Nanotechnology}, Aug 2014, pp. 499--504.

\bibitem{chang2016demonstration}
Y.-F. Chang, B.~Fowler, Y.-C. Chen, F.~Zhou, C.-H. Pan, T.-C. Chang, and J.~C.
  Lee, ``Demonstration of synaptic behaviors and resistive switching
  characterizations by proton exchange reactions in silicon oxide,''
  \emph{Scientific reports}, vol.~6, p. 21268, 2016.

\bibitem{alibart2012high}
F.~Alibart, L.~Gao, B.~D. Hoskins, and D.~B. Strukov, ``High precision tuning
  of state for memristive devices by adaptable variation-tolerant algorithm,''
  \emph{Nanotechnology}, vol.~23, no.~7, p. 075201, 2012.

\bibitem{naous2016stochasticity}
R.~Naous, M.~Al-Shedivat, and K.~N. Salama, ``Stochasticity modeling in
  memristors,'' \emph{IEEE Transactions on Nanotechnology}, vol.~15, no.~1, pp.
  15--28, 2016.

\bibitem{yang2010high}
J.~J. Yang, M.-X. Zhang, J.~P. Strachan, F.~Miao, M.~D. Pickett, R.~D. Kelley,
  G.~Medeiros-Ribeiro, and R.~S. Williams, ``High switching endurance in tao x
  memristive devices,'' \emph{Applied Physics Letters}, vol.~97, no.~23, p.
  232102, 2010.

\bibitem{amat2018memristive}
E.~Amat, A.~Rubio \emph{et~al.}, ``Memristive crossbar memory lifetime
  evaluation and reconfiguration strategies,'' \emph{IEEE Transactions on
  Emerging Topics in Computing}, vol.~6, no.~2, pp. 207--218, 2018.

\bibitem{fantini2013intrinsic}
A.~Fantini, L.~Goux, R.~Degraeve, D.~Wouters, N.~Raghavan, G.~Kar, A.~Belmonte,
  Y.-Y. Chen, B.~Govoreanu, and M.~Jurczak, ``Intrinsic switching variability
  in hfo 2 rram,'' in \emph{Memory Workshop (IMW), 2013 5th IEEE
  International}.\hskip 1em plus 0.5em minus 0.4em\relax IEEE, 2013, pp.
  30--33.

\bibitem{kim2016voltage}
K.~M. Kim, J.~J. Yang, J.~P. Strachan, E.~M. Grafals, N.~Ge, N.~D. Melendez,
  Z.~Li, and R.~S. Williams, ``Voltage divider effect for the improvement of
  variability and endurance of tao x memristor,'' \emph{Scientific reports},
  vol.~6, p. 20085, 2016.

\bibitem{jo2010nanoscale}
S.~H. Jo, T.~Chang, I.~Ebong, B.~B. Bhadviya, P.~Mazumder, and W.~Lu,
  ``Nanoscale memristor device as synapse in neuromorphic systems,'' \emph{Nano
  letters}, vol.~10, no.~4, pp. 1297--1301, 2010.

\bibitem{beolknowm}
\BIBentryALTinterwordspacing
K.~Inc., ``Knowm memristors,'' Tech. Rep., 2015. [Online]. Available:
  \url{https:\/\/knowm.org\/downloads\/Knowm\_Memristors.pdf}
\BIBentrySTDinterwordspacing

\bibitem{xie2015interconnect}
L.~Xie, H.~A. Du~Nguyen, M.~Taouil, S.~Hamdioui, and K.~Bertels, ``Interconnect
  networks for memristor crossbar,'' in \emph{Nanoscale Architectures
  (NANOARCH), 2015 IEEE/ACM International Symposium on}.\hskip 1em plus 0.5em
  minus 0.4em\relax IEEE, 2015, pp. 124--129.

\bibitem{du2017interconnect}
H.~A. Du~Nguyen, L.~Xie, J.~Yu, M.~Taouil, and S.~Hamdioui, ``Interconnect
  networks for resistive computing architectures,'' in \emph{2017 12th
  International Conference on Design \& Technology of Integrated Systems In
  Nanoscale Era (DTIS)}.\hskip 1em plus 0.5em minus 0.4em\relax IEEE, 2017, pp.
  1--6.

\bibitem{mazumder2012memristors}
P.~Mazumder, S.-M. Kang, and R.~Waser, ``Memristors: devices, models, and
  applications,'' \emph{Proceedings of the IEEE}, vol. 100, no.~6, pp.
  1911--1919, 2012.

\bibitem{bayat2015phenomenological}
F.~M. Bayat, B.~Hoskins, and D.~B. Strukov, ``Phenomenological modeling of
  memristive devices,'' \emph{Applied Physics A}, vol. 118, no.~3, pp.
  779--786, 2015.

\bibitem{biolek2018modeling}
D.~Biolek, Z.~Kolka, V.~Biolkov{\'a}, Z.~Biolek, M.~Potrebi{\'c}, and
  D.~To{\v{s}}i{\'c}, ``Modeling and simulation of large memristive networks,''
  \emph{International Journal of Circuit Theory and Applications}, vol.~46,
  no.~1, pp. 50--65, 2018.

\bibitem{singh2018comparative}
J.~Singh and B.~Raj, ``Comparative analysis of memristor models and memories
  design,'' 2018.

\bibitem{pickett2009switching}
M.~D. Pickett, D.~B. Strukov, J.~L. Borghetti, J.~J. Yang, G.~S. Snider, D.~R.
  Stewart, and R.~S. Williams, ``Switching dynamics in titanium dioxide
  memristive devices,'' \emph{Journal of Applied Physics}, vol. 106, no.~7, p.
  074508, 2009.

\bibitem{strukov2008missing}
D.~B. Strukov, G.~S. Snider, D.~R. Stewart, and R.~S. Williams, ``The missing
  memristor found,'' \emph{nature}, vol. 453, no. 7191, p.~80, 2008.

\bibitem{prodromakis2011versatile}
T.~Prodromakis, B.~P. Peh, C.~Papavassiliou, and C.~Toumazou, ``A versatile
  memristor model with nonlinear dopant kinetics,'' \emph{IEEE transactions on
  electron devices}, vol.~58, no.~9, pp. 3099--3105, 2011.

\bibitem{joglekar2009elusive}
Y.~N. Joglekar and S.~J. Wolf, ``The elusive memristor: properties of basic
  electrical circuits,'' \emph{European Journal of Physics}, vol.~30, no.~4, p.
  661, 2009.

\bibitem{biolek2009spice}
Z.~Biolek, D.~Biolek, and V.~Biolkova, ``Spice model of memristor with
  nonlinear dopant drift.'' \emph{Radioengineering}, vol.~18, no.~2, 2009.

\bibitem{6353604}
S.~Kvatinsky, E.~G. Friedman, A.~Kolodny, and U.~C. Weiser, ``Team: Threshold
  adaptive memristor model,'' \emph{IEEE Transactions on Circuits and Systems
  I: Regular Papers}, vol.~60, no.~1, pp. 211--221, Jan 2013.

\bibitem{biol}
D.~Biolek, Z.~Kolka, V.~Biolkova, and Z.~Biolek, ``Memristor models for spice
  simulation of extremely large memristive networks,'' in \emph{2016 IEEE
  International Symposium on Circuits and Systems (ISCAS)}, May 2016, pp.
  389--392.

\bibitem{messaris2018data}
I.~Messaris, A.~Serb, S.~Stathopoulos, A.~Khiat, S.~Nikolaidis, and
  T.~Prodromakis, ``A data-driven verilog-a reram model,'' \emph{IEEE
  Transactions on Computer-Aided Design of Integrated Circuits and Systems},
  2018.

\bibitem{12}
L.~Chua, ``Memristor-the missing circuit element,'' \emph{IEEE Transactions on
  Circuit Theory}, vol.~18, no.~5, pp. 507--519, Sep 1971.

\bibitem{batas2011memristor}
D.~Batas and H.~Fiedler, ``A memristor spice implementation and a new approach
  for magnetic flux-controlled memristor modeling,'' \emph{IEEE Transactions on
  Nanotechnology}, vol.~10, no.~2, pp. 250--255, 2011.

\bibitem{5433753}
Ã.~Rak and G.~Cserey, ``Macromodeling of the memristor in spice,'' \emph{IEEE
  Transactions on Computer-Aided Design of Integrated Circuits and Systems},
  vol.~29, no.~4, pp. 632--636, April 2010.

\bibitem{benderli2009spice}
S.~Benderli and T.~Wey, ``On spice macromodelling of tio 2 memristors,''
  \emph{Electronics letters}, vol.~45, no.~7, pp. 377--379, 2009.

\bibitem{keshmiri2014study}
V.~Keshmiri, ``A study of the memristor models and applications,'' 2014.

\bibitem{5976989}
H.~Kim, M.~P. Sah, C.~Yang, T.~Roska, and L.~O. Chua, ``Neural synaptic
  weighting with a pulse-based memristor circuit,'' \emph{IEEE Transactions on
  Circuits and Systems I: Regular Papers}, vol.~59, no.~1, pp. 148--158, Jan
  2012.

\bibitem{strachan2013state}
J.~P. Strachan, A.~C. Torrezan, F.~Miao, M.~D. Pickett, J.~J. Yang, W.~Yi,
  G.~Medeiros-Ribeiro, and R.~S. Williams, ``State dynamics and modeling of
  tantalum oxide memristors,'' \emph{IEEE Transactions on Electron Devices},
  vol.~60, no.~7, pp. 2194--2202, 2013.

\bibitem{memristormodel}
H.~Abdalla and M.~D. Pickett, ``Spice modeling of memristors,'' in \emph{2011
  IEEE International Symposium of Circuits and Systems (ISCAS)}, May 2011, pp.
  1832--1835.

\bibitem{simmons1963generalized}
J.~G. Simmons, ``Generalized formula for the electric tunnel effect between
  similar electrodes separated by a thin insulating film,'' \emph{Journal of
  applied physics}, vol.~34, no.~6, pp. 1793--1803, 1963.

\end{thebibliography}


%








\end{document}